\documentclass[a4paper,fleqn,usenatbib]{mnras}

\usepackage{mathptmx}
\usepackage[T1]{fontenc}
\usepackage{ae,aecompl}

\usepackage{graphicx}	% Including figure files
\usepackage{amsmath}	% Advanced maths commands
\usepackage{amssymb}	% Extra maths symbols

\title[Spectral properties of 441 radio pulsars]{Spectral properties of 441 radio pulsars}

\author[F.~Jankowski et al.]{F.~Jankowski,$^{1,2}$\thanks{E-mail: fjankowsk@gmail.com} W.~van~Straten,$^{1,2,3}$ E.~F.~Keane,$^{4,1,2}$ M.~Bailes,$^{1,2}$ E.~Barr,$^{5,1}$\newauthor
S.~Johnston$^{6}$ and M.~Kerr$^{7}$
\\
$^{1}$Centre for Astrophysics and Supercomputing, Swinburne University of Technology, PO Box 218, Hawthorn, VIC 3122, Australia\\
$^{2}$ARC Centre of Excellence for All-Sky Astrophysics (CAASTRO)\\
$^{3}$Institute for Radio Astronomy \& Space Research, Auckland University of Technology, Private Bag 92006, Auckland 1142, New Zealand\\
$^{4}$SKA Organisation, Jodrell Bank Observatory, Cheshire, SK11 9DL, UK\\
$^{5}$Max-Planck-Institut f\"ur Radioastronomie, Auf dem H\"ugel 69, D-53121 Bonn, Germany\\
$^{6}$CSIRO Astronomy and Space Science, Australia Telescope National Facility, Box 76 Epping, NSW, 1710, Australia\\
$^{7}$Space Science Division, Naval Research Laboratory, Washington, DC 20375-5352, USA}

\date{Accepted XXX. Received YYY; in original form ZZZ}

\pubyear{2017}

\begin{document}
\label{firstpage}
\pagerange{\pageref{firstpage}--\pageref{lastpage}}
\maketitle

%\begin{abstract}
%This is a simple template for authors to write new MNRAS papers.
%The abstract should briefly describe the aims, methods, and main results of the paper.
%It should be a single paragraph not more than 250 words (200 words for Letters).
%No references should appear in the abstract.
%\end{abstract}

\begin{abstract}
We present a study of the spectral properties of 441 pulsars observed with the Parkes radio telescope near the centre frequencies of 728, 1382 and 3100~MHz. The observations at 728 and 3100~MHz were conducted simultaneously using the dual-band 10-50cm receiver. These high-sensitivity, multi-frequency observations provide a systematic and uniform sample of pulsar flux densities. We combine our measurements with spectral data from the literature in order to derive the spectral properties of these pulsars. Using techniques from robust regression and information theory we classify the observed spectra in an objective, robust and unbiased way into five morphological classes: simple or broken power law, power law with either low or high-frequency cut-off and log-parabolic spectrum. While about $79 \%$ of the pulsars that could be classified have simple power law spectra, we find significant deviations in 73 pulsars, 35 of which have curved spectra, 25 with a spectral break and 10 with a low-frequency turn-over. We identify 11 gigahertz-peaked spectrum (GPS) pulsars, with 3 newly identified in this work and 8 confirmations of known GPS pulsars; 3 others show tentative evidence of GPS, but require further low-frequency measurements to support this classification. The weighted mean spectral index of all pulsars with simple power law spectra is $-1.60 \pm 0.03$. The observed spectral indices are well described by a shifted log-normal distribution. The strongest correlations of spectral index are with spin-down luminosity, magnetic field at the light-cylinder and spin-down rate. We also investigate the physical origin of the observed spectral features and determine emission altitudes for three pulsars.
\end{abstract}


%%%%%%%%%%%%%%%%%%%%%%%%%%%%%%%%%%%%%%%%%%%%%%%%%%

%%%%%%%%%%%%%%%%% BODY OF PAPER %%%%%%%%%%%%%%%%%%

\section{Introduction}
\label{sec:Introduction}

Although pulsars were discovered nearly 50 years ago \citep{1968Hewish}, the exact mechanism by which they emit electro-magnetic radiation is far from being understood. The pulsar emission is often described using models that include a magnetosphere filled with an electron-positron plasma that co-rotates with the pulsar \citep{1969Goldreich}. However, important details such as the location of the emission sites, including the polar cap close above the stellar surface \citep{1971Sturrock, 1975Ruderman, 2004Baring}, the slot gap \citep{1979Arons}, or the outer gap \citep{1986Cheng} and the exact emission mechanism are still unclear. Potential mechanisms for the radio emission are coherent synchrotron-curvature radiation of particle bunches \citep{1975Ruderman, 2004Gil} and a form of plasma emission, apart from more exotic models \citep{2016Melrose}. Studying the radio spectra of pulsars could aid in developing an understanding of the emission mechanism, but accurate spectral data are sadly lacking for the majority of pulsars. The exception are a few pulsars that have been studied across a broader frequency range (e.g. \citealt{2011Keith, 2015Dai}). A recent version of the ATNF pulsar catalogue, version 1.54 \citep{2005Manchester}, shows that pulsar flux densities are relatively well known near 1.4~GHz and also 400~MHz, where most of the pulsars were discovered, but are poorly known at other radio frequencies. Out of the 2536 currently known pulsars, about $66 \: \%$ have a recorded flux density measurement at 1.4~GHz; above 2~GHz, the fraction is only about $6 \: \%$; and, between 600 to 900~MHz, there are measurements for only $22 \: \%$ of the pulsars. The low frequency fluxes near 400~MHz are better known with about $30 \: \%$ of the pulsars having measurements and the frequencies below that are subject of current study by low frequency projects such as the LWA, LOFAR and MWA \citep{2015Stovall, 2016Bilous, 2016Bell}. The problem is not only that little data are available at intermediate and high ($\geq$ 2~GHz) frequencies, but also that the data in the catalogue were taken with different telescopes and different generations of receivers and backends, each with their own systematic errors. In addition, a large fraction of about $61 \: \%$ at 1.4~GHz and nearly $80 \: \%$ at 800~MHz of these flux density measurements are from the original discovery papers, which are typically not absolute flux density calibrated and simply estimates using the signal-to-noise ratio (S/N) of the detection converted into a flux density using the radiometer equation and supposedly known parameters of the observing system and sky temperature in that direction. That is done because observations of a calibrator source and the noise diode before each observation add a significant amount of overhead time to a survey. It was found that there can be significant differences between multiple measurements done this way (e.g. \citealt{2013Levin}). It is therefore crucial to obtain absolute flux density calibrated measurements.

Efforts to measure the radio spectra of a large number of pulsars began in earnest with \citet{1973Sieber}, followed by \citet{1980Malofeev} and \citet{1981Izvekova} at low frequencies near 100~MHz and below. It was found that the majority of pulsars have steep spectra that can be described using a simple power law with spectral index $\alpha$. There was also a realisation that some pulsar spectra deviate from this at low frequencies and show a turn-over, which is usually attributed to either synchrotron self-absorption or thermal free-free absorption, while some show a high-frequency cut-off in the form of a spectral steepening or a break in the spectrum \citep{1973Sieber}. For example \citet{1995Lorimer} studied 280 pulsars and analysed their spectra, finding a mean spectral index of $-1.6$. \citet{1996Xilouris} summarised the frequency dependence of various parameters which characterise the pulsar emission, such as the shape of the spectrum, the modulation index, which describes the temporal variability in flux density, the separation of pulse components, the pulse width and the radius-to-frequency mapping, among others. They found a mean spectral index of $-1.87$ for simple power law spectra and $-1.55$ before and $-2.72$ after a spectral break at 1~GHz for broken power law spectra. Later \citet{2000Maron} extended the work of \citet{1995Lorimer} using mainly the same set of pulsars, but broadened the frequency coverage. They derived a mean spectral index of $-1.8$ and realised that only $10 \%$ of the pulsars studied show broken power law spectra. \citet{2013Bates} showed that the intrinsic spectral index distributions for three different pulsar surveys are Gaussian with a mean index of $-1.4$.

Spectral analysis also led to the recent discovery of a new class of pulsars that have gigahertz-peaked spectra (GPS): convex spectra where the flux density maximum occurs at a frequency around 1 GHz \citep{2007Kijak, 2011Kijak, 2015Dembska, 2017Kijak}. It has been proposed that the GPS phenomenon occurs because of thermal free-free absorption in dense ionized material surrounding a pulsar, for example in a pulsar wind nebula, molecular cloud, or supernova remnant \citep{2015Lewandowski, 2016Rajwade}, or when the radio emission from a pulsar passes through the dense wind of a binary companion. In the case of the pulsar J1302--6350 (B1259--63), the only known radio pulsar with a main-sequence Be star companion, the pulsed flux density and its spectrum was found to vary with orbital phase \citep{1996Johnston, 2011KijakBinary, 2015DembskaBinary}. There is considerable discussion about the spectral shape exhibited by GPS pulsars and their spectra have been modelled by broken power law, log-parabolic \citep{2013Bates, 2014Dembska} and free-free absorption models by different authors. Throughout the paper we define a GPS as any spectrum that peaks at a frequency near 1~GHz, regardless of spectral shape.

A major difficulty is that the observed flux densities of pulsars can vary significantly over time because of a combination of intrinsic fluctuations in the pulsar emission and strong diffractive (DISS) and refractive interstellar scintillation (RISS). DISS is the constructive and destructive interference of radio waves emitted by a pulsar at the observer's location, which have undergone scattering in the turbulent interstellar medium. The interference pattern changes over time because of the relative motion between the pulsar, scattering screen and observer and gives rise to strong fluctuations in observed flux density with a timescale of minutes to hours (e.g. \citealt{1995Armstrong}). RISS is responsible for weak long-term fluctuations in observed flux density when the effect of DISS is accounted for. It occurs because of focussing and defocussing of the pulsar emission by the scattering medium and has a timescale of days to months \citep{1982Sieber, 1986Romani, 1990Rickett, 1999Bhat_1}. Experimentally \citet{2000Stinebring} showed that high dispersion measure (DM) pulsars at large distances have nearly constant observed flux densities over years, indicating that the pulsar emission is stable when individual pulses are integrated for at least a few hours, and diffractive scintillation has been accounted for.

Another motivation to obtain accurate flux density measurements over a wide frequency range and corresponding spectral indices for a large fraction of the pulsar population is in order to make accurate predictions for pulsar surveys and observations with the Square Kilometre Array (SKA) \citep{2015Keane}, SKA path-finders and other radio telescopes, such as MeerKAT, FAST, or the recently refurbished Molonglo Observatory Synthesis Telescope \citep{2017Bailes}. The flux density measurements contribute directly to the design of surveys via pulsar population synthesis (e.g. \citealt{2014Bates}), or to optimise observing strategies.

In this work we concentrate on the stable radio spectrum as it appears after a minimum observation length of a few minutes. We used the total pulse profile to estimate the mean flux density, i.e. we do not consider the properties of individual pulses or profile components. Unless otherwise stated, we quote all uncertainties at the $1 \sigma$ level. In \S\ref{sec:Observations} we describe the target selection and our observing programme at the Parkes telescope. In \S\ref{sec:Analysis} we describe the data analysis methods including the absolute flux density calibration, flux density extraction and the modelling of the pulsar spectra. In \S\ref{sec:Results} we present our results and show the objective spectral classification and correlation analysis and in \S\ref{sec:Discussion} we discuss these especially in relation to findings from the literature. Finally, in \S\ref{sec:Conclusions} we give our overall conclusions.

\section{Observations}
\label{sec:Observations}

\subsection{Target selection}
\label{sec:TargetSelection}

We selected the target pulsars as follows: We extracted the available flux density data from version 1.54 of the ATNF pulsar catalogue and extrapolated to the centre frequencies 728 and 3100~MHz of the 10-50cm receiver using a simple power law. For pulsars with a single flux density measurement we used the median spectral index of $-1.76$ for the extrapolation. The sky temperature $T_\text{sky}$ at each pulsar position was derived from the 408~MHz all-sky atlas of \citet{1982Haslam} and extrapolated to the centre frequencies using a power law scaling with exponent of $-2.6$ \citep{1987Lawson}. We used the radiometer equation to determine the expected S/N of each pulsar (e.g. \citealt{2012Lorimer}). The parameters of the receivers used at Parkes were taken from the ATNF receiver fleet table\footnote{\url{http://www.parkes.atnf.csiro.au/observing/documentation/user_guide/}}. Pulsars that were extensively studied as part of other Parkes projects such as P456 - Parkes Pulsar Timing Array \citep{2013Manchester} or P574 - Pulsar timing with the Parkes Radio Telescope for the Fermi mission \citep{2010Weltevrede} were not considered, as they already had sufficient data. As one of the aims of the project was to inform the target selection at the refurbished Molonglo radio telescope, we chose only the pulsars from the above set which could be observed at Molonglo with a S/N of at least 20 in 10 minutes.

\subsection{Observing setup}
\label{sec:Observing setup}

We observed at Parkes simultaneously with two bands centred at 728 and 3100 MHz using the 10-50cm receiver and used the CASPSR backend at 728 MHz and DFB4 or DFB3 at 3100 MHz. For the observations with the multi-beam receiver centred at 1382 MHz we also used CASPSR. One reason to use CASPSR was because of its advanced radio frequency interference (RFI) mitigation technique that uses spectral kurtosis to identify and eliminate RFI based on its deviation from a Gaussian distribution in the raw voltage stream \citep{2010Nita}. This is particularly valuable for long period pulsars. Before each pulsar observation we carried out a calibration observation $0.25^\circ$ offset in right ascension from the pulsar position, which corresponds to about 1.1 FWHM central beam widths at 20 cm, lasting 50 seconds. In this observation the noise diode inside the receiver was fired at a frequency of $11.123 \: \text{Hz}$ and the data folded. Afterwards the pulsar observation was performed. Additionally we observed an absolute flux density calibrator, the radio galaxy Hydra A, once every week of observations.

\subsection{Observations}
\label{sec:Observations1}

The P875 project was granted 76 hours in total in semester 2014APR and 2014OCT. We combined the data with long-term pulsar observations from the Parkes project P574 \citep{2010Weltevrede}, especially in order to characterise the effect of strong refractive interstellar scintillation. The set comprises 8.5 years of data beginning in July 2007 with a maximum number of 248 observing epochs. The raw data can be downloaded from the Parkes pulsar data archive \citep{2011Hobbs}.

\section{Analysis}
\label{sec:Analysis}

\subsection{Data analysis and calibration procedure}
\label{sec:CalibrationProcedure}

\begin{table}
\caption{We used a continuum source, the radio galaxy Hydra A (3C218) as absolute flux density reference. Shown are its flux density at 1400 MHz and its spectral index $\alpha$ as computed by different authors over the frequencies $\nu$. Uncertainties are shown where available. $\dagger$This is the two-point spectral index computed between 2.7 and 5 GHz.}
\label{tab:Calibrators}
\centering
\begin{tabular}{llll}
\hline
$S_{1400}$			& $\alpha$					& $\nu$	& reference\\
$[\text{Jy}]$		&							& [MHz]	&\\
\hline
$43.1$				& $-0.910 \pm 0.011$			& 405 -- 10700	& \citet{1977Baars}\\
$45.05 \pm 0.42$		& $-0.88 \pm 0.08^\dagger$	& 2700, 5000		& \citet{1981Kuehr}\\
$44.72 \pm 0.38$		& $-0.915 \pm 0.008$			& 468 -- 8870	& own fit, data from\\
					&							&				& \citet{1981Kuehr}\\
--					& $-0.89 \pm 0.03$			& 330 -- 1415	& \citet{2004Lane}\\
\hline
\end{tabular}
\end{table}

\begin{table}
\caption{Parameters of the data for the projects P875 and P574, where $\nu$ is the centre frequency, band the name of the frequency band, which only roughly corresponds to its centre wavelength, $B$ the bandwidth and $B_\text{flat}$ the bandwidth of the flat part of the band after removal of the band edges. $S_\text{sys}$ and $T_\text{sys}$ are the SEFD and system temperature that we determined from our P875 measurements, see \S\ref{sec:ParkesSystemTemperatureEstimation}}
\label{tab:MatchingGstarAndATNFData}
\centering
\begin{tabular}{llllllll}
\hline
band		& backend	& $\nu$	& $B$	& $B_\text{flat}$	& $S_\text{sys}$	& $T_\text{sys}$\\
		&			& [MHz]	& [MHz]	& [MHz]	& [Jy]	& [K]\\
\hline
P875\\
50cm		& CASPSR		& 728	& 64		& 50		& 64		& 40.7\\
20cm		& CASPSR		& 1382	& 400	& 300	& 35		& 22.3\\
10cm		& DFB4, 3	& 3100	& 1024	& 912	& 52		& 33.1\\
P574\\
50cm		& DFB3		& 732	& 64		& 50		& --		& --\\
20cm		& DFB1 -- 4	& 1369	& 256	& 228	& --		& --\\
10cm		& DFB2 -- 4	& 3094	& 1024	& 912	& --		& --\\
\hline
\end{tabular}
\end{table}

The polarisation and absolute flux density of the observations were calibrated using the \texttt{psrpl}\footnote{\url{http://psrchive.sourceforge.net/manuals/psrpl/}} calibration pipeline, which is based on \texttt{PSRCHIVE} tools \citep{2004Hotan}. For the polarisation calibration we used the ideal feed assumption and for the absolute flux density calibration we used the radio galaxy Hydra A (3C218). While there are small discrepancies in published measurements, which could arise by different pointing positions along the two radio lobes, its flux density at 1.4~GHz and spectral index is well known, see Table~\ref{tab:Calibrators}. We adopted the value reported by \citet{1977Baars}. In the data processing we removed the band edges, where the gain falls off more than $25 \%$. The exact parameters of the data are shown in Table~\ref{tab:MatchingGstarAndATNFData}. We excised RFI in the frequency, time and pulsar rotational phase domains by searching for strong outliers which deviated at least $5 \sigma$ from the smoothed mean values in each of these. In the worst case we zero-weighted corrupted integrations.

\subsection{Flux density extraction}
\label{sec:FluxDensityExtraction}

The flux densities were extracted using a custom program that uses the python bindings to \texttt{PSRCHIVE}. We determined the peak and pulse averaged total flux density from the Stokes I component, for which we used the standard deviation of the baseline fluctuations as uncertainty. We extracted the flux densities for each observation of each pulsar individually and calculated the error-weighted mean and standard deviation after removing non-detections. The large bandwidth of the receivers and the high S/N of the observations allowed us to split them into frequency sub-bands and extract multiple flux density points. We determined the maximum number of sub-bands iteratively for each pulsar and all its observations for a S/N threshold of 10 per sub-band.

\subsection{Flux density fluctuations due to scintillation}
\label{sec:FluxDensityFluctuationsDueToScintillation}

Interstellar scintillation can affect the measured flux densities of pulsars drastically. It is therefore crucial to take its effect into account in order to determine reliable estimates of the pulsar flux densities and their uncertainties. Therefore, we estimate the influence of scintillation on our flux density data in two ways: 1) using a theoretical simulation that incorporates the \texttt{NE2001} galaxy model \citep{2002Cordes} and 2) directly from our flux density time series data. Here we explain our methods, but show the results in \S\ref{sec:FluxDensityVariability}.

\subsubsection{Estimating the influence of scintillation using a theoretical simulation}
\label{sec:EstimatingTheInfluenceTheoretically}

We simulate the influence of strong diffractive and refractive, or weak scintillation on our observed pulsar flux density measurements theoretically using the \texttt{NE2001} galaxy model and usual equations from the literature. We summarise the details of the simulation in appendix \ref{sec:InfluenceOfScinitillationOnObservedPulsarFluxDensities}.

\subsubsection{Flux density variability derived directly from our data}
\label{sec:FluxDensityVariabilityDerivedFromOurData}

We also derive the variability directly from our flux density time series data. For that we use only the pulsars for which we have at least six measurement epochs. We are especially interested in its temporal variability with respect to DM and observing frequency to determine the flux density uncertainty for the pulsars for which we have only a small number of observations. Although our data have been cleaned extensively and we believe that they do not contain any significant errors, we employ analysis methods that allow for a small number of corrupted data points without affecting the overall result. For that purpose we define a robust modulation index as:
\begin{equation}
	m_{\text{r}, \nu} = \frac{ \sigma_{\text{r}, \nu} }{ \text{med} (S_{\nu}) },
	\label{eq:RobustModulationIndex}
\end{equation}
where med is the median computed over all flux density measurements and $\sigma_{\text{r}, \nu}$ is the robust standard deviation computed using the interquartile range (IQR):
\begin{equation}
	\sigma_{\text{r}, \nu} = 0.9183 \: \text{IQR} =  0.9183 \: \left (q_{75} (S_{\nu}) - q_{25} (S_{\nu}) \right),
	\label{eq:InterQuartileRange}
\end{equation}
where $q_{75}$ and $q_{25}$ are the 75 and 25 percentile of the all the flux densities measured for a pulsar at that centre frequency. The factor in front appears because we are assuming a Rician flux density distribution with non-centrality (shape) and scale parameter of unity. This reflects the fact that the flux density amplitude distribution has a long tail due to interstellar scintillation. This is also the reason why the median and mean flux density value can be largely different. We derive it numerically as $1/(q_{75} - q_{25})$. For an exponential distribution with a rate parameter of unity (e.g. \citealt{1998Cordes}), the factor is very similar: $0.9102$. For a standard Gaussian distribution it is $0.7413$. Using the robust modulation index is conceptually similar to using the central $67 \%$ of the data only, which is a standard technique (e.g. \citealt{2000Stinebring}). The increased robustness comes at a cost of slightly increased uncertainties. For our data set we find that the normal and robust modulation index are in good agreement within the uncertainties, which are estimated using the bootstrap resampling method using $10^4$ samples for each pulsar.

\subsubsection{Combining the data}
\label{sec:CombiningTheData}

We derive the combined flux density points from the individual measurements using robust techniques. We define a detection as an observation with S/N of at least 6 after data cleaning. Each flux density point has an uncertainty derived from the baseline fluctuations. We then compute the combined value as the weighted median over all $N$ detections at a certain frequency. Its uncertainty is:
\begin{equation}
	u_S^2 = u_{\text{sys}}^2 +
	\begin{cases}
		\frac{\sigma_{\text{r}, \nu}^2}{N} + \left( \frac{6}{5} \frac{1}{N} - \frac{1}{5} \right) \: u_{\text{scint}}^2 (\text{DM}, \nu)							& \text{if} \ 1 \leq N < 6\\
		\frac{\sigma_{\text{r}, \nu}^2}{N}	& \text{if} \ N \geq 6\\
  \end{cases},
 \label{eq:TotalFluxDensityUncertainty}
\end{equation}
where $u_{\text{sys}}$ is the combined systematic uncertainty arising due to the finite accuracy of the absolute flux density calibration, fluctuations in system temperature and other unknown factors, whose relative value we assume to be $5 \%$. The statistical uncertainty consists of the standard error derived from the robust standard deviation $\sigma_{\text{r}, \nu}$ computed over all measurements at that frequency. We assume a Rician flux density amplitude distribution in all cases. For a single epoch this reduces to the uncertainty of that flux density point. For pulsars with less than six measurement epochs we add to that in quadrature the uncertainty due to scintillation:
\begin{equation}
	u_{\text{scint}} = m_{\text{r},\nu} \left( \text{DM}, \nu \right) \: \bar{S}_\nu,
	\label{eq:Uscint}
\end{equation}
where $m_{\text{r},\nu}$ is the robust modulation index, which in our empirical model is a function of DM and frequency and $\bar{S}_\nu$ is the weighted median flux density. We derive the scaling relationship based on our data in \S\ref{sec:FluxDensityVariability}. We assume that the fluctuations in flux density are adequately represented in the robust standard deviation for a sample size of six or more measurement epochs. We choose this minimum sample size, as the measured width of the distribution approaches the true width of the underlying distribution with a mean square deviation of less than 0.4 \citep{2013Yakovleva}. In addition, we carried out a Monte Carlo simulation with $10^{5}$ realisations that shows that the scale parameter, i.e. the width of the distribution, can be estimated to better than $40 \%$ in about $97 \%$ of the cases for a sample size of six in a one parametric estimation, where the shape and location parameter are held fixed. If a pulsar was never detected at a particular frequency we compute $3 \sigma$ upper limits from the weighted median over all non-detections and include a $5 \%$ systematic uncertainty.

In the spectral plots in Fig.~\ref{fig:ExamplesBestFitSpectra}, \ref{fig:GPSNew}, \ref{fig:GPSUnclear} and \ref{fig:GPSConfirmations} we show two error bars on our data: the inner and thicker one in lighter blue represents the statistical uncertainty due to scatter in the measurements $\frac{\sigma_{\text{r}, \nu}}{\sqrt{N}}$, whereas the outer error bar shows the total uncertainty, Eq.~\ref{eq:TotalFluxDensityUncertainty}.

\subsection{Robust regression and objective spectral classification}
\label{sec:RobustRegressionAndObjectiveSpectralClassification}

\begin{table}
\caption{Publications from which we took the data for the combined flux density data set. $\nu$ is the centre frequency of the flux density points.}
\label{tab:LiteratureFluxDensityData}
\begin{tabular}{lllll}
\hline
reference			& \#pulsars	& $\nu$ [MHz]\\
\hline
this work				& 441		& 728 -- 3100\\
-- (P875 only)			& 288		& 728 -- 3100\\
-- (P574 only)			& 154		& 732 -- 3094\\
\citet{1973Sieber}		& 27			& 10 -- 10690\\
\citet{1978Bartel}		& 18			& 14800, 22700\\
\citet{1981Izvekova}		& 86			& 39 -- 102.5\\
\citet{1995Lorimer}		& 280		& 408 -- 1606\\
\citet{1997vanOmmen}		& 82			& 800, 950\\
\citet{2000Maron},		& 281		& 39 -- 87000\\
\multicolumn{2}{l}{\url{http://astro.ia.uz.zgora.pl/olaf/paper1/}}\\
\citet{2000Malofeev}		& 235		& 102.5\\
\citet{2005Karastergiou}	& 48			& 3100\\
\citet{2006Johnston}		& 32			& 8400\\
\citet{2007Kijak}		& 11			& 325 -- 1060\\
\citet{2011Keith}		& 9			& 17000, 24000\\
\citet{2011Bates}		& 18			& 6500\\
\citet{2011Kijak}		& 15			& 610 -- 4850\\
\citet{2013Zakharenko}	& 74			& 20, 25\\
\citet{2016Bilous},		& 194		& 149\\
\url{http://astron.nl/psrcensus/}		&		&\\
\citet{2015Dai}			& 24 (MSPs)	& 730 -- 3100\\
\citet{2016Bell}			& 17			& 154\\
\citet{2016Basu}			& 1			& 325 -- 1280\\
\citet{2017Han}			& 228		& 1270 -- 1460\\
\citet{2017Murphy}		& 60			& 72 -- 231\\
\citet{2017Kijak}		& 15			& 325, 610\\
ATNF pulsar catalogue,	& 2536		& 40 -- 8000\\
\citet{2005Manchester},	&			&\\
various					&			&\\
\hline
\end{tabular}
\end{table}

We combine our flux density data that we split into frequency sub-bands with spectral data from the literature, see Table~\ref{tab:LiteratureFluxDensityData}. We compiled as much literature data as we could find, but do not claim that our database contains all pulsar flux density measurements ever obtained in the radio, which is simply infeasible. Nevertheless, we are confident that our combined data set represents the vast majority of pulsar spectral data available today. The literature data nicely extends our own measurements at frequencies below 400~MHz and above 3~GHz. To ensure uniqueness of data points, we gave preference to the data from individual publications rather than including them from summary publications such as the ATNF pulsar catalogue. For measurements where no uncertainty was given by the original authors we assumed a relative uncertainty of $50 \%$, which is a conservative value and was also adopted in earlier work \citep{1973Sieber}. We show upper limits from this work and from the literature as such in the spectral plots, but exclude them from the spectral fit. That is because our fitting algorithm in its current form does not handle censored data.

For the spectral model fitting we choose a Gaussian likelihood defined as:
\begin{equation}
	L = \prod_i^N \frac{1}{ \sqrt{2 \pi} \sigma_{y,i} } \exp \left( - \frac{ \left( f(x_i, \mathbf{a}) - y_i \right)^2 }{ 2 \sigma_{y,i}^2 } \right),
	\label{eq:GaussianLikelihood}
\end{equation}
where $f$ is the model function and $\mathbf{a}$ the model parameters, $y_i$ the measured flux densities at the frequencies $x_i$ and $\sigma_{y,i}$ the measurement uncertainties of the flux density points. The frequencies $x_i$ are assumed to be known without error. The negative log-likelihood then results in the weighted $\chi^2$ cost function:
\begin{equation}
	\chi^2 = -\log L = \sum_i^N \frac{1}{2} \left( \frac{f(x_i, \mathbf{a}) - y_i}{ \sigma_{y,i}} \right)^2 + C,
	\label{eq:Chi2}
\end{equation}
where $C$ is a constant that is usually neglected. As we are fitting a spectral model to a combined set of data, not just to our measured data, but also to data from the literature, we have to deal with all sorts of potential flaws in the data. For example the uncertainty of a flux density measurement could be largely underestimated or the measurement could be incompatible with the rest of the other data points because of differences or errors in the absolute flux density calibration. Outliers with unusually small uncertainties could drastically influence our weighted $\chi^2$ fit and the resulting best-fitting parameters because of the quadratic dependence on differences in flux density. In order to minimise the influence of outliers and points with underestimated uncertainties, we resort to using techniques from robust regression, in particular the Huber loss function, which is defined as:
\begin{equation}
	\rho = 
	\begin{cases}
		\frac{1}{2} t^2		& \text{if} \ |t| < k\\
		k |t| - \frac{1}{2} k^2	& \text{if} \ |t| \geq k\\
	\end{cases},
	\label{eq:HuberLossFunction}
\end{equation}
where $t = f(x, \mathbf{a}) - y$ and $k$ is a constant that defines at which distance the loss function starts to penalise outliers \citep{1964Huber}. We use a value of $k = 1.345$, for which Huber showed that the loss function operating on data from a Gaussian distribution is asymptotically $95 \%$ as efficient as the ordinary least squares estimator on the same set. Instead of minimising the negative log-likelihood in Eq.~\ref{eq:Chi2}, we minimise the Huber loss function $\rho$ and define our robust cost function as:
\begin{equation}
	\beta = - \log \tilde{L} = \sum_i^N \:
	\begin{cases}
		\frac{1}{2} \left( \frac{ f_i - y_i }{ \sigma_{y,i} } \right)^2		& \text{if} \ \left| \frac{ f_i - y_i }{ \sigma_{y,i} } \right| < k\\
		k \left| \frac{ f_i - y_i }{ \sigma_{y,i} } \right| - \frac{1}{2} k^2	& \text{otherwise}\\
	\end{cases},
	\label{eq:RobustChi2}
\end{equation}
where the normal weighted $\chi^2$ estimator in Eq.~\ref{eq:Chi2} is reproduced for $\left| \frac{ f_i - y_i }{ \sigma_{y,i} } \right| < k$, up to a constant factor, and data points that fall outside this range are penalised as outliers. We want to point out that no data points are discarded, they just get assigned less weight in the fit. We use the \texttt{MIGRAD} minimisation algorithm implemented in the software package \texttt{MINUIT} \citep{1975James} and in particular the python bindings to it in \texttt{iminuit}\footnote{\url{https://github.com/iminuit/iminuit}} to find the model parameters of the best fit that minimises the robust cost function $\beta$, i.e. the negative log-likelihood. We rely on the Aikake Information Criterion (AIC) to determine the best-fitting model and implement it as follows, including the correction for finite sample sizes:
\begin{multline}
	\text{AIC}_\text{c} = -2 \log L_\text{max} + 2 K + \frac{2K (K+1)}{N-K-1} = \\
	= 2 \beta_\text{min} + 2K + \frac{2K (K+1)}{N-K-1} + C,
	\label{eq:AIC}
\end{multline}
where $L_\text{max}$ is the maximum of the likelihood, $\beta_\text{min}$ the minimised robust cost function, $K$ the number of free and varying parameters in the model, $N$ the number of data points in the fit and $C$ a constant \citep{2007Liddle, 2014Ivezic}. We drop the subscript c denoting that the AIC is corrected for finite sample sizes and use the term AIC to refer to Eq.~\ref{eq:AIC} from here on. The model that results in the lowest AIC is the one that best fit the data without over-fitting. We classify the pulsar spectra based upon this best fitting model. We also compute the relative likelihood of each model $i$ with corresponding $\text{AIC}_i$ to the best-fitting model with $\text{AIC}_\text{min}$, the so-called Aikake weight, as:
\begin{equation}
	l_i = \exp \left(- \frac{1}{2} \left| \text{AIC}_i - \text{AIC}_\text{min} \right| \right),
	\label{eq:RelativeLikelihood}
\end{equation}
from which we calculate the probability of the best-fitting model as:
\begin{equation}
	p_\text{best} = 1 / \sum_{i}^R l_i,
	\label{eq:ProbabilityOfBestModel}
\end{equation}
where the summation runs over all $R$ models tested \citep{2010Burnham}. This is the probability that the best model is indeed the best-fitting model to the data among the ones that we tested, but does not indicate an overall probability compared with an arbitrary other model. It gives an indication how much better the model fits the data compared with the rest.

\subsection{Spectral models}
\label{sec:SpectralModels}

We selected five spectral models from the literature that have distinctive spectral shapes, which is the key aspect that we can test using our measurements. They have been used successfully in the past to describe the spectra of pulsars, with the simple and broken power law being the most commonly applied ones. While these are morphological, the spectral index could potentially be related with other pulsar parameters (\S\ref{sec:CorrelationsOfSpectralIndexWithPulsarParameters}). Our model selection is sufficiently diverse to fully represent the spectral shapes seen in the data set. However, we cannot test every conceivable model and we limit ourselves to those five classes. In particular, we fit the following analytical models to the data, where $\nu$ is the centre frequency and $\nu_0 = 1.3 \: \text{GHz}$ a constant reference frequency:
\begin{enumerate}
  \item simple power law of the form:
  \begin{equation}
    S_\nu = b \: x^{\alpha},
    \label{eq:SimplePowerlaw}
  \end{equation}
  where $x = \frac{\nu}{\nu_0}$, with $\alpha$ the spectral index and $b$ a constant. The fit parameters are $\alpha$ and $b$.
  
  \item broken power law of the form:
  \begin{equation}
    S_\nu = b \:
    \begin{cases}
      x^{\alpha_1}						& \text{if} \: \: x \leq x_\text{b}\\
      x^{\alpha_2} \: x_\text{b}^{\alpha_1 - \alpha_2}	& \text{otherwise}\\
    \end{cases},
    \label{eq:BrokenPL}
  \end{equation}
  where $x = \frac{\nu}{\nu_0}$, $x_\text{b} = \frac{\nu_\text{b}}{\nu_0}$, with $\nu_b$ being the frequency of the spectral break, $\alpha_1$ the spectral index before and $\alpha_2$ the one after the break. The fit parameters are $b, \alpha_1, \alpha_2$ and $\nu_b$.
  
  \item log-parabolic spectrum (LPS) of the form:
  \begin{equation}
    \log_{10} S_\nu = a x^2 + b x + c,
    \label{eq:LPS}
  \end{equation}
  where $x = \log_{10} \left( \frac{\nu}{\nu_0} \right)$, $a$ is the curvature parameter, $b$ is the spectral index for $a = 0$ and $c$ is a constant. This model has been used to describe the spectra of radio galaxies (e.g. \citealt{1977Baars}) and curved pulsar spectra \citep{2013Bates, 2014Dembska}. The fit parameters are $a$, $b$ and $c$.
 
  \item power law with high-frequency cut-off of the form:
  \begin{equation}
    S_\nu = b \: x^{-2} \: \left( 1 - \frac{x}{x_\text{c}} \right), \: x < x_\text{c}
    \label{eq:PLWithHardCutOff}
  \end{equation}
  where $x = \frac{\nu}{\nu_0}$, $x_\text{c} = \frac{\nu_\text{c}}{\nu_0}$ with $b$ a constant and $\nu_c$ the cut-off frequency. The fit parameters are $b$ and $\nu_c$. The functional form is based on the work by \citet{2013Kontorovich} and provides us with a direct test of their model for the coherent pulsar emission. Specifically, the radio emission should be created by electrons accelerated in the electric field of the pulsar and the cut-off frequency of the spectrum should depend on the maximum value of the electric field $E_\text{max}$ as:
  \begin{equation}
   \nu_\text{c} = \sqrt{\frac{ e E_\text{max} }{ 2 m_e h }} = \sqrt{\frac{\pi e B}{m_e c P}},
   \label{eq:HardCutOffFrequency}
  \end{equation}
where $e$ is the charge and $m_e$ is the mass of the electron, $h$ is Planck's constant and $c$ is the speed of light in vacuum. The second identity is true under an assumption about the maximum electric field in the centre of the polar cap, where $P$ is the period and $B$ is the magnetic field of the pulsar.
 
  \item power law with low-frequency turn-over of the form:
  \begin{equation}
    S_\nu = b \: x^{\alpha}  \exp \left( \frac{ \alpha }{ \beta } \: x_\text{c}^{-\beta} \right),
    \label{eq:PLwithLowFreqTurnOver}
  \end{equation}
  where $x = \frac{\nu}{\nu_0}$, $x_\text{c} = \frac{\nu}{\nu_\text{c}}$, $\alpha$ is the spectral index, $\nu_c$ is the turn-over frequency, $b$ a constant and $0 < \beta \leq 2.1$ determines the smoothness of the turn-over. Fit parameters are $\alpha$, $\nu_c$, $b$ and $\beta$. This functional form is motivated by a synchrotron self-absorption process proposed to be responsible for the low-frequency turn-over \citep{1981Izvekova}, but can describe the spectra expected both due to synchrotron self and thermal free-free absorption. For the special case $\beta = 2.1$ it is equivalent to a free-free absorption model (e.g. \citealt{2016Rajwade, 2017Kijak}).
\end{enumerate}

\subsection{Quality of the spectral classification -- classification categories}
\label{sec:ClassificationCategories}

In the following sections we discuss the spectral classifications of individual pulsars. They get classified by the best-fitting spectral model -- the one with the lowest AIC. However, this classification has different qualities, for example the best-fitting spectral model might only be slightly preferred by the data over all the others tested. To quantify that in an objective way we define the following categories (as opposed to classification) based on the value of $p_\text{best}$ and the significance $s$ in standard deviations $\sigma$ of the spectral feature at hand (for a log-parabolic spectrum that is the curvature coefficient $a$ for example):

\begin{description}
	\item{weak}: $p_\text{best} < 0.5$ or $s < 2$.
	
	\item{candidate}: $0.5 \leq p_\text{best} < 0.7$ and $s \geq 2$.
	
	\item{strong}: $p_\text{best} \geq 0.7$ and $s \geq 2$.
	
	\item{clear}: $p_\text{best} \geq 0.8$ and $s \geq 3$.
\end{description}

\section{Results}
\label{sec:Results}

\subsection{Parkes system temperature estimation}
\label{sec:ParkesSystemTemperatureEstimation}

As part of the absolute flux density calibration procedure the system equivalent flux density (SEFD) is estimated, which we denote as $S_\text{sys}$ and is the sum over both polarisations of the receiver. We determined it from our Hydra A flux density calibrator observations and list it and the resulting system temperature $T_\text{sys}$ for each receiver used in Table~\ref{tab:MatchingGstarAndATNFData}. The gain used to compute $T_\text{sys}$ was taken from the ATNF receiver fleet table. Note that the measured system temperature of the 10cm part of the 10-50cm receiver and especially the one of the multi-beam receiver are slightly lower than what is quoted in the receiver fleet table. They are lower by about 2 and $6 \: \text{K}$ respectively. The value measured for the 50cm part agrees well.

\subsection{Flux density variability}
\label{sec:FluxDensityVariability}

\begin{figure}
  \centering
  \includegraphics[width=\columnwidth]{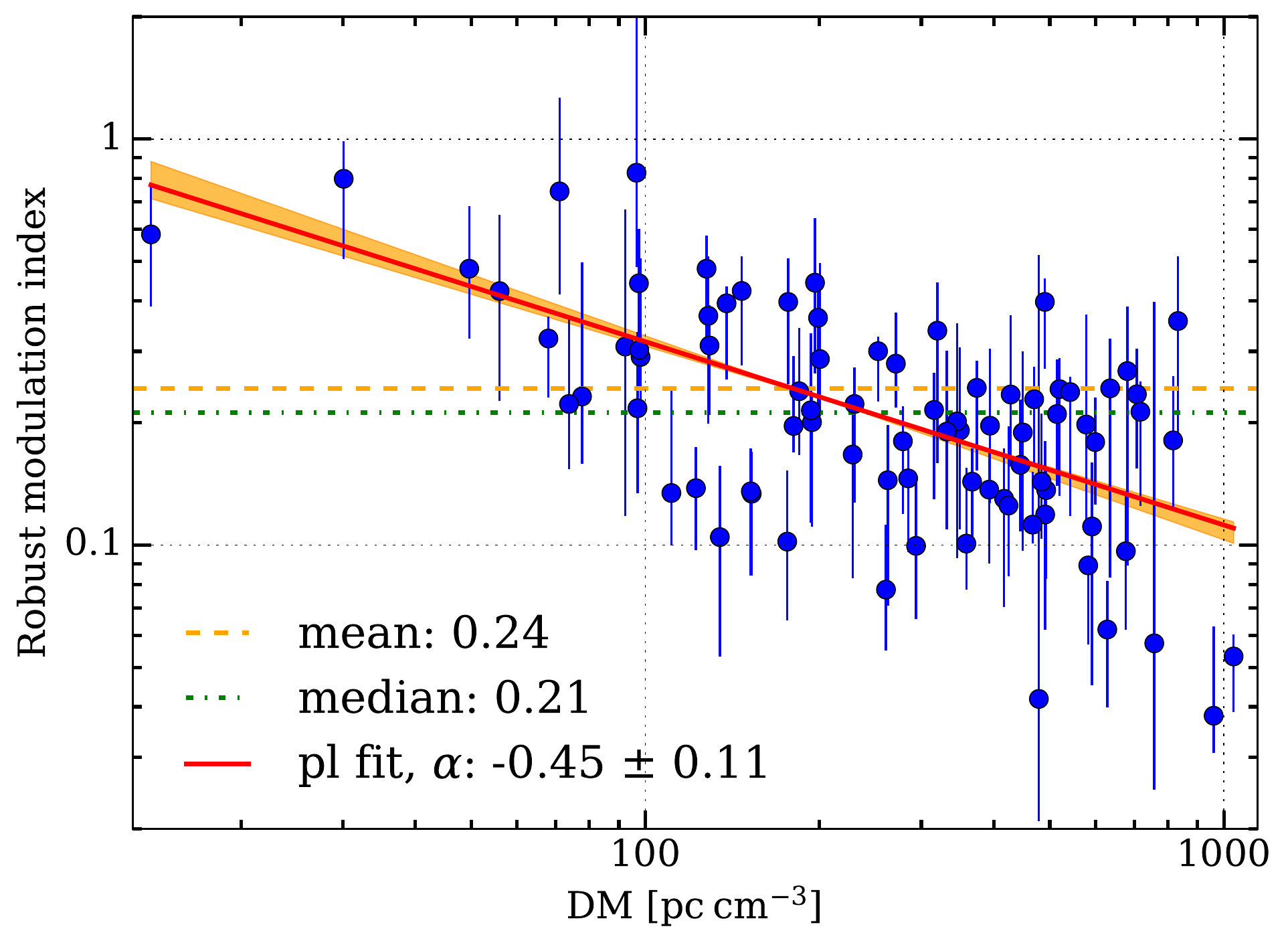}
  \caption{Measured robust modulation index versus DM of the pulsars with at least six observing epochs at a centre frequency of 3.1~GHz. The red solid line shows the best error-weighted fit of Eq.~\ref{eq:PowerlawModVsDM} to the data. The orange band is the uncertainty introduced by making the bootstrap errors symmetric in the fitting procedure.}
  \label{fig:ModulationIndexVsDM}
\end{figure}

\begin{figure}
  \centering
  \includegraphics[width=\columnwidth]{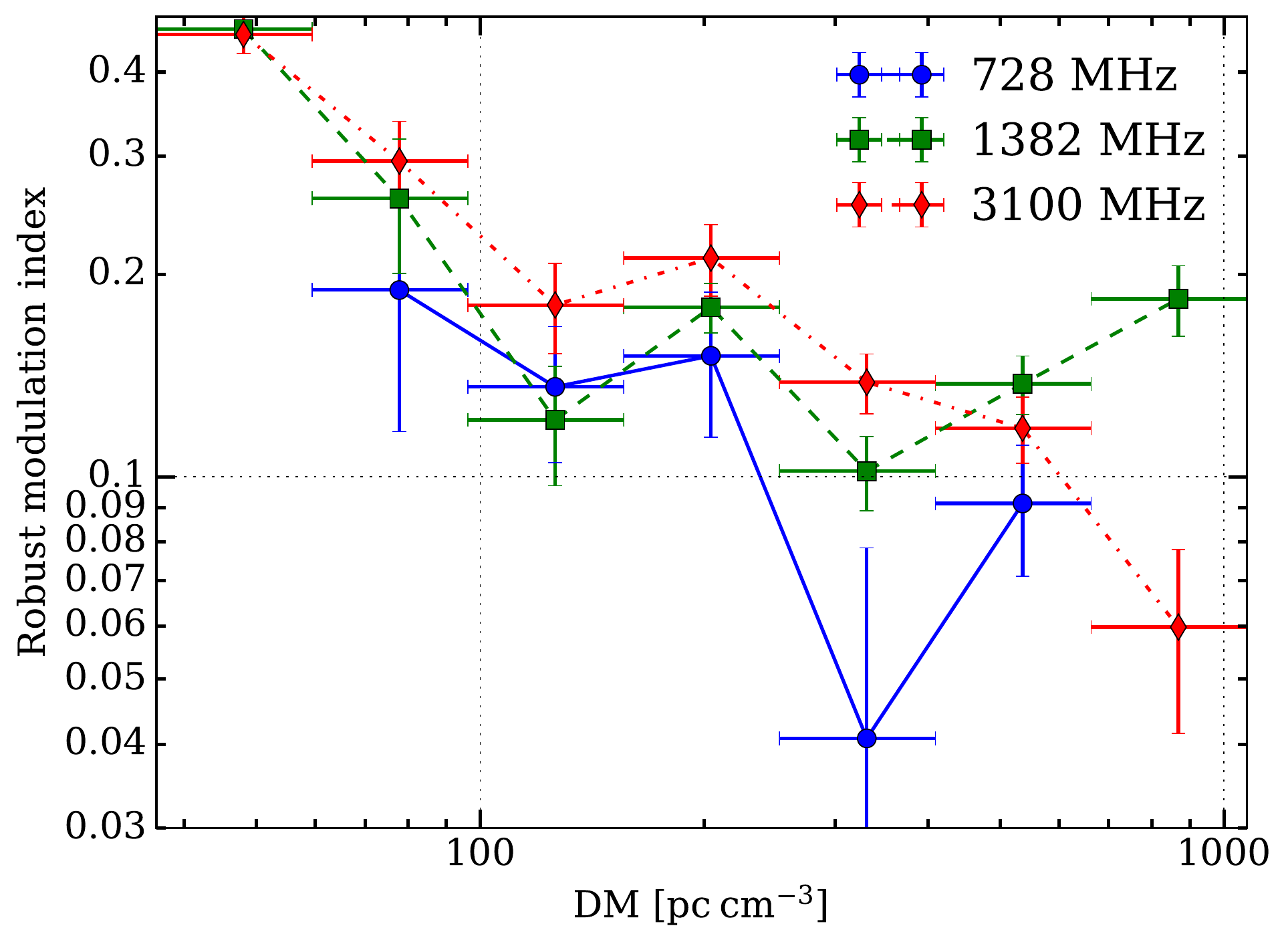}
  \caption{Comparison of the scaling with DM of the robust modulation index determined from our data for the three centre frequencies.}
  \label{fig:ModulationIndexFreqComparison}
\end{figure}

\begin{table}
\caption{Best-fitting parameters from the weighted fit of Eq.~\ref{eq:PowerlawModVsDM} to the robust modulation index versus DM data for all pulsars with at least six measurement epochs for the three centre frequencies.}
\label{tab:WmodVsDMScaling}
\centering
\begin{tabular}{llll}
\hline
$\nu$ [MHz]	& 728		& 1382	& 3100\\
\#pulsars	& 37			& 150	& 82\\
\hline
$a$			& $-0.47 \pm 0.27$	& $-0.22 \pm 0.05$	& $-0.45 \pm 0.11$\\
$b$			& $0.21 \pm 0.02$	& $0.25 \pm 0.01$	& $0.23 \pm 0.01$\\
\hline
\end{tabular}
\end{table}

\subsubsection{Results from the theoretical simulation}

We conducted a theoretical simulation using the \texttt{NE2001} galaxy model to quantify the effect of scintillation on our data, see \S\ref{sec:FluxDensityFluctuationsDueToScintillation}. We find that the vast majority of pulsars have a total modulation index close to zero. Specifically, about $83$, $75$ and $60 \: \%$ of the pulsars have $m_\text{tot} \leq 0.3$ at 728, 1382 and 3100 MHz. The modulation is less than that for the band integrated flux densities.

\subsubsection{Results from the data driven approach}

We also derived the variability in flux density directly from our flux density time series data. In Fig.~\ref{fig:ModulationIndexVsDM} we show the robust modulation index at 3.1~GHz plotted against DM for all pulsars with at least six measurement epochs at that centre frequency. The behaviour at the other frequencies is very similar. It is apparent that the modulation index decreases with increasing DM in each case, as shown earlier for a much smaller number of pulsars by \citet{1982Sieber, 2000Stinebring}. We fit a power law of the form:
\begin{equation}
	m_{\text{r},\nu} = b \: \left( \frac{d}{d_0} \right)^{a}
	\label{eq:PowerlawModVsDM}
\end{equation}
to the data in an error-weighted manner using the same procedure as described in \S\ref{sec:RobustRegressionAndObjectiveSpectralClassification}, where $d$ is the DM of the pulsars, $d_0 = 200 \: \text{pc} \: \text{cm}^{-3}$ is a constant reference DM, $b$ is a constant and $a$ is the power law scaling exponent. We make the bootstrap uncertainties of the robust modulation index symmetric by taking the mean of the asymmetric lower and upper error. To estimate how the procedure affects the fit, we also perform the fit using uncertainties estimated as the maximum and minimum of both. The best fit is shown as the red solid line in Fig.~\ref{fig:ModulationIndexVsDM} and the uncertainty introduced by this procedure is shown by the orange band. It is less than $5 \: \%$ in all cases. The best-fitting parameters are listed in Table~\ref{tab:WmodVsDMScaling}. The modulation index decreases with DM following a power law. The slope is nearly the same at 728 and 3100 MHz, while it is roughly half of that at 1.4 GHz. That means that the power law scaling is shallower there, where we have the largest number of measurements. The uncertainty of the slope is the highest at 728 MHz, where we have the smallest sample size.

We also compare the scaling of the robust modulation index with DM at the different observing frequencies, see Fig.~\ref{fig:ModulationIndexFreqComparison}. In order to reduce the scatter that is present in the data we binned them in equal DM bins in logarithmic space and computed the weighted means for each. As uncertainty we estimated the standard error in each DM bin. We find that the modulation is the highest at 3.1~GHz, followed by 1.4~GHz and 728~MHz, except in the highest DM bin centred at $868 \: \text{pc} \: \text{cm}^{-3}$, where the value at 1.4~GHz exceeds the other two. Above a DM of $126 \: \text{pc} \: \text{cm}^{-3}$ the modulation index is nearly constant with a weighted mean value of around $0.14$ at 1.4~GHz. At the other two frequencies it is still decreasing with DM.

\subsubsection{Comparison of the two approaches}
\label{sec:ComparisonOfTheTwoApproaches}

\begin{figure}
	\centering
	\includegraphics[width=\columnwidth]{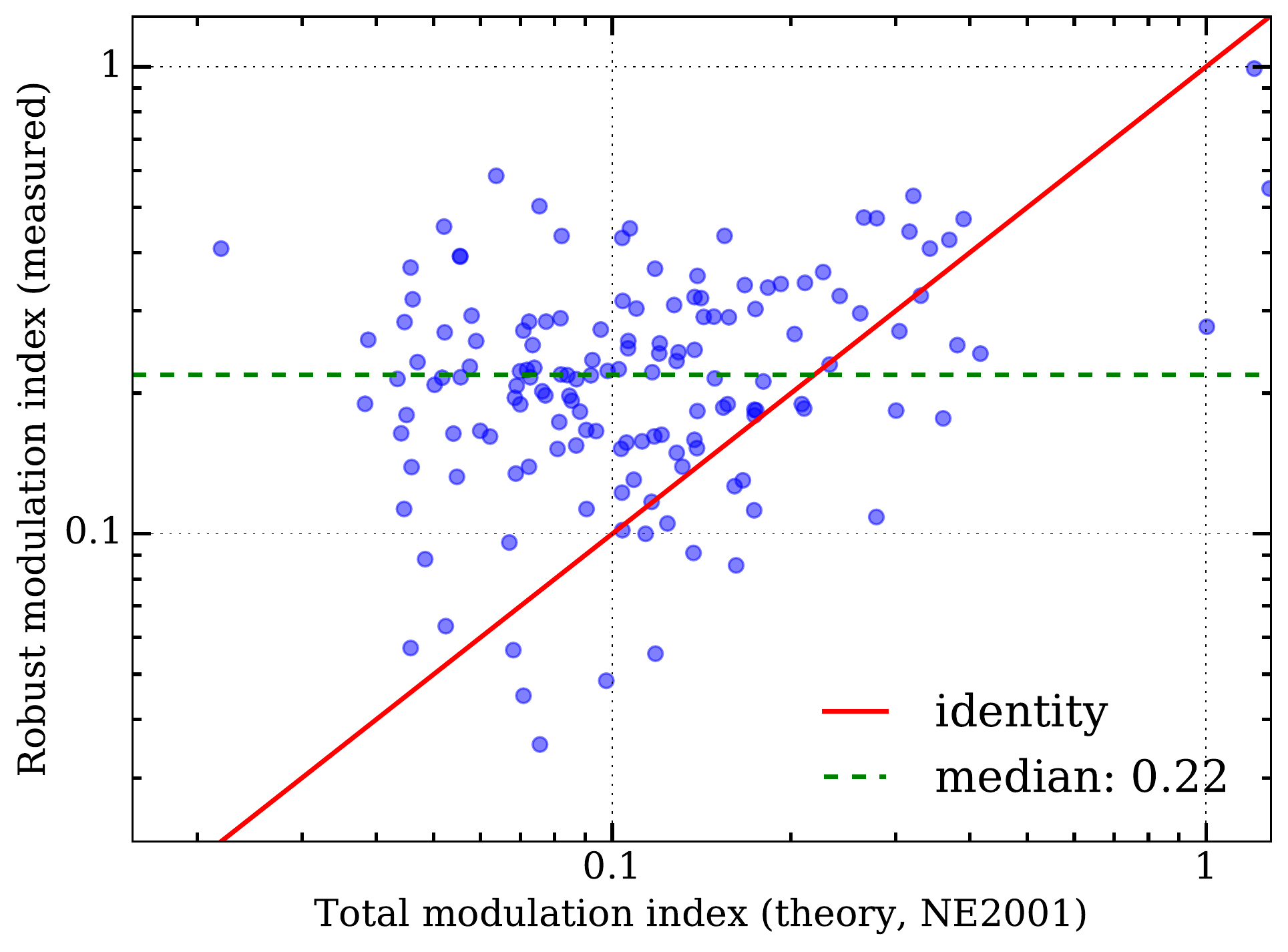}
	\includegraphics[width=\columnwidth]{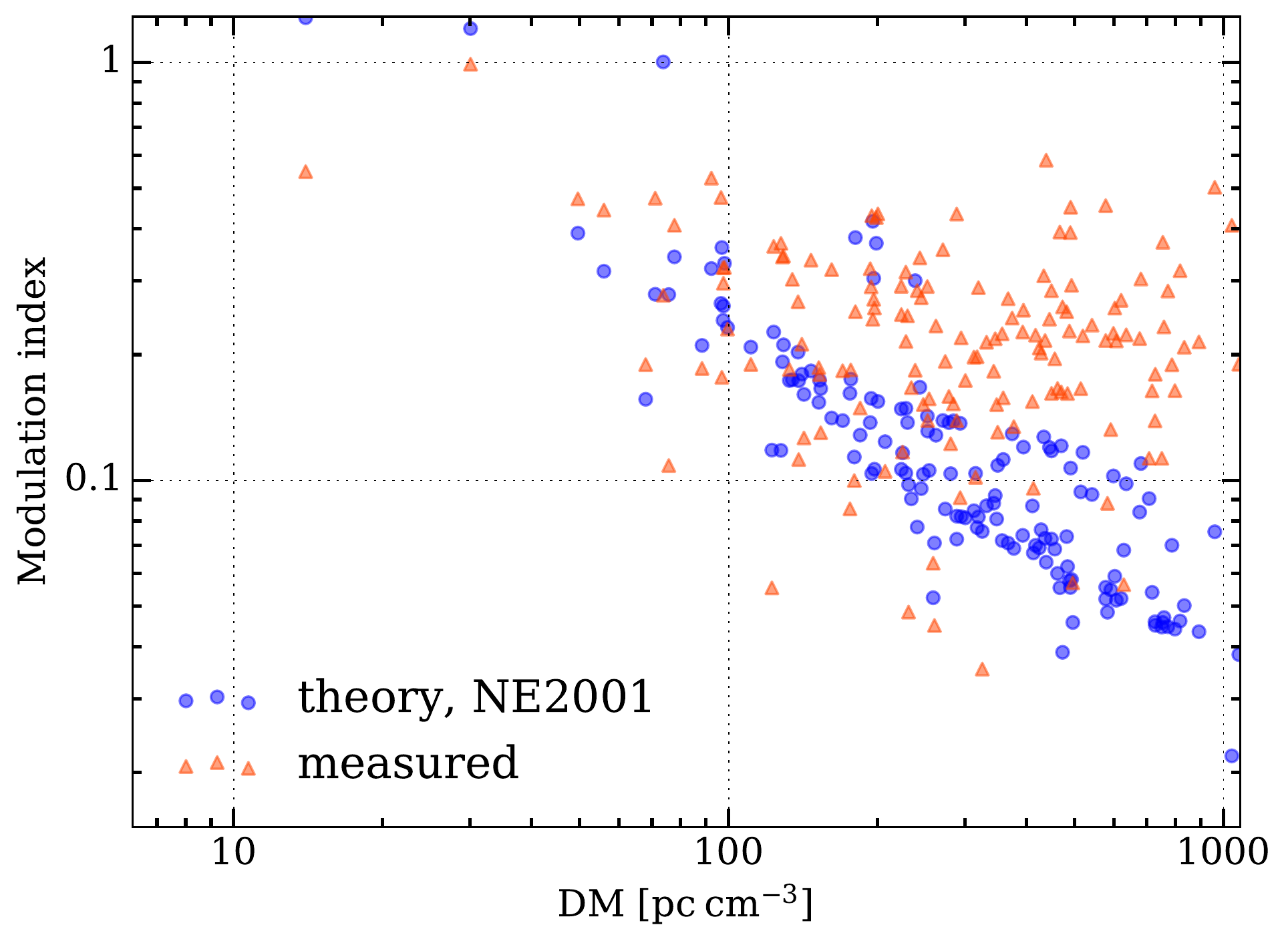}
	\caption{Comparison between the modulation indices at 1.4~GHz measured from our flux density time series data and the theoretical prediction based on the \texttt{NE2001} galaxy model. Top: Direct comparison between the modulation indices, with the red solid line indicating the identity. Bottom: Modulation index versus DM of the pulsars.}
	\label{fig:ModulationComparison}
\end{figure}

We find that both approaches are largely consistent at 728~MHz and 3.1~GHz. Both modulation indices are in good agreement there, with the measured ones showing a larger scatter than the theoretical ones. However, at 1.4~GHz there is considerably more variability in the data than expected from the theoretical simulation. The variability at 1.4~GHz is presented in Fig.~\ref{fig:ModulationComparison}, where we show a comparison of the robust modulation indices measured from our flux density time series data with the theoretically expected modulation indices calculated using the \texttt{NE2001} model. In the top panel we directly compare the modulation indices and in the bottom panel we compare the modulation indices plotted against DM of the pulsars. Top panel: There is a large scatter around the identity line and the measured modulation indices reach a plateau with a median of about $0.22$ for the majority of pulsars. This is most likely radiometer noise in combination with fluctuations in absolute flux density calibration and residual pulse-to-pulse fluctuations (see also \citealt{2000Stinebring}). In this noise-limited region the theoretical simulation significantly underestimates the true scatter in the data. Bottom panel: The measured modulation indices exceed the theoretically expected ones in nearly all cases and the theoretical ones are effectively a lower limit for the flux density variability for a realistic data set that is limited in observation time and S/N. This is most obvious at 1.4~GHz.

We conclude that our empirical scaling relationship (Eq.~\ref{eq:PowerlawModVsDM}) based on the parameters determined from our data is therefore a more realistic and more conservative estimate for the flux density variability than the theoretically expected one. Therefore, we use it as an estimate for the uncertainty $u_{\text{scint}} (\text{DM}, \nu) = m_{\text{r},\nu} \left( \text{DM}, \nu \right) \: \bar{S}_\nu$ in flux density due to scintillation.

\subsection{Calibrated flux densities}
\label{sec:CalibratedFluxDensities}

One of the main results of this paper are the band-integrated, calibrated flux densities at 728, 1382 and 3100~MHz for all the pulsars in our data set, which are listed in Table~\ref{tab:FluxDensities} in the appendix. The table contains the data for 441 pulsars. In case a pulsar was never detected with a S/N of at least 6 at a certain frequency we quote a $3 \sigma$ upper limit for its flux density. The table also contains the spectral classification, the range of frequencies over which the spectral classification was performed, the spectral index for pulsars that have simple power law spectra and the robust modulation index in case we had at least six measurement epochs at that centre frequency.

\subsection{Comparison with data from the literature}
\label{sec:ComparisonWithLiteratureData}

\begin{figure}
	\centering
	\includegraphics[width=\columnwidth]{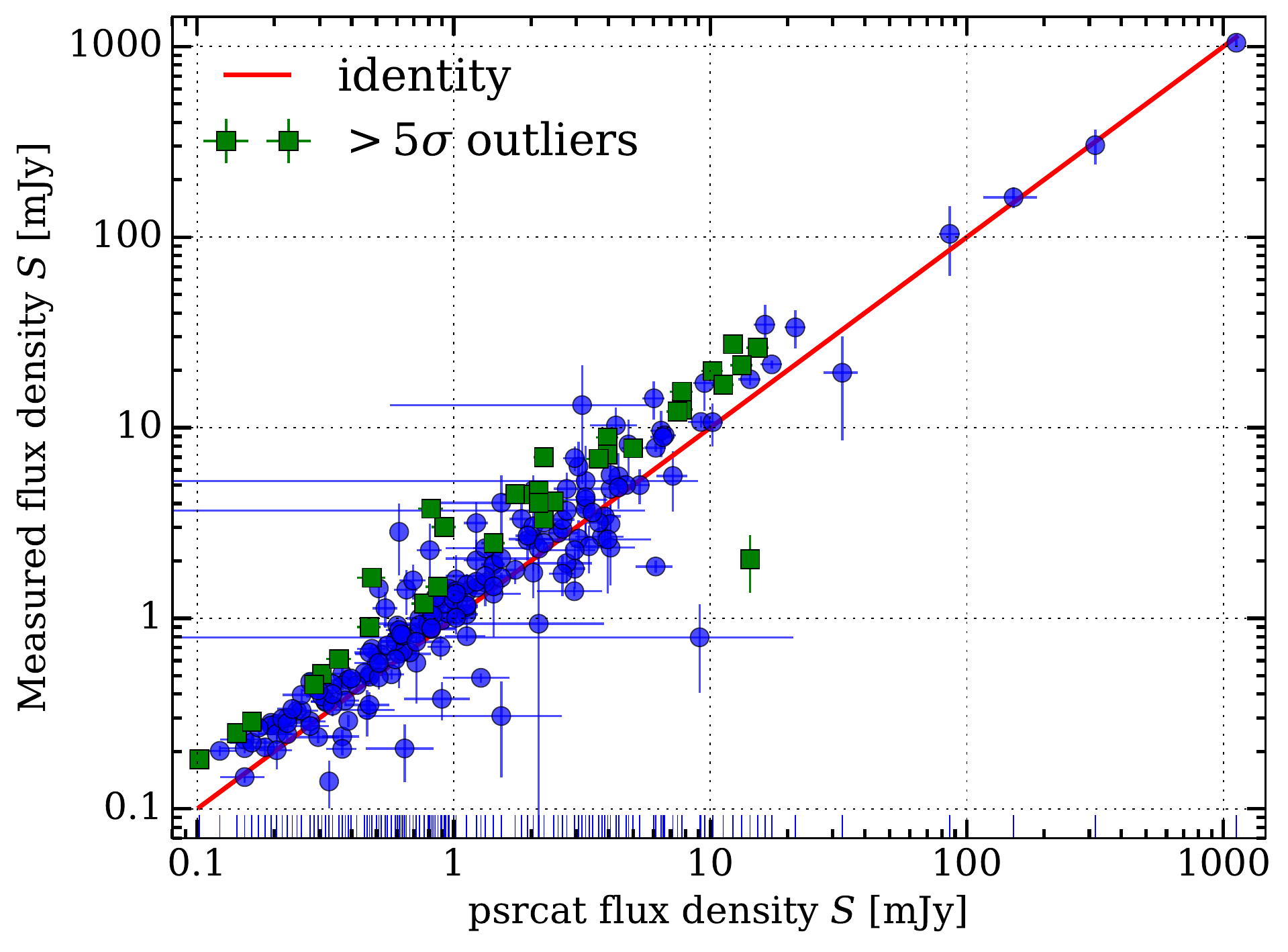}
	\caption{Comparison between the measured flux densities from this work and directly matched flux densities within 100~MHz from the ATNF pulsar catalogue at a centre frequency of 1.4~GHz. The red solid line indicates the identity.}
	\label{fig:PsrcatComparison}
\end{figure}

A comparison between the flux density measurements from this work and the ones from the ATNF pulsar catalogue is shown in Fig.~\ref{fig:PsrcatComparison}. We show only the direct matches with the catalogue, i.e. measurements that are within 100~MHz of a particular centre frequency. While there are 219 matches at 1.4~GHz, there are only 10 and 20 at 728 and 3100~MHz respectively. Effectively all our measurements at those centre frequencies are new ones. Green rectangles show the flux density points that deviate by at least $5 \sigma$ from the identity line. Our data and the pulsar catalogue data are generally in good agreement, with the majority of data points located close to the identity line. The agreement at 1.4~GHz is good, with only 33 measurements deviating significantly. At 3.1~GHz only two measurements deviate significantly, but the uncertainties are generally larger. Unfortunately, sometimes no uncertainties are given in the catalogue. Of the 33 $5 \sigma$ outliers at 1.4~GHz, 30 are from publications of the Parkes Multibeam Pulsar Survey (PMPS). 23 are from \citet{2004Hobbs}, four from \citet{2006Lorimer}, two from \citet{2003Kramer} and one from \citet{2001Manchester}, and all are based on the same flux density estimation method. We had already noted that in multiple cases their flux densities are systematically about a factor of two lower than both our data and other data from the literature. The discrepancy is maybe not surprising, as PMPS flux densities were estimated using the radiometer equation and observations of high-DM pulsars used as standard candles \citep{2001Manchester}. We notice that three of their calibration pulsars have flux densities at 1.4~GHz that are roughly half of what we measure for them. The other outliers are each from different older publications. Apart from these outliers, the rms relative difference is $31 \%$. Nonetheless, more of our measurements at 1.4~GHz lay above the identity line than below, which indicates that our measurements are slightly biased high. That could be a result of a decrease in the flux density of Hydra A, or a slight offset in telescope pointing -- the spectral index of Hydra A changes rapidly across its extent \citep{2004Lane}. It is not due to the exclusion of observations below a S/N of 6 in the flux density estimation; even if these are included we see the same slight bias.

\subsection{Modelling the pulsar spectra}
\label{sec:ModellingThePulsarSpectra}

\begin{table}
\caption{Fraction of pulsar spectra that can best be characterised by the given spectral model.}
\label{tab:BestFitSpectralModels}
\centering
\begin{tabular}{llllll}
\hline
set				& \#pulsars	& \%\\
\hline
total			& 441\\
classified		& 349		& 79.1\\
not classified	& 92			& 20.9\\
\hline
best model							& \#pulsars	& \%\\
\hline
simple power law 					& 276		& 79.1\\
log-parabolic spectrum				& 35			& 10.0\\
broken power law						& 25			& 7.1\\
pl with low-frequency turn-over		& 10			& 2.9\\
pl with hard high-frequency cut-off	& 3			& 0.9\\
\hline
\end{tabular}
\end{table}

\begin{figure*}
	\centering
	\includegraphics[width=0.49\textwidth]{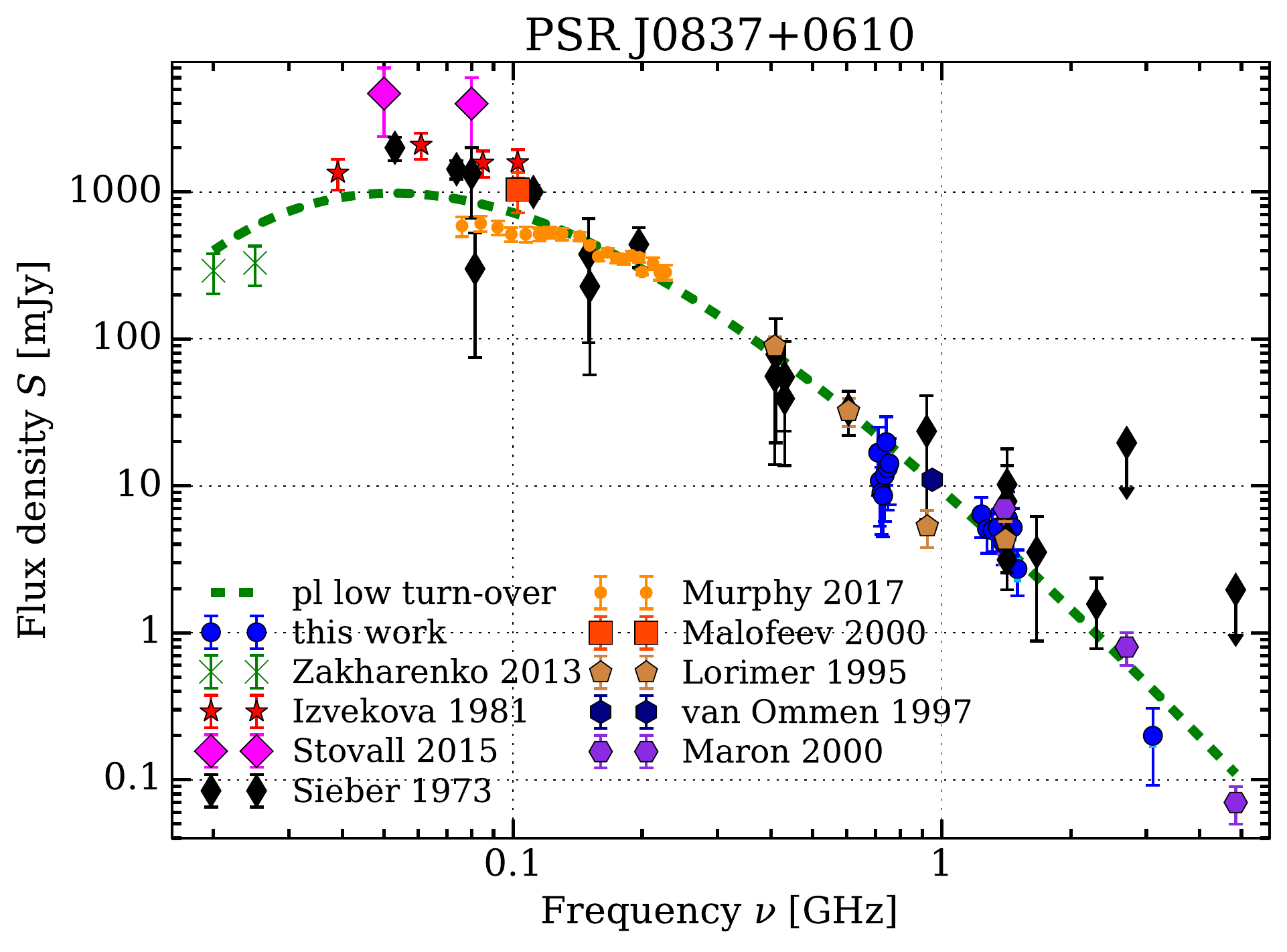}
	\includegraphics[width=0.49\textwidth]{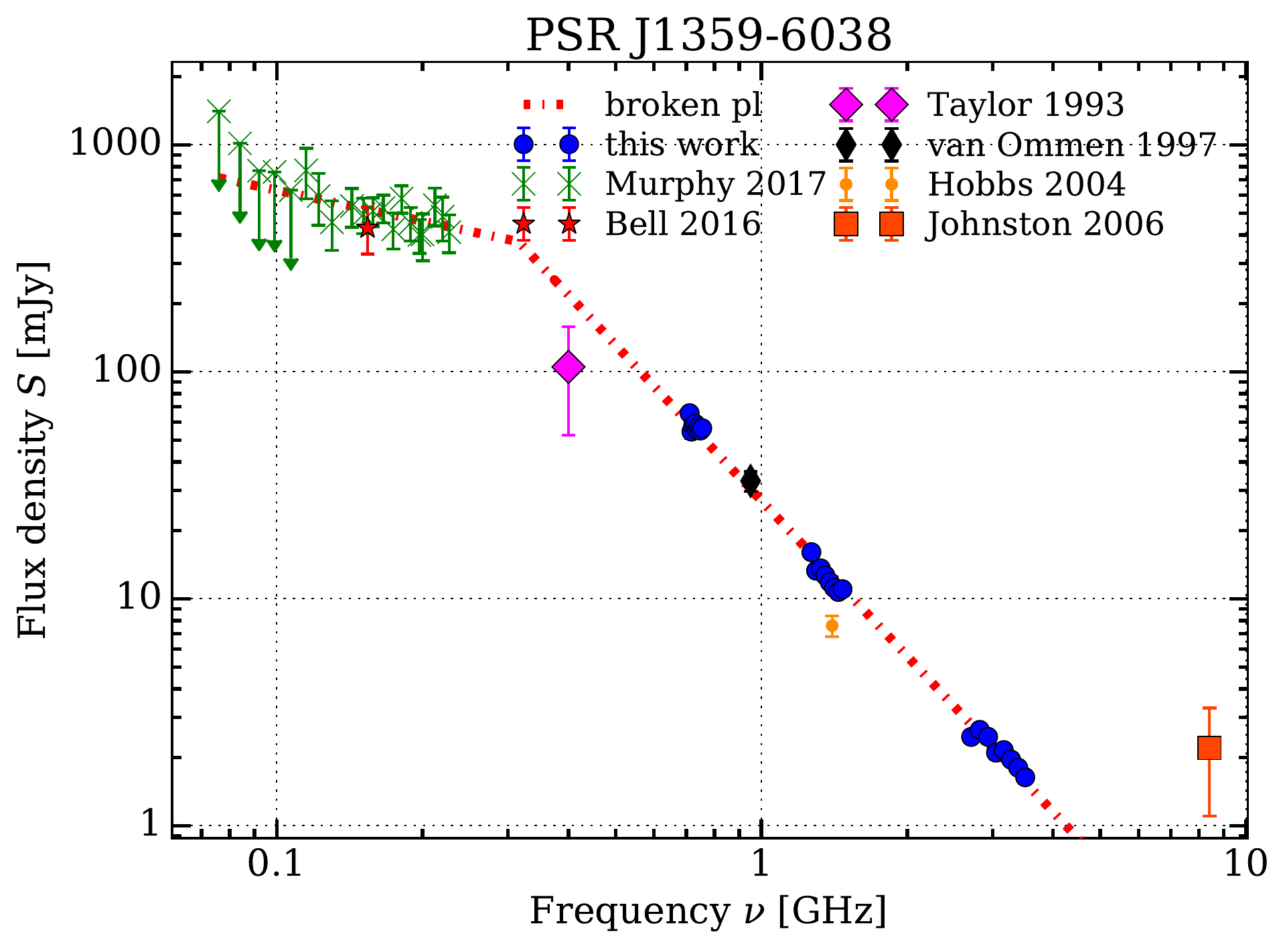}
	\includegraphics[width=0.49\textwidth]{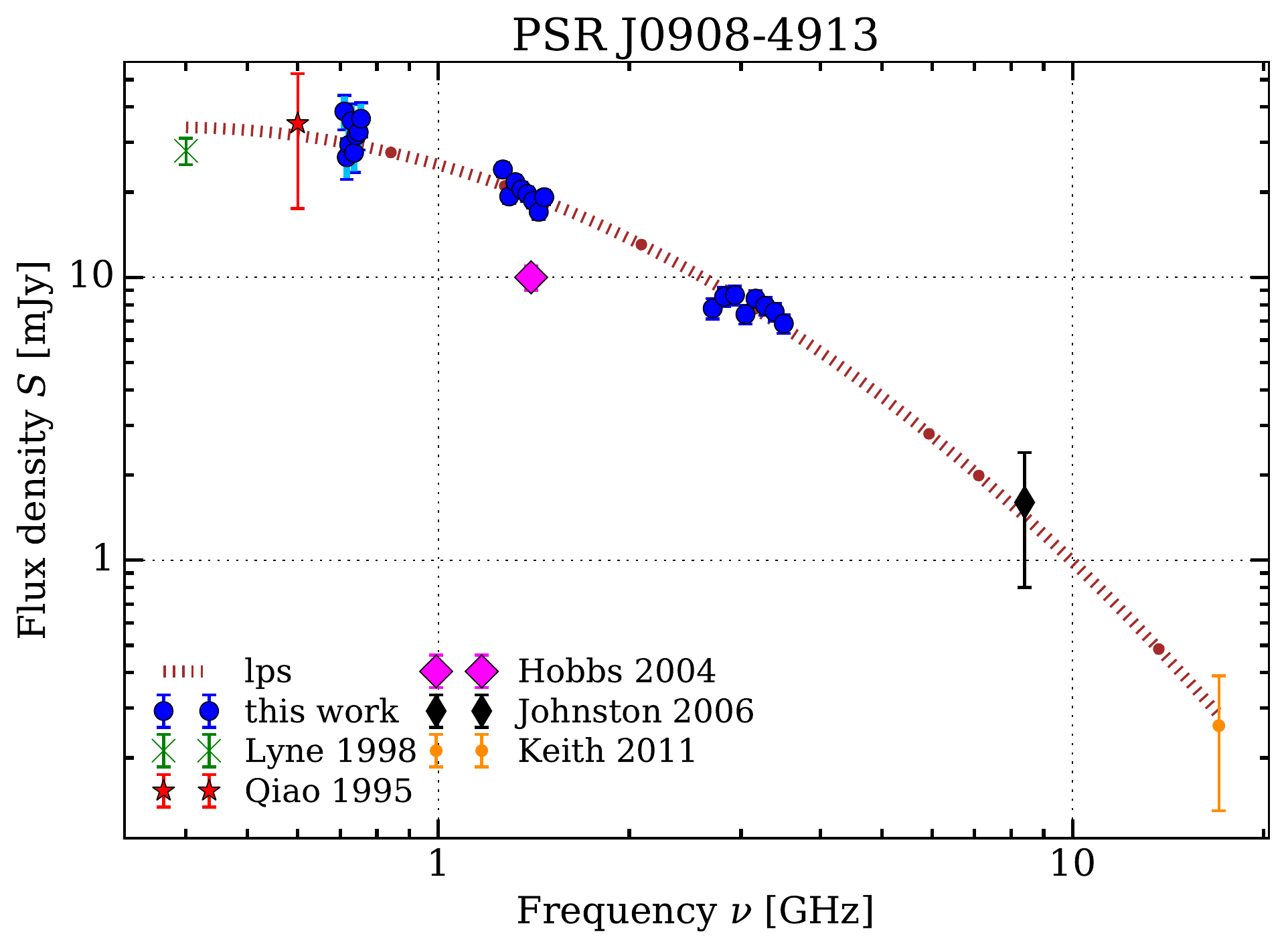}
	\includegraphics[width=0.49\textwidth]{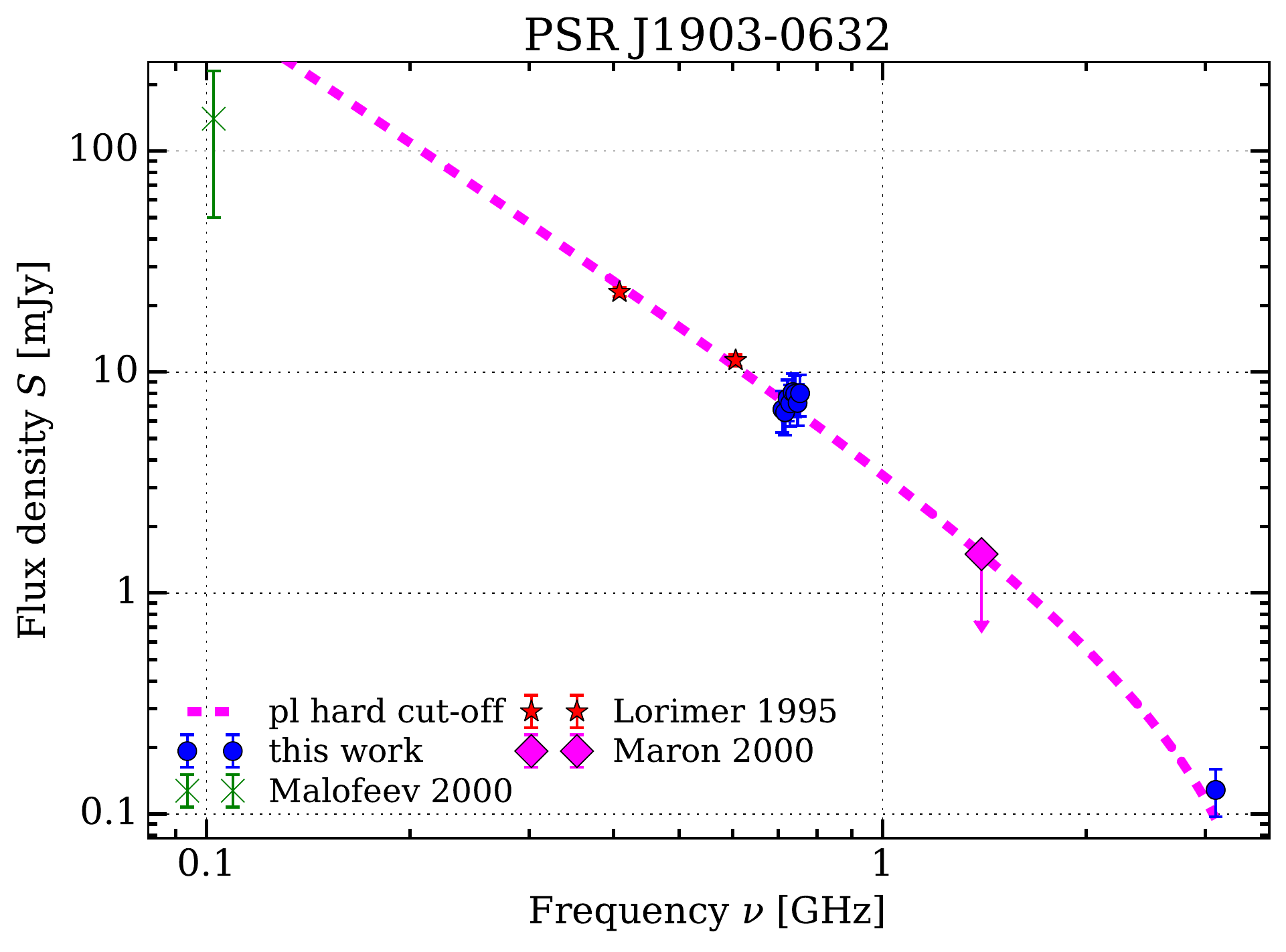}
	\caption{Example spectra where each spectral model fits best, from top left in clockwise direction: power law with low-frequency turn-over, broken power law, log-parabolic spectrum and power law with high-frequency hard cut-off. In this and all other spectral plots we show two error bars on our data: the inner one in lighter blue represents the statistical uncertainty due to scatter in the measurements, whereas the outer error bar shows the total uncertainty, which additionally includes scintillation and the systematic uncertainty (Eq.~\ref{eq:TotalFluxDensityUncertainty}).}
	\label{fig:ExamplesBestFitSpectra}
\end{figure*}

We used our flux density measurements that are split into frequency sub-bands combined with data from the literature and fit different spectral models to the combined data set in a robust manner as described in \S\ref{sec:RobustRegressionAndObjectiveSpectralClassification}. We decide to use the following requirements for the characterisation of pulsar spectra: the spectra need to consist of at least four flux density measurements at four different centre frequencies and the data points must span at least a factor of two in frequency. This choice provides a balance between sufficient spectral coverage for the model fitting and the amount of pulsars that can be classified. In particular, it also ensures that we cannot characterise a spectrum using only observations at 10~cm that are split into frequency sub-bands. The pulsars for which the requirements are not fulfilled are excluded from the spectral analysis. We characterise the spectra based on the best-fitting model, that is the one with the lowest AIC. We can successfully characterise the spectra of $349$ pulsars, which is about $79 \%$ of the total data set after the removal of non-detections. The results are shown in Table~\ref{tab:BestFitSpectralModels} where we give an overview of the fraction of pulsar spectra that can be best characterised by a given spectral model.

We find that the majority of pulsar spectra, about $79 \%$, can best be characterised as a simple power law in the frequency range studied. The pulsars with log-parabolic spectra account for about $10 \%$ and the broken power law spectra for about $7 \%$. The spectral models with a low-frequency turn-over and hard high-frequency cut-off are rare cases accounting for about $3$ and less than $1 \%$ each. We list the pulsars that have spectra that deviate significantly from a simple power law in separate tables depending on the spectral classification: LPS pulsars in Table~\ref{tab:LPSPulsars}, pulsars with broken power law spectra in Table~\ref{tab:BrokenPLPulsars}, the ones with a hard high-frequency cut-off in Table~\ref{tab:HardCutoffPulsarsMaximumEandB} and the ones with power law spectra that have a low-frequency turn-over in Table~\ref{tab:LowFreqCutOffPulsars}. For each spectral model apart from a simple power law we show one example where it fits the data best in Fig.~\ref{fig:ExamplesBestFitSpectra}.

\subsubsection{How does the frequency coverage affect the spectral classification?}

The classification depends naturally on the spectral coverage, i.e whether spectral features can be determined from the data. Fortunately, the combination of our measurements with literature data provides reasonable to very good coverage in terms of the number of data points (median 13, maximum 90) and fractional frequency coverage (median 7.8, maximum 1600) for all pulsars that fulfilled our classification requirements. We examine the classification of pulsars for which we have good low or high-frequency coverage separately. We have good low-frequency coverage, which we define as having at least two data points below 600~MHz for $119$ pulsars and good high-frequency coverage with at least one data point above 4~GHz for $88$ pulsars. For the ones with good low-frequency coverage the simple power law is still the most common spectrum ($56 \%$), followed by the broken power law with $16 \%$ and the LPS with $14 \%$. Only $10$ pulsars have a power law spectrum with low-frequency turn-over. For the pulsars with good high-frequency coverage simple power law spectra account for $56 \%$, LPS for $18 \%$ and broken power law spectra for $17 \%$. That means that with good low-frequency coverage a spectral break is significantly favoured in comparison with the whole data set. At high frequencies a spectral break is slightly favoured over spectral curvature with both showing an increase by a factor of two or more in fraction.

\subsection{Simple power law spectra and spectral indices}
\label{sec:SimplePowerlawSpectra}

\begin{figure}
	\centering
	\includegraphics[width=\columnwidth]{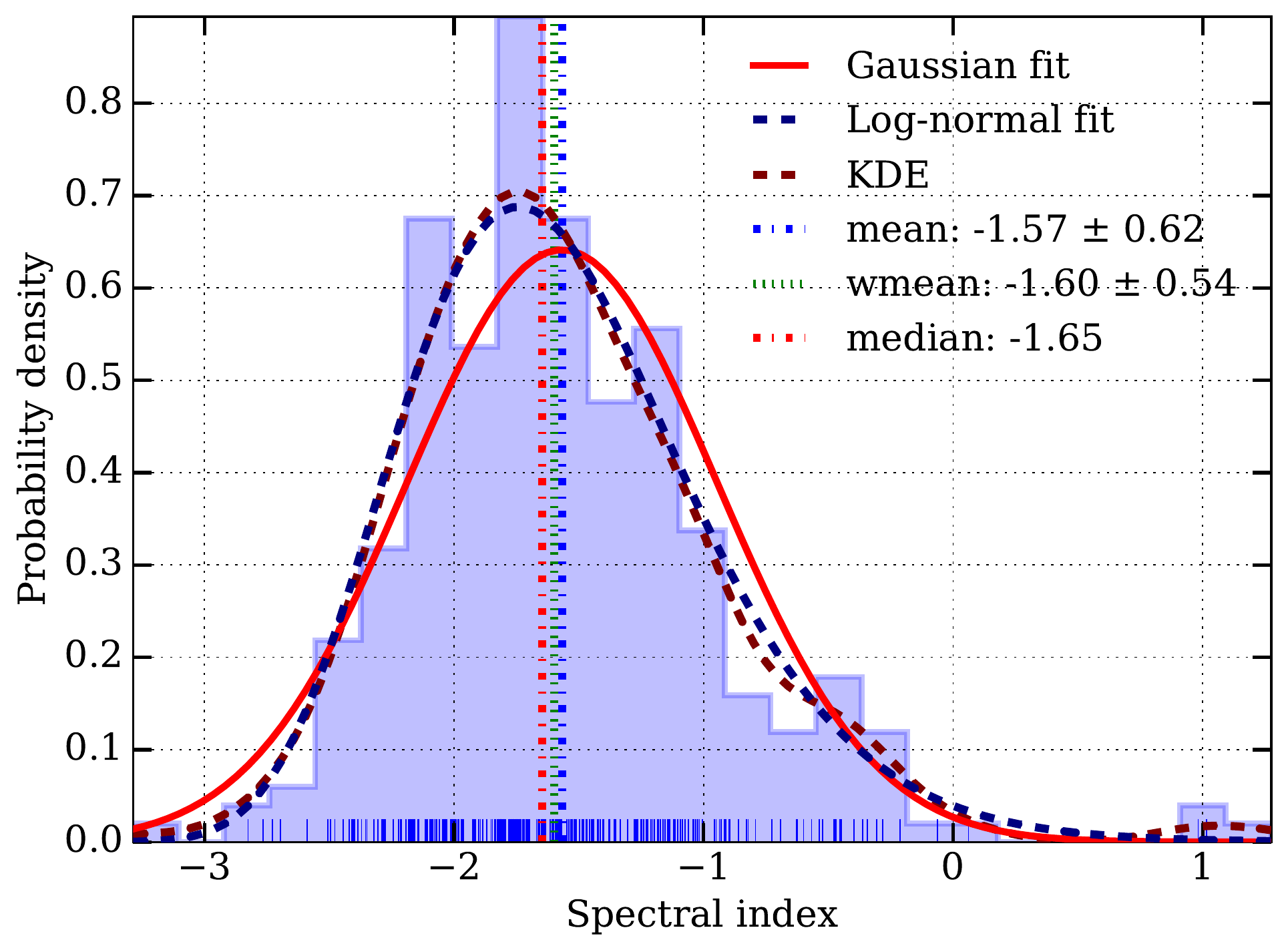}
	\caption{Histogram of the spectral indices $\alpha$ for all pulsars that were classified to have simple power law spectra.}
	\label{fig:SpectralIndexHistogram}
\end{figure}

The majority of pulsars, $276$ in total, have spectra that follow a simple power law. The best-fitting spectral indices are shown in Table~\ref{tab:FluxDensities} in the appendix. A histogram of the resulting spectral indices is shown in Fig.~\ref{fig:SpectralIndexHistogram}. The mean spectral index is $-1.57 \pm 0.62$, the one weighted by the uncertainty of each individual index is $-1.60 \pm 0.54$ and the median spectral index is $-1.65$. The uncertainty given here is the (weighted) standard deviation. When quoting the standard error instead, the weighted mean spectral index is $-1.60 \pm 0.03$. We also show a Gaussian and a shifted log-normal fit to the data, together with a kernel density estimation (KDE) using a Gaussian kernel. The KDE largely agrees with the log-normal fit. The spectral index distribution deviates from a Gaussian and has a tail that extends towards positive values. We used the Shapiro--Wilk (S--W) test for normality to check quantitatively whether the observed spectral indices are sampled from a Gaussian or a log-normal distribution, for which we transformed the data into logarithmic space according to $z = \log_{10} \left( \alpha + 5 \right)$ and applied the S--W test to the transformed data. We find that we have to reject the null hypothesis in both cases. However, for the log-normal distribution it is close with values $0.99$ ($0.02$). A detailed investigation of the unbinned cumulative distribution function and Q-Q plots for both distributions leads us to conclude that a log-normal distribution fits the data nearly perfectly, while a Gaussian shows deviations. When we constrain the spectral indices to the $257$ that are well determined with uncertainties of less than $0.5$, the S--W test indicates with values $0.99$ ($0.17$), that we cannot reject the null hypothesis that the data was sampled from a log-normal distribution. The best-fitting values are $0.2$, $-4.6$ and $3.0$ for the shape, location and scale parameter, which corresponds to a skew of $0.6$ and excess kurtosis of $0.7$. We still have to reject the null hypothesis for a Gaussian distribution.

\begin{table*}
\caption{Pulsars with flat, or positive simple power law spectra. We show their DM, spectral index $\alpha$, characteristic age $\tau$, spin-down luminosity $\dot{E}$ and the classification category.}
\label{tab:PlWithFlatSpectra}
\centering
\begin{tabular}{lllllll}
\hline
PSRJ	& $p_\text{best}$	& DM					& $\alpha$	& $\tau$	& $\dot{E}$	& category\\
	& 			& [$\text{pc} \: \text{cm}^{-3}$]	& 		& [yr]		& [$\text{erg} \: \text{s}^{-1}$]	&\\
\hline
J1028--5819 & 1.00 & 96.5 & $1.3 \pm 0.8$ & $9.00 \cdot 10^{4}$ & $8.30 \cdot 10^{35}$ & weak\\
J1650--4921 & 1.00 & 229.9 & $0.1 \pm 0.2$ & $1.36 \cdot 10^{6}$ & $1.90 \cdot 10^{34}$ & weak\\
J1653--4249 & 0.97 & 416.1 & $1.0 \pm 0.6$ & $2.02 \cdot 10^{6}$ & $8.30 \cdot 10^{32}$ & weak\\
J1832--0644 & 1.00 & 578.0 & $1.0 \pm 0.5$ & $3.18 \cdot 10^{5}$ & $3.60 \cdot 10^{33}$ & strong\\
\hline
\end{tabular}
\end{table*}

We investigate the pulsars at the extremes of the spectral index distribution individually and in particular the ones with flat, or positive spectral indices. We list these in Table~\ref{tab:PlWithFlatSpectra}. All pulsars except for PSR J1832--0644 have only a weak classification. Interestingly, PSR J1028--5819 is close to having a GPS with a peak at roughly 2~GHz. However, a simple power law is preferred because of the large uncertainty of the data at 3.1~GHz. In other aspects the pulsar is special because it has an extremely narrow pulse profile with a FWHM of only $0.3 \: \text{ms}$ and a corresponding duty cycle of about $0.3 \%$ at 1.4~GHz, which is among the narrowest profiles known. It was suggested that this is because we are grazing the emission beam \citep{2008Keith}. Moreover, profile components at the edges of beams seem to exhibit flatter spectral indices in general \citep{1988Lyne, 1994Kramer, 2015Dai} and this is what we could see here. For PSR J1832--0644 the spectral index is determined from our data at 3.1~GHz and literature data at 1.4~GHz from PMPS \citep{2002Morris} only. The positive spectral index is likely caused by a difference in absolute flux density scale, see \S\ref{sec:ComparisonWithLiteratureData}. The pulsars with the steepest spectra are PSR J1059--5742 ($-3.3 \pm 0.4$), which has a well determined spectrum from 0.4 to 3.1~GHz, and PSR J1833--0338 ($-2.8 \pm 0.1$) with an equally well determined spectrum from 0.1 to 3.1~GHz and a hint of a spectral break below 400~MHz. Both have a candidate classification.

We also tested whether the Galactic plane affected the measured spectral indices and in particular if a hot spot in the Galactic plane led to an underestimate of the low frequency flux densities, resulting in a shallow spectral index. We do not find any correlation of spectral index with Galactic latitude or longitude. Generally speaking, most of the pulsars are young ($\tau \leq 3.2 \cdot 10^{5} \: \text{yr}$) or energetic ($\dot{E} \geq 10^{34} \: \text{erg} \: \text{s}^{-1}$), indicating that these have flatter spectral indices on average.

\subsection{Log-parabolic spectra}
\label{sec:LogParabolicSpectra}

\begin{table*}
\caption{Pulsars that have log-parabolic spectra, where $a$ is the curvature coefficient, $b$ the spectral index for the case $a = 0$, $\nu_\text{p}$ the peak frequency, $f_\text{scat}$ the expected pulse width at 1.4~GHz due to scatter broadening in the ISM as a fraction of period and the classification category. We quote uncertainties at the $1 \sigma$ level. We also show their DMs, whether they are in binary systems and associations with other sources, where O is an optical observation of a white dwarf companion, X an X-ray, $\gamma$ a $\gamma$-ray source, S denotes a supernova remnant and P a pulsar wind nebula. The MSPs with $P \leq 30 \: \text{ms}$ are marked with $^\dagger$. $^\ddagger$ The pulse profile at 1.4~GHz is strongly scatter broadened and the flux densities at and below that frequency are most likely underestimated, leading to a spurious classification.}
\label{tab:LPSPulsars}
\centering
\begin{tabular}{lllllllllll}
\hline
PSRJ	& $p_\text{best}$ & DM 					& assoc	& binary?		& $a$	& $b$	& $\nu_\text{p}$ & category & $f_\text{scat}$ & comment\\
	& 			& [$\text{pc} \: \text{cm}^{-3}$]	&	& 	& 	&	& [MHz] & & &\\
\hline
J0659+1414 & 0.55 & 14.0 & S,$\gamma$ & No & $0.5 \pm 0.3$ & $-0.7 \pm 0.2$ & -- & weak & 0.05\\
J0711--6830$^\dagger$ & 0.46 & 18.4 & -- & No & $-1.6 \pm 0.5$ & $-1.6 \pm 0.1$ & $400 \pm 100$ & weak & 0.28\\
J0820--4114 & 0.59 & 113.4 & -- & No & $-0.7 \pm 0.3$ & $-2.2 \pm 0.3$ & $40 \pm 60$ & candidate & 0.28\\
J0823+0159 & 0.47 & 23.7 & -- & Yes & $-0.9 \pm 0.3$ & $-2.3 \pm 0.2$ & $60 \pm 80$ & weak & 0.03\\
J0907--5157 & 0.41 & 103.7 & -- & No & $-0.2 \pm 0.1$ & $-1.1 \pm 0.1$ & -- & weak & 0.08\\
J0908--4913 & 0.73 & 180.4 & -- & No & $-0.8 \pm 0.2$ & $-0.8 \pm 0.1$ & $400 \pm 100$ & strong & 0.04\\
J0934--5249 & 0.42 & 100.0 & -- & No & $-1.7 \pm 0.6$ & $-3.0 \pm 0.3$ & $200 \pm 100$ & weak & 0.02\\
J0959--4809 & 0.62 & 92.7 & -- & No & $-0.8 \pm 0.3$ & $-2.1 \pm 0.3$ & $80 \pm 90$ & candidate & 0.18\\
J1019--5749 & 0.49 & 1039.4 & -- & No & $-2.6 \pm 1.5$ & $0.5 \pm 0.6$ & $1600 \pm 500$ & weak & 9.76 & $\ddagger$\\
J1024--0719$^\dagger$ & 0.64 & 6.5 & X,$\gamma$ & No & $0.6 \pm 0.3$ & $-1.3 \pm 0.1$ & -- & candidate & 0.12\\
J1055--6028 & 0.93 & 635.9 & $\gamma$ & No & $-3.3 \pm 0.8$ & $-1.2 \pm 0.2$ & $900 \pm 100$ & clear & 0.70 & potential GPS\\
J1057--5226 & 0.51 & 30.1 & X,$\gamma$,O & No & $-0.2 \pm 0.1$ & $-2.2 \pm 0.1$ & -- & weak & 0.08\\
J1410--6132 & 0.80 & 960.0 & -- & No & $-2.4 \pm 0.5$ & $0.9 \pm 0.3$ & $2000 \pm 300$ & clear & 18.79 & $\ddagger$\\
J1512--5759 & 0.41 & 628.7 & -- & No & $-1.6 \pm 0.3$ & $-1.6 \pm 0.1$ & $410 \pm 80$ & weak & 0.57\\
J1635--5954 & 0.49 & 134.9 & -- & No & $-1.8 \pm 0.9$ & $-1.9 \pm 0.3$ & $400 \pm 300$ & weak & 0.03\\
J1658--4958 & 0.50 & 193.4 & -- & No & $-2.2 \pm 1.0$ & $-1.9 \pm 0.3$ & $500 \pm 200$ & weak & 0.04\\
J1703--3241 & 0.47 & 110.3 & -- & No & $-1.1 \pm 0.5$ & $-1.5 \pm 0.2$ & $300 \pm 200$ & weak & 0.04\\
J1705--3950 & 0.73 & 207.1 & -- & No & $-1.7 \pm 0.6$ & $-0.0 \pm 0.2$ & $1300 \pm 200$ & strong & 0.03 & new GPS\\
J1723--3659 & 0.74 & 254.2 & -- & No & $-1.4 \pm 0.5$ & $-0.7 \pm 0.2$ & $700 \pm 200$ & strong & 0.05 & known GPS\\
J1727--2739 & 0.50 & 147.0 & -- & No & $-2.3 \pm 1.2$ & $-1.7 \pm 0.2$ & $500 \pm 300$ & weak & 0.08\\
J1731--4744 & 0.72 & 123.3 & -- & No & $-0.4 \pm 0.2$ & $-1.7 \pm 0.1$ & -- & strong & 0.03\\
J1745--3040 & 0.74 & 88.4 & -- & No & $-0.9 \pm 0.2$ & $-1.4 \pm 0.1$ & $220 \pm 90$ & strong & 0.03\\
J1752--2806 & 0.54 & 50.4 & -- & No & $-1.3 \pm 0.1$ & $-2.6 \pm 0.1$ & $120 \pm 30$ & candidate & 0.01\\
J1812--1733 & 0.46 & 518.0 & -- & No & $-1.2 \pm 0.7$ & $-1.5 \pm 0.3$ & $300 \pm 300$ & weak & 0.28\\
J1824--1945 & 0.62 & 224.6 & -- & No & $-0.3 \pm 0.1$ & $-2.0 \pm 0.1$ & -- & candidate & 0.01\\
J1825--1446 & 0.64 & 357.0 & -- & No & $-0.7 \pm 0.2$ & $-0.1 \pm 0.1$ & $1100 \pm 200$ & candidate & 0.06 & known GPS\\
J1826--1334 & 0.80 & 231.0 & $\gamma$,X,P & No & $-1.1 \pm 0.3$ & $0.0 \pm 0.1$ & $1400 \pm 100$ & clear & 0.08 & known GPS\\
J1830--1059 & 0.81 & 161.5 & -- & No & $-2.2 \pm 0.6$ & $-0.6 \pm 0.2$ & $900 \pm 100$ & clear & 0.02 & potential GPS\\
J1832--0827 & 0.58 & 300.9 & -- & No & $-0.8 \pm 0.3$ & $-1.2 \pm 0.1$ & $300 \pm 100$ & candidate & 0.01\\
J1835--0643 & 0.77 & 472.9 & -- & No & $-1.2 \pm 0.4$ & $-1.9 \pm 0.2$ & $200 \pm 100$ & strong & 0.13\\
J1835--1020 & 0.60 & 113.7 & -- & No & $-1.2 \pm 0.3$ & $-0.8 \pm 0.1$ & $600 \pm 200$ & candidate & 0.02 & known GPS\\
J1836--1008 & 0.60 & 317.0 & -- & No & $-1.0 \pm 0.4$ & $-2.5 \pm 0.1$ & $70 \pm 80$ & candidate & 0.03\\
J1843--0211 & 0.52 & 441.7 & -- & No & $-1.8 \pm 0.7$ & $-1.2 \pm 0.3$ & $600 \pm 200$ & candidate & 0.03 & potential GPS\\
J1847--0402 & 0.68 & 142.0 & -- & No & $0.5 \pm 0.2$ & $-2.0 \pm 0.1$ & -- & candidate & 0.04\\
J1857+0212 & 0.67 & 506.8 & -- & No & $0.6 \pm 0.3$ & $-1.4 \pm 0.1$ & -- & candidate & 0.09\\
\hline
\end{tabular}
\end{table*}

The pulsars listed in Table~\ref{tab:LPSPulsars} have log-parabolic spectra with significant curvature. In the table we show the best-fitting parameters $a$ and $b$, the derived peak frequency $\nu_\text{p}$, the probability $p_\text{best}$ and the classification category. We also show their DMs, whether they are in binary systems and potential associations with optical, X-ray and $\gamma$-ray sources taken from the pulsar catalogue and identify the millisecond pulsars (MSPs) with a $^\dagger$. We estimate the expected scatter broadening of the pulse profiles in the ISM as a fraction of pulse period using the empirical formula of \citet{2004Bhat}:
\begin{equation}
	\log \tau_s = -6.46 + 0.154 \log \text{DM} + 1.07 \left( \log \text{DM} \right)^2 - 3.86 \log \nu,
	\label{eq:ScatterBroadening}
\end{equation}
where $\nu$ is the observing frequency in GHz and $\tau_s$ is the scattering time assuming thin screen scattering given in ms. We denote the expected ratio of pulse width at 1.4~GHz and pulse period as $f_\text{scat}$.

The table contains three separate classes of LPS pulsars: 1) Four pulsars with slightly concave spectra with positive curvature coefficients. Their classification categories are either weak or candidate and all of them have flux density measurements with relatively large uncertainties below 400~MHz. We expect that their spectral classifications will shift towards simple power laws once low-frequency data are available. 2) 21 pulsars with spectral peaks at frequencies up to about 500~MHz and 3) 10 pulsars whose spectra peak between about 0.6 to $2 \: \text{GHz}$, indicating that they belong to the class of GPS pulsars. The two pulsars with the highest DMs of close to $1000 \: \text{pc} \: \text{cm}^{-3}$, PSRs J1019--5749 and J1410--6132, show large amounts of scatter broadening of their profiles at 1.4~GHz, covering $50$ to $70 \%$ of pulse longitude, and large expected $f_\text{scat}$. Their flux densities at and below 1.4~GHz are most likely underestimated and the LPS classification a result of this. Interferometric techniques are needed to determine their flux densities accurately below 1.4~GHz, see e.g. \citet{2015Dembska}. We exclude them from further discussion. The remaining pulsars include four known, one newly identified and three potential GPS pulsars. We discuss the GPS pulsars separately in \S\ref{sec:GigahertzPeakedSpectra} and describe a small selection of LPS pulsars below.

PSR J0823+0159: This is the only pulsar in a binary system. We have good frequency coverage from 25~MHz to 4.8~GHz. A curved spectrum is slightly preferred with a peak at around 60~MHz. PSR J1024--0719: It is an MSP and studied as part of the PPTA with good spectral coverage from 100~MHz to 5~GHz. A LPS is preferred with a concave spectral shape and a candidate category. It seems that the spectrum curves up at low and high frequencies. PSR J1512--5759: The 50~cm data have a positive spectral index and a LPS is weakly preferred with a broken power law being second.

With the aim of understanding the physical origin of the LPS phenomenon, we list the available information about the environments in which the pulsars are located and search for associations with sources at other frequencies, see Table~\ref{tab:LPSPulsars}. The table contains two MSPs and only one pulsar that is known to be in a binary system. PSR J0823+0159 is the only case in which the LPS could be due to absorption in the wind of a companion. However, J0823+0159's companion is a DA white dwarf in a wide orbit \citep{2000Koester, 2005vanKerkwijk}, which should not have a significant wind. In addition, we tested for correlations of the curvature parameter $a$ and the parameter $b$ with DM, spin frequency $\tilde{\nu}$ and spin-down rate $\dot{\tilde{\nu}}$ for all LPS pulsars. We do not find any significant dependence of $a$ and $b$ on DM. Apart from that we see an increase of $b$ with increasing spin frequency, except for the MSPs, and with increasing absolute spin-down rate for all pulsars that have a measured $\dot{\tilde{\nu}}$. The dependence is similar to what we find for the spectral indices of pulsars with simple power law spectra, see \S\ref{sec:CorrelationsOfSpectralIndexWithPulsarParameters}. The curvature parameter $a$ however appears to be uncorrelated with $\tilde{\nu}$ and $\dot{\tilde{\nu}}$.

\subsection{Broken power law spectra}
\label{sec:BrokenPowerLawSpectra}

\begin{table*}
\caption{Pulsars that have broken power law spectra, where $\nu_\text{br}$ is the frequency of the spectral break, $\alpha_1$ and $\alpha_2$ the spectral index before and after the break. We also show the classification category and mark MSPs with $^\dagger$.}
\label{tab:BrokenPLPulsars}
\centering
\begin{tabular}{lllllll}
\hline
PSRJ	& $p_\text{best}$ & $\nu_\text{br}$ & $\alpha_1$ & $\alpha_2$ & category	& comment\\
	&			& [MHz]			&	&	&	&\\
\hline
J0437--4715$^\dagger$ & 1.00 & $1900 \pm 400$ & $-0.85 \pm 0.01$ & $-2.5 \pm 0.6$ & strong\\
J0543+2329 & 0.91 & $800 \pm 90$ & $-0.3 \pm 0.2$ & $-1.5 \pm 0.1$ & clear\\
J0742--2822 & 0.87 & $1400 \pm 2$ & $-2.11 \pm 0.08$ & $-1.59 \pm 0.09$ & clear\\
J0820--1350 & 0.97 & $500 \pm 100$ & $-1.1 \pm 0.2$ & $-2.4 \pm 0.1$ & clear\\
J0835--4510 & 1.00 & $880 \pm 50$ & $-0.55 \pm 0.03$ & $-2.24 \pm 0.09$ & clear\\
J0837--4135 & 0.95 & $740 \pm 20$ & $-0.1 \pm 0.1$ & $-1.8 \pm 0.2$ & clear\\
J0840--5332 & 0.48 & $730 \pm 20$ & $-1.1 \pm 0.2$ & $-3.2 \pm 0.4$ & weak\\
J0856--6137 & 0.95 & $736 \pm 3$ & $-2.3 \pm 0.1$ & $-0.5 \pm 0.4$ & clear\\
J0942--5552 & 0.96 & $1100 \pm 200$ & $-1.0 \pm 0.1$ & $-2.3 \pm 0.1$ & clear\\
J1001--5507 & 0.50 & $340 \pm 80$ & $-0.0 \pm 0.6$ & $-1.8 \pm 0.1$ & candidate\\
J1045--4509$^\dagger$ & 0.45 & $920 \pm 70$ & $-1.1 \pm 0.1$ & $-2.18 \pm 0.07$ & weak\\
J1136+1551 & 1.00 & $300 \pm 5$ & $0.1 \pm 0.1$ & $-2.12 \pm 0.05$ & clear\\
J1243--6423 & 0.45 & $1700 \pm 400$ & $-3.8 \pm 0.5$ & $-1.0 \pm 1.0$ & weak\\
J1327--6222 & 0.78 & $717 \pm 3$ & $-0.7 \pm 0.1$ & $-2.3 \pm 0.1$ & strong\\
J1359--6038 & 0.86 & $320 \pm 60$ & $-0.5 \pm 0.6$ & $-2.28 \pm 0.05$ & clear\\
J1453--6413 & 0.99 & $320 \pm 30$ & $-0.4 \pm 0.2$ & $-2.5 \pm 0.1$ & clear\\
J1522--5829 & 0.68 & $1400 \pm 6$ & $-2.8 \pm 0.4$ & $-0.0 \pm 0.4$ & candidate\\
J1651--5255 & 0.46 & $1000 \pm 300$ & $-1.0 \pm 2.0$ & $-2.6 \pm 0.2$ & weak\\
J1743--3150 & 0.66 & $610 \pm 40$ & $0.8 \pm 0.3$ & $-2.2 \pm 0.3$ & candidate & known GPS\\
J1751--3323 & 0.43 & $1279 \pm 3$ & $0.6 \pm 0.3$ & $-1.0 \pm 0.2$ & weak & new GPS\\
J1806--1154 & 0.62 & $900 \pm 100$ & $1.0 \pm 2.0$ & $-2.9 \pm 0.3$ & candidate & new GPS\\
J1852--0635 & 1.00 & $1279 \pm 6$ & $1.1 \pm 0.2$ & $-0.77 \pm 0.02$ & clear & known GPS\\
J1900--2600 & 0.95 & $800 \pm 200$ & $-1.34 \pm 0.08$ & $-2.5 \pm 0.2$ & clear\\
J2048--1616 & 0.98 & $950 \pm 6$ & $-0.57 \pm 0.09$ & $-2.6 \pm 0.1$ & clear\\
J2053--7200 & 0.42 & $950 \pm 20$ & $-1.0 \pm 6.0$ & $-4.0 \pm 6.0$ & weak\\
\hline
\end{tabular}
\end{table*}

The pulsars with broken power law spectra are listed in Table~\ref{tab:BrokenPLPulsars}. The most prominent examples are PSR J0437--4715, the brightest and closest MSP, whose spectrum breaks around 2~GHz and is well determined from 70~MHz to 17~GHz, the Vela pulsar (PSR J0835--4510), which is well studied from 70~MHz to 24.4~GHz and seems to have a flat spectrum below 900~MHz; and the bright pulsar J0742--2822, whose spectrum flattens slightly above 1.4~GHz. Other examples are PSR J1045--4509, an MSP studied as part of the PPTA and the pulsar J1522--5829, whose spectrum seems to be flat above 1.4~GHz.

\subsection{Power law spectra with high-frequency cut-off and magnetic field in polar cap}
\label{sec:MagneticFieldInPolarCap}

\begin{table}
\caption{Pulsars that have power law spectra with a hard high-frequency cut-off at a frequency $\nu_c$, their classification category, the magnetic field at the neutron star surface and at the light-cylinder radius, at the magnetic poles. We list the inferred magnetic fields $B_\text{pc}$ at the polar cap centres from our spectral fit and their corresponding altitudes $z$.}
\label{tab:HardCutoffPulsarsMaximumEandB}
\centering
\begin{tabular}{llll}
\hline
PSRJ								& J1707--4053		& J1751--4657		& J1903--0632\\
\hline
$p_\text{best}$					& 0.32				& 0.52				& 0.59\\
category							& weak				& candidate			& candidate\\
$\nu_c$	[MHz]					& $8100 \pm 500$		& $3900 \pm 400$		& $3900 \pm 400$\\
$B_\text{surf}$ [$10^{12}$ G]		& 2.14				& 1.98				& 2.45\\
$B_\text{LC}$ [G]				& 100.0				& 44.6				& 280.0\\
$B_\text{pc}$ [$10^{11}$ G]		& $6.9 \pm 0.4$		& $2.0 \pm 0.2$		& $1.2 \pm 0.1$\\
$z$ [km]							& $14.6 \pm 0.9$		& $21 \pm 2$			& $27 \pm 3$\\
$z/R_\text{LC}$ [$\%$]			& $0.05 \pm 0.01$	& $0.06 \pm 0.01$	& $0.13 \pm 0.01$\\
\hline
\end{tabular}
\end{table}

For the pulsars that have spectra with a hard high-frequency cut-off we determine the magnetic field strength at the centre of the polar cap from our fits to the spectral data by inverting Eq.~\ref{eq:HardCutOffFrequency}, which yields:
\begin{equation}
	B_\text{pc} = \frac{m_e c}{\pi e} \: P \: \nu_\text{c}^2,
	\label{eq:BFieldHardCutOffPulsars}
\end{equation}
where $m_e$ and $e$ are the mass and charge of the electron, $c$ is the speed of light, $P$ is the period of the pulsar and $\nu_\text{c}$ is the cut-off frequency. The pulsars are listed in Table~\ref{tab:HardCutoffPulsarsMaximumEandB} together with the magnetic fields at the surface $B_\text{surf}$ and at the light-cylinder radius $B_\text{LC}$ calculated assuming spin-down due to magnetic dipole radiation and the magnetic field at the centre of the polar cap $B_\text{pc}$ determined from the spectral fit. We calculate the values at the magnetic poles for a canonical neutron star and $90 \degr$ inclination, which are a factor of two higher than the values at the magnetic equator that are conventionally quoted. Under the simplistic assumption that the magnetic field strength falls off as a magnetic dipole field, i.e. proportional to $z^{-3}$, where $z$ is the distance from centre of the star, we can use the magnetic field strength at the centre of the polar cap to determine the height of the polar cap centre $z$. These values are also given in Table~\ref{tab:HardCutoffPulsarsMaximumEandB}.

The measured magnetic field strengths are between about $1$ and $7 \cdot 10^{11} \text{G}$ and the inferred altitudes of the polar cap centres vary from $15$ to $27 \: \text{km}$ ($0.05$ to $0.1 \% \: R_\text{LC}$). This suggests that the pulsar emission is produced extremely close to the surface. The derived values are much smaller than what is traditionally considered. Usually the polar cap radio emission is thought to be created at an altitude lower than $10 \%$ of the light cylinder radius, at a height of less than about $100$ to $400 \: \text{km}$ as determined by the most recent measurements that were conducted with focus on low frequencies \citep{2012Hassall, 2016Perera}. However, radius-to-frequency mapping predicts that low frequency emission is created higher in the magnetosphere than that at higher frequency \citep{1978Cordes}. As we are probing much higher frequencies of up to  8~GHz, the expected altitude should be lower than a few hundreds of kilometres. Interestingly, other authors have also derived very low emission altitudes of about 29~km ($0.24 \% \: R_\text{LC}$) from pulse separation measurements at 42 and 74~MHz \citep{2016Tsai}, which is a completely different method than our spectral fitting. Nevertheless, the derived emission altitudes seem unreasonably low, which could point to problems in the spectral classification (the highest category is candidate), or to problems in the theoretical emission model of \citet{2013Kontorovich}, its underlying assumptions, or to deviations from a simple magnetic dipole scaling of the field strength near the neutron star surface. Further high-frequency data will help to better constrain the spectra and may resolve this issue.

\subsection{Power law spectra with low-frequency turn-over}
\label{sec:PLSpectraWithLowFrequencyTurnOver}

\begin{table}
\caption{Top: Pulsars that have power law spectra with a low-frequency turn-over, where $\nu_\text{c}$ is the turn-over frequency, $\alpha$ the spectral index and $\beta$ the free fit parameter that determines the smoothness of the turn-over. $^\text{g}$ New, known, or potential GPS pulsar. $^\dagger$ MSP. Bottom: The same, but for the special case where the smoothness parameter $\beta$ is fixed to 2.1, which corresponds to a turn-over due to free-free absorption. This is separate from the rest of the analysis presented in this paper. Pulsar spectra are not classified based on this special case and the given $p_\text{best}$ and classification categories should not be compared with the other tables (see text).}
\label{tab:LowFreqCutOffPulsars}
\centering
\begin{tabular}{lllll}
\hline
PSRJ	& $p_\text{best}$	& $\nu_\text{c}$		& $\alpha$ 	& $\beta$\\
	&					& [MHz]				&			&\\
\hline
J0630--2834 & 0.98 & $70 \pm 6$ & $-1.7 \pm 0.1$ & $1.8 \pm 0.4$\\
J0809--4753 & 0.66 & $120 \pm 10$ & $-2.7 \pm 0.3$ & $1.1 \pm 0.4$\\
J0837+0610 & 0.77 & $53 \pm 7$ & $-3.4 \pm 0.7$ & $0.5 \pm 0.2$\\
J0922+0638 & 0.96 & $45 \pm 7$ & $-1.8 \pm 0.2$ & $0.9 \pm 0.4$\\
J0942--5657 & 0.83 & $100 \pm 10$ & $-2.7 \pm 0.1$ & $2.1 \pm 0.3$\\
J0953+0755 & 1.00 & $91 \pm 6$ & $-2.6 \pm 0.2$ & $0.54 \pm 0.08$\\
J1644--4559$^\text{g}$ & 1.00 & $601 \pm 7$ & $-3.14 \pm 0.02$ & $2.10 \pm 0.01$\\
J1651--4246 & 0.80 & $60 \pm 20$ & $-2.5 \pm 0.2$ & $0.8 \pm 0.3$\\
J1803--2137$^\text{g}$ & 0.97 & $900 \pm 100$ & $-1.2 \pm 0.4$ & $1.6 \pm 0.6$\\
J1913--0440 & 0.49 & $150 \pm 20$ & $-2.2 \pm 0.3$ & $1.4 \pm 0.6$\\
\hline
\multicolumn{5}{c}{Special case of fixed $\beta = 2.1$ (free-free absorption)}\\
PSRJ		& $p_\text{best}$	& $\nu_\text{c}$	& $\alpha$	& category\\
		&					& [MHz]			&			&\\
\hline
J0711--6830$^\dagger$ & 0.60 & $530 \pm 50$ & $-2.2 \pm 0.2$ & candidate\\
J0809--4753 & 0.59 & $123 \pm 8$ & $-2.37 \pm 0.07$ & candidate\\
J0908--4913 & 0.55 & $580 \pm 60$ & $-1.2 \pm 0.1$ & candidate\\
J0922+0638 & 0.91 & $41 \pm 3$ & $-1.62 \pm 0.06$ & clear\\
J0934--5249 & 0.52 & $360 \pm 40$ & $-3.6 \pm 0.4$ & candidate\\
J0942--5657 & 0.95 & $100 \pm 10$ & $-2.7 \pm 0.1$ & clear\\
J0943+1631 & 0.53 & $59 \pm 8$ & $-1.6 \pm 0.4$ & candidate\\
J1001--5507 & 0.42 & $190 \pm 20$ & $-1.75 \pm 0.09$ & weak\\
J1017--5621 & 0.33 & $500 \pm 100$ & $-2.3 \pm 0.2$ & weak\\
J1057--5226 & 0.39 & $50 \pm 10$ & $-2.04 \pm 0.07$ & weak\\
J1512--5759 & 0.52 & $710 \pm 40$ & $-2.6 \pm 0.1$ & candidate\\
J1614--5048 & 0.43 & $800 \pm 100$ & $-2.6 \pm 0.2$ & weak\\
J1635--5954 & 0.50 & $630 \pm 90$ & $-2.7 \pm 0.5$ & weak\\
J1644--4559$^\text{g}$ & 1.00 & $601 \pm 7$ & $-3.14 \pm 0.02$ & clear\\
J1651--4246 & 0.47 & $90 \pm 6$ & $-2.19 \pm 0.03$ & weak\\
J1658--4958 & 0.54 & $760 \pm 90$ & $-3.2 \pm 0.6$ & candidate\\
J1705--3950$^\text{g}$ & 0.55 & $1100 \pm 100$ & $-0.9 \pm 0.4$ & candidate\\
J1723--3659$^\text{g}$ & 0.48 & $650 \pm 60$ & $-1.2 \pm 0.2$ & weak\\
J1727--2739 & 0.53 & $800 \pm 100$ & $-2.8 \pm 0.6$ & candidate\\
J1743--3150$^\text{g}$ & 0.79 & $480 \pm 20$ & $-2.9 \pm 0.2$ & strong\\
J1751--3323$^\text{g}$ & 0.45 & $1000 \pm 100$ & $-1.4 \pm 0.3$ & weak\\
J1803--2137$^\text{g}$ & 0.99 & $800 \pm 30$ & $-0.9 \pm 0.1$ & clear\\
J1806--1154$^\text{g}$ & 0.48 & $650 \pm 70$ & $-3.4 \pm 0.4$ & weak\\
J1832--0827 & 0.46 & $500 \pm 60$ & $-1.5 \pm 0.2$ & weak\\
J1835--1020$^\text{g}$ & 0.71 & $560 \pm 50$ & $-1.3 \pm 0.2$ & strong\\
J1836--1008 & 0.49 & $390 \pm 50$ & $-2.9 \pm 0.2$ & weak\\
J1843--0211$^\text{g}$ & 0.51 & $800 \pm 100$ & $-2.3 \pm 0.3$ & candidate\\
J1913--0440 & 0.72 & $160 \pm 20$ & $-2.1 \pm 0.1$ & strong\\
\hline
\end{tabular}
\end{table}

The pulsars that have power law spectra with a low-frequency turn-over are listed in Table~\ref{tab:LowFreqCutOffPulsars}. Notable examples are the pulsars J0630--2834, J0837+0610 and J0953+0755, whose spectra are very well determined from literature data and our own measurements from about 20~MHz up to 10, 4.9 and 14.8~GHz. They are among the pulsars with the most data points in our set. Their spectral classification is clear and they have turn-over frequencies of 50 to 100~MHz. A further example is the pulsar J1803--2137, which has a turn-over near 1~GHz. None of these pulsars is in a binary system, but many are visible in X-rays, suggesting a connection of the turn-over with absorption in a PWN. Most of the pulsars have strong to clear classifications, or are very close to it. Half of them have smooth turn-overs with $\beta$ near or below unity and half have sharper transitions with $\beta \geq 1.5$.

\subsubsection{Examining evidence for free-free absorption}
\label{sec:FreeFreeAbsorption}

The thermal free-free absorption model as implemented by \citet{2016Rajwade} or \citet{2017Kijak} is a special case of this model, where the smoothness parameter $\beta$ is fixed to a constant value of 2.1. As such, it has three free parameters, one less than the more generic low-frequency turn-over model. For completeness and to enable comparison with previous work we test this case separately from the rest of the analysis presented in this paper. That is, we compare the AIC of the free-free absorption model to those of the four other models (as described in \S\ref{sec:SpectralModels}) and present the parameters of the spectra that are best characterised by it in the bottom part of Table~\ref{tab:LowFreqCutOffPulsars}. However, the values of $p_\text{best}$ and the classification categories in the bottom part of Table~\ref{tab:LowFreqCutOffPulsars} should not be compared with $p_\text{best}$ and the categories presented for other models (in Tables~\ref{tab:LPSPulsars}, \ref{tab:BrokenPLPulsars} and \ref{tab:HardCutoffPulsarsMaximumEandB}). The free-free absorption model is not included among the models compared in the rest of this work; therefore its AIC is not included in the calculation of $p_\text{best}$ (Eq.~\ref{eq:ProbabilityOfBestModel}) as presented in other tables. Consequently, the bottom of Table~\ref{tab:LowFreqCutOffPulsars} contains some pulsars that have higher values of $p_\text{best}$ for other models (e.g. PSRs J1705--3950 and J1723--3659 in Table~\ref{tab:LPSPulsars}); although the free-free absorption model becomes the preferred model for these pulsars when it is included among those compared, the preference is less significant. On the other hand, the table also contains a few pulsars that have higher values of $p_\text{best}$ for the free-free absorption model (e.g. PSRs J1743--3150 and J1835--1020 in Tables~\ref{tab:BrokenPLPulsars} and \ref{tab:LPSPulsars}).

Generally the table contains most of the pulsars that have low-frequency turn-over spectra, all newly identified GPS pulsars, one of the potential GPS pulsars and five, but not all, of the known GPS pulsars (see \S\ref{sec:GigahertzPeakedSpectra}). In addition, it contains a small number of pulsars that otherwise have mainly weak log-parabolic, simple, or broken power law classifications. This suggests that the turn-over in these spectra could be due to free-free absorption, at least for the spectra with relatively high turn-over frequencies near 500~MHz or above. However, we do not classify them as free-free absorption for the following reason: there is considerable covariance between the $\beta$ parameter and the other fit parameters, in the sense that for other fixed values of $\beta$ between 0.5 and 2.1 a similar set of pulsars is objectively characterised as a power law spectrum with low-frequency turn-over (and fixed $\beta$), but with shifted values of spectral index, turn-over frequency and normalisation. This is because the spectral coverage at low frequencies is insufficient or too uncertain to unambiguously resolve the sharpness of the turn-over for many of these pulsars. For this reason, we prefer the more generic turn-over model, where $\beta$ is a free fit parameter, and classify the spectra based on that.

\subsection{Gigahertz-peaked spectra}
\label{sec:GigahertzPeakedSpectra}

11 pulsars in our data set show GPS, by which we mean spectra with peaks between about 0.6 to 2.0~GHz, regardless of best-fitting spectral model. 3 others are potential GPS pulsars. The distribution of best-fitting models is dominated by the LPS, with 8 pulsars having a LPS, 4 a broken power law and 2 a power law with low-frequency turn-over. In addition, we independently measured flux densities for all known GPS pulsars from the literature, except PSR J1740+1000. In particular, these are PSRs J1056--6258, J1743--3150, J1803--2137, J1809--1917, J1825--1446, J1826--1334 and J1852--0635 \citep{2007Kijak, 2011Kijak, 2014Dembska, 2015Dembska, 2016Basu}. Low-frequency measurements by \citet{2016Bilous} recently suggested that PSR J1740+1000, which was initially thought to show GPS, has a simple power law spectrum with a large amount of flux density variability. During the refereeing process of this paper, \citet{2017Kijak} identified another five GPS pulsars (PSRs J1644--4559, J1723--3659, J1835--1020, J1841--0345 and J1901+0510), of which we have measurements for three. In the following we discuss the most interesting spectra.

\subsubsection{New identifications}
\label{sec:NewIdentification}

\begin{figure}
	\centering
	\includegraphics[width=\columnwidth]{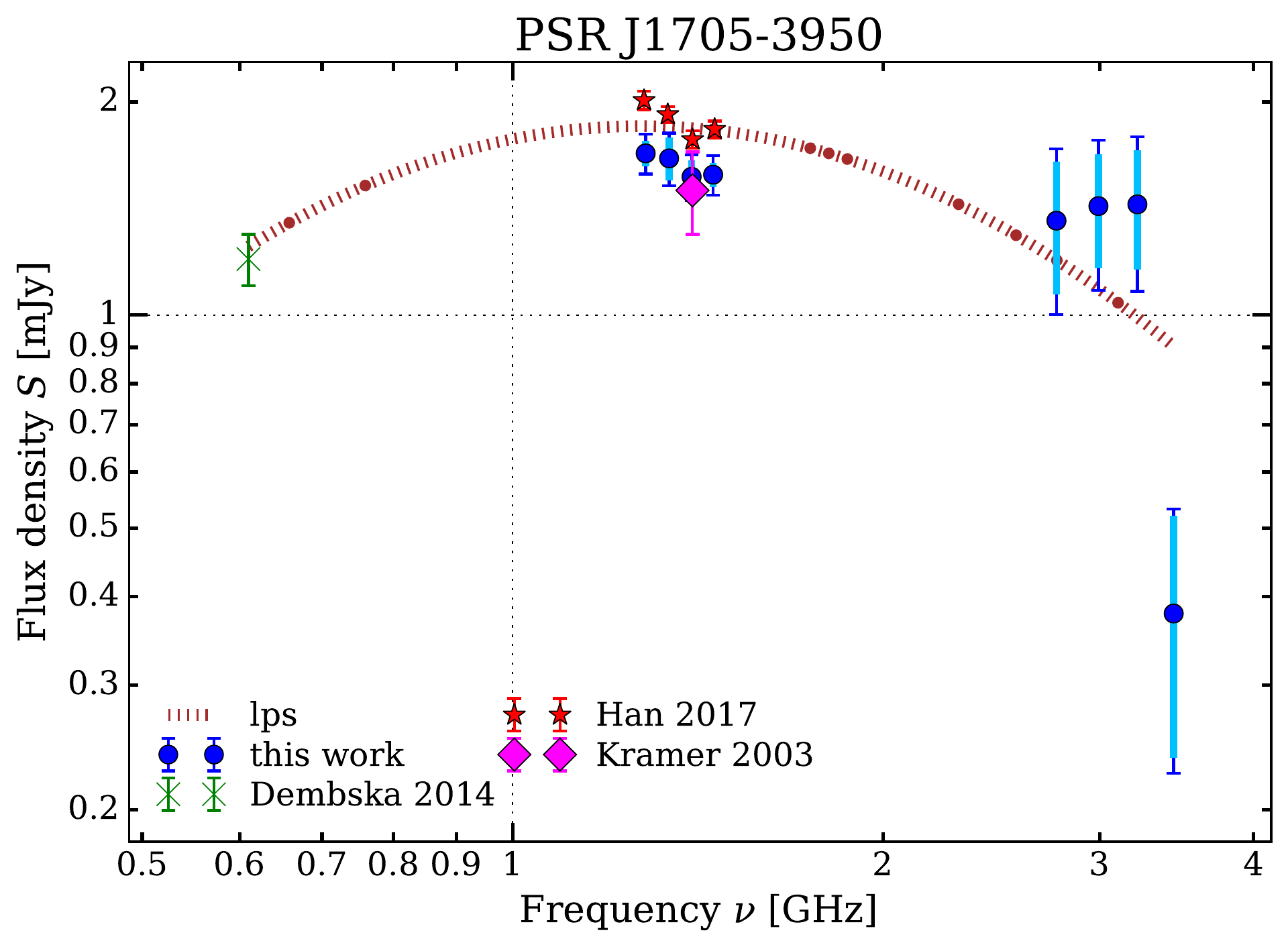}
	\includegraphics[width=\columnwidth]{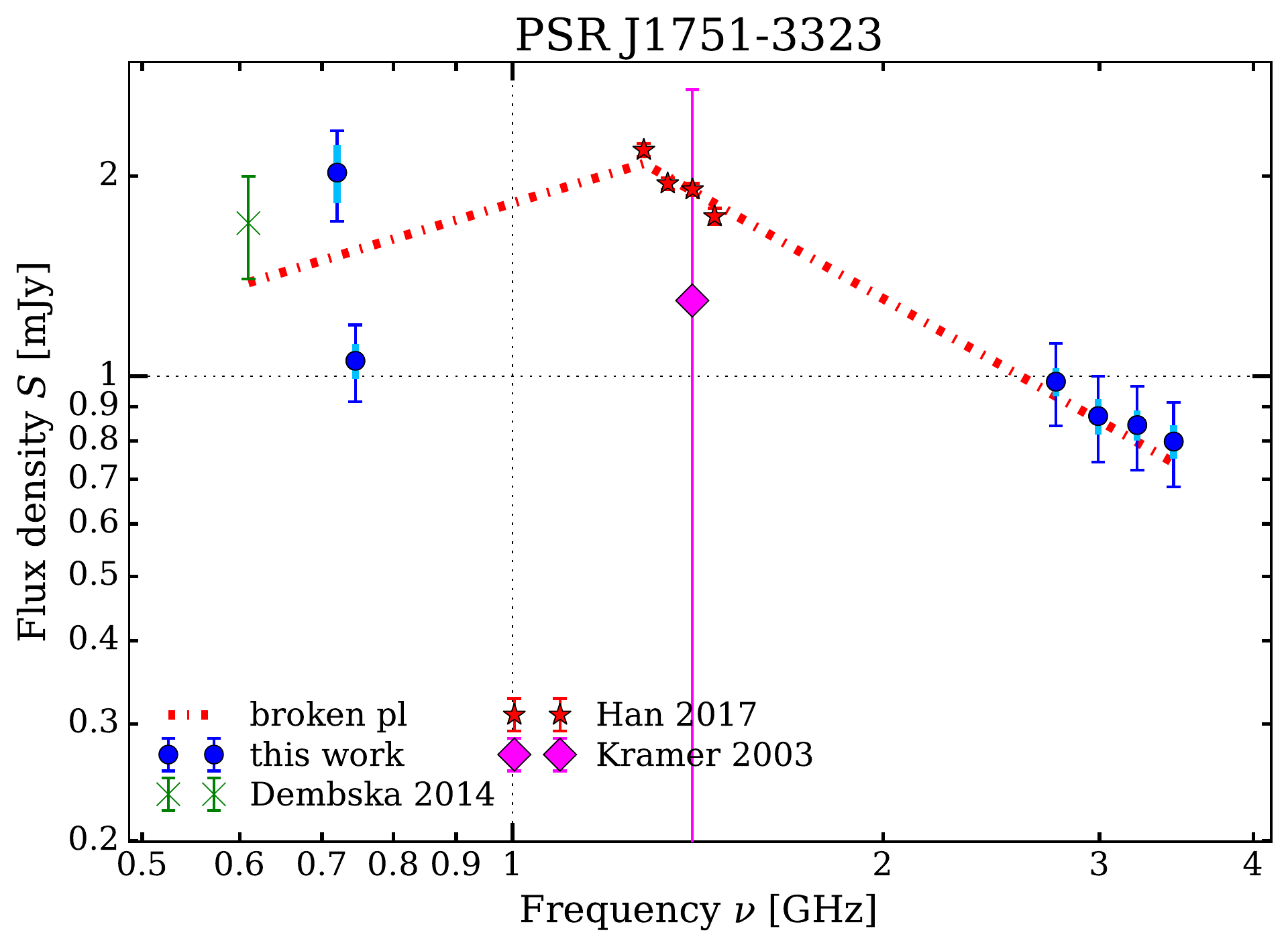}
	\includegraphics[width=\columnwidth]{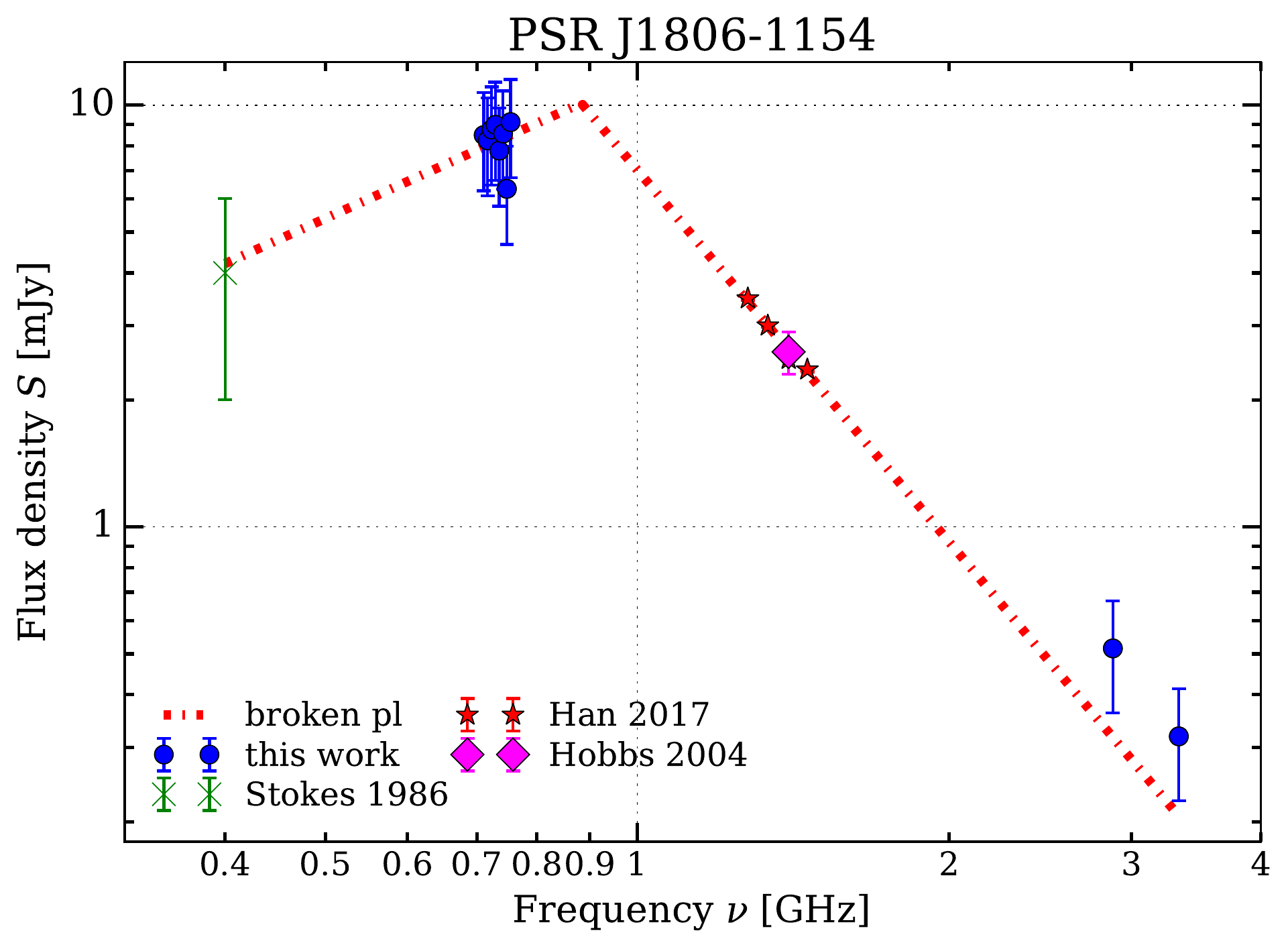}
	\caption{Spectra of the newly identified GPS pulsars.}
	\label{fig:GPSNew}
\end{figure}

We show the spectra of the newly identified GPS pulsars in Fig.~\ref{fig:GPSNew}. PSR J1705--3950: It is a newly identified GPS pulsar with a candidate LPS classification. \citet{2014Dembska} declared it as a potential GPS pulsar, but could not determine the spectral turn-over. PSR J1751--3323: Our and the data from \citet{2017Han} determine the spectral break around 1.3~GHz nicely. Its classification as a broken power law is only weak, but the second most-preferred model is a LPS with a peak around 1.1~GHz. The combined probability is $0.74$ with respect to all other models tested. PSR J1806--1154: The pulsar has a broken power law spectrum with a break around 900~MHz.

\subsubsection{Potential GPS pulsars}
\label{sec:PotentialGPSPulsars}

\begin{figure}
	\centering
	\includegraphics[width=\columnwidth]{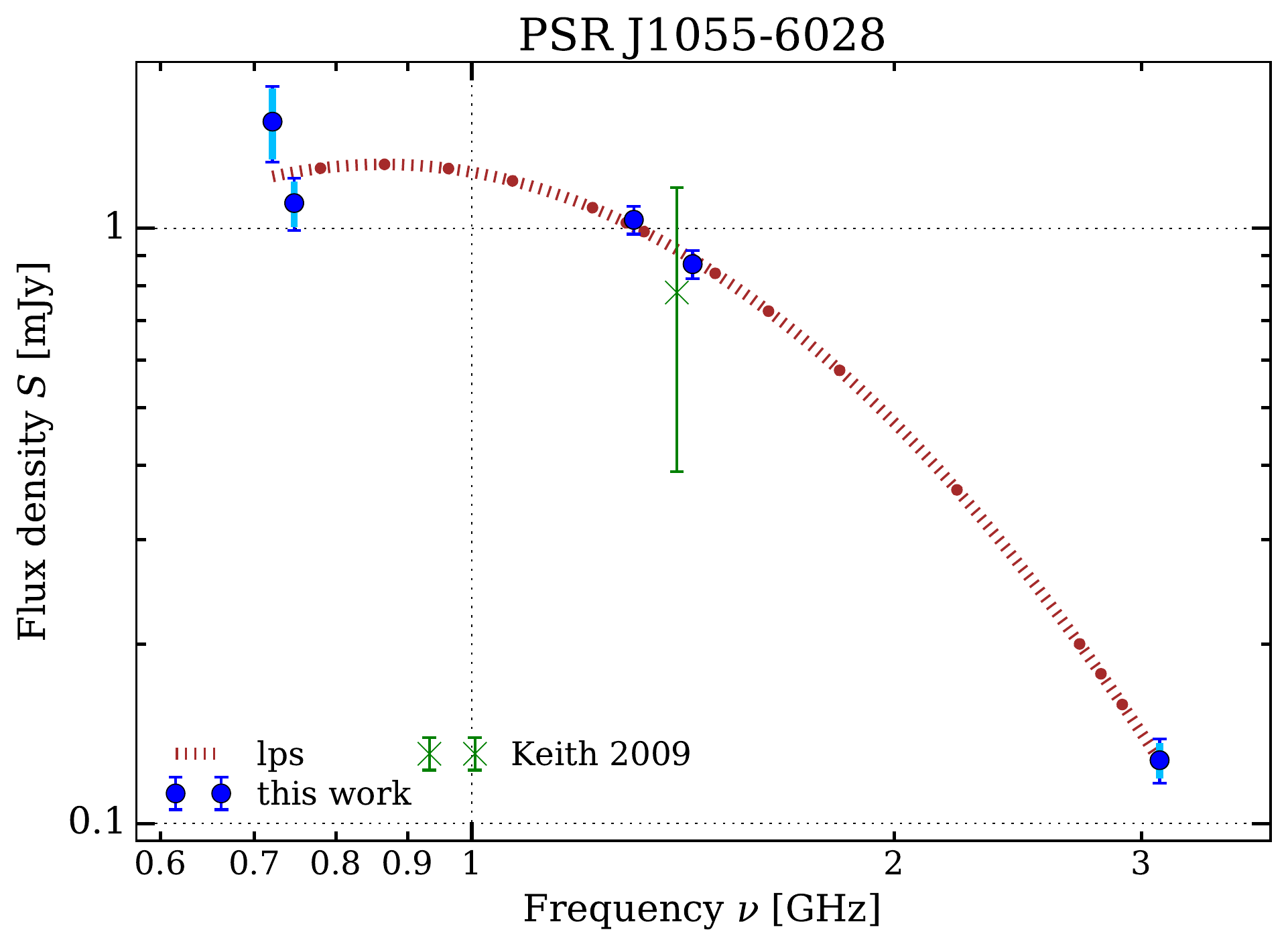}
	\includegraphics[width=\columnwidth]{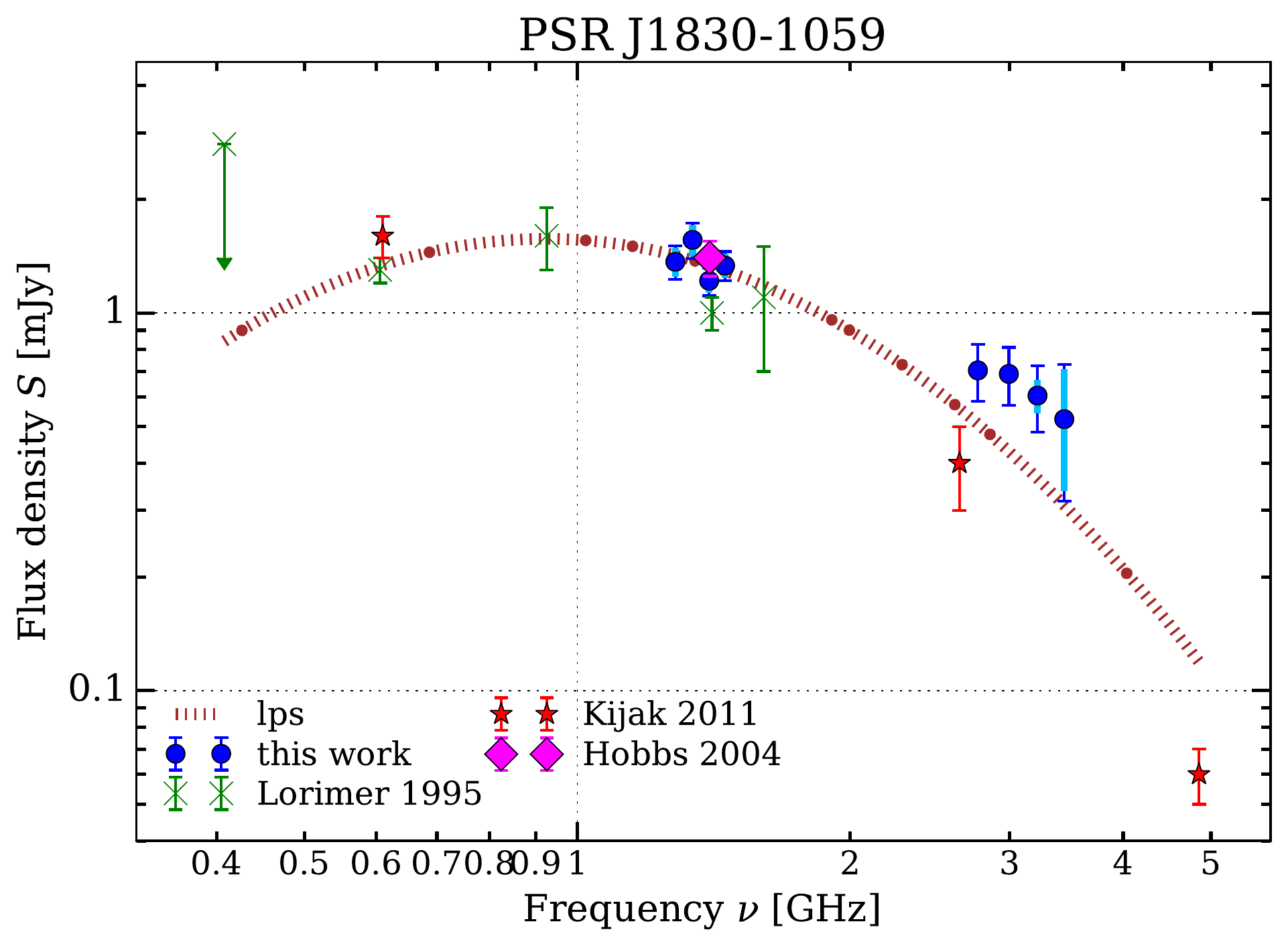}
	\includegraphics[width=\columnwidth]{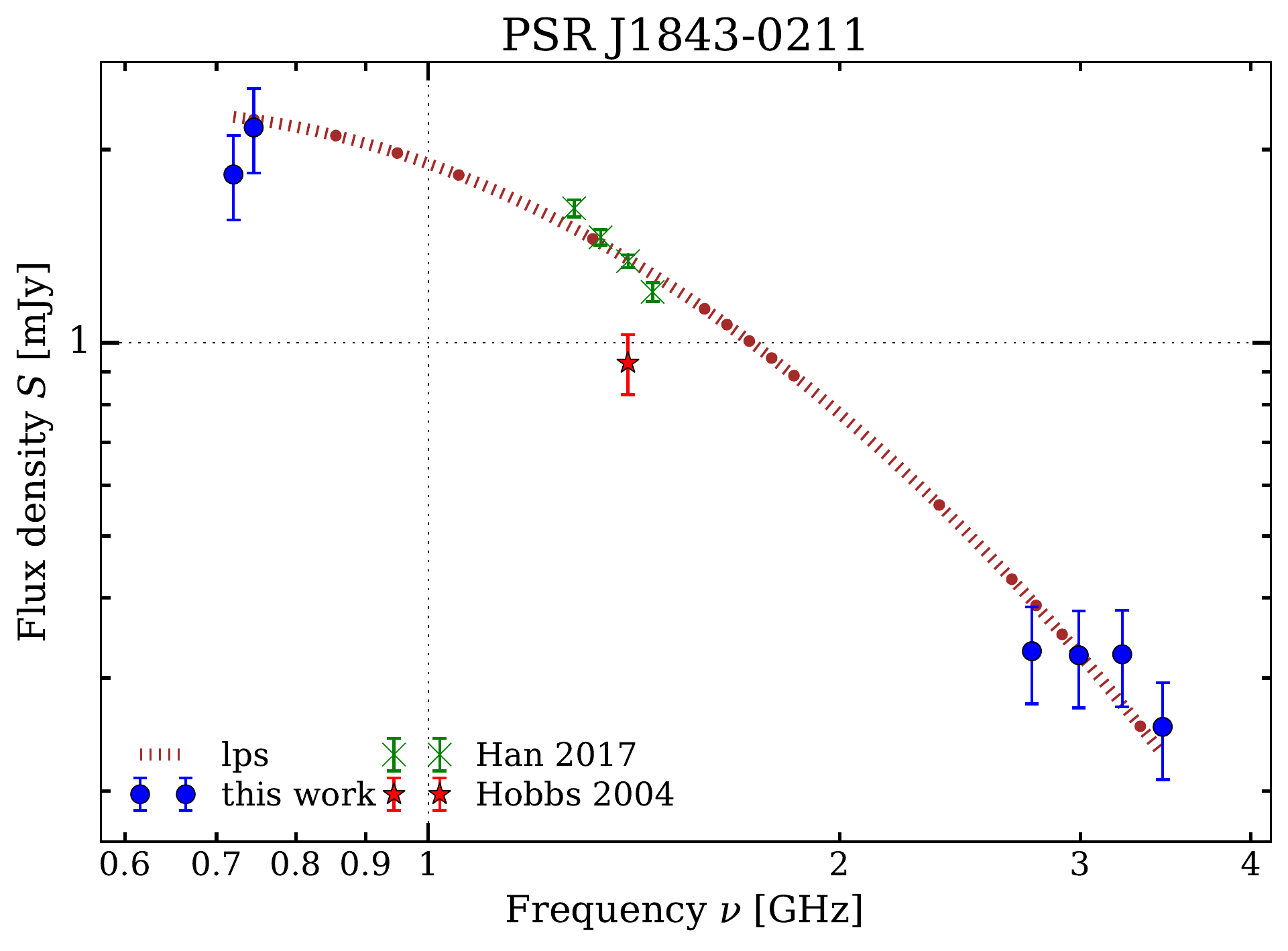}
	\caption{Spectra of the potential GPS pulsars. Further low-frequency measurements are needed to confirm their spectral turn-overs.}
	\label{fig:GPSPotential}
\end{figure}

In addition to the new identifications we find three further pulsars that seem to exhibit GPS. In fact, two of them have a clear classification, indicating that significant spectral curvature is present and that the spectral model is clearly preferred over all others tested. However, we consider them potential GPS pulsars until further low-frequency measurements are available that confirm their turn-overs beyond doubt. We present their spectra in Figure~\ref{fig:GPSPotential}. They are: PSR J1055--6028, which has a clear LPS classification and a peak around 900~MHz. We resolve its turn-over partially. Its pulse profile at 1.4~GHz has a width of about 3.7~ms, occupies less than $10 \%$ pulse longitude, and a single-sided exponential scattering tail is visible. The estimate of the scatter-broadening based on Eq.~\ref{eq:ScatterBroadening} is too high. PSR J1830--1059 is another potential GPS pulsar with a clear LPS classification. \citet{2011Kijak} declared its spectrum as unclear and speculated that it might be a broken power law. \citet{2017Kijak} follows the same argument, but fits its spectrum using a free-free absorption model. PSR J1843--0211: The pulsar has a LPS spectrum with a peak around 600~MHz and a candidate classification.

\subsubsection{Different and complex spectra}
\label{sec:DifferentAndComplexSpectra}

\begin{figure}
	\centering
	\includegraphics[width=\columnwidth]{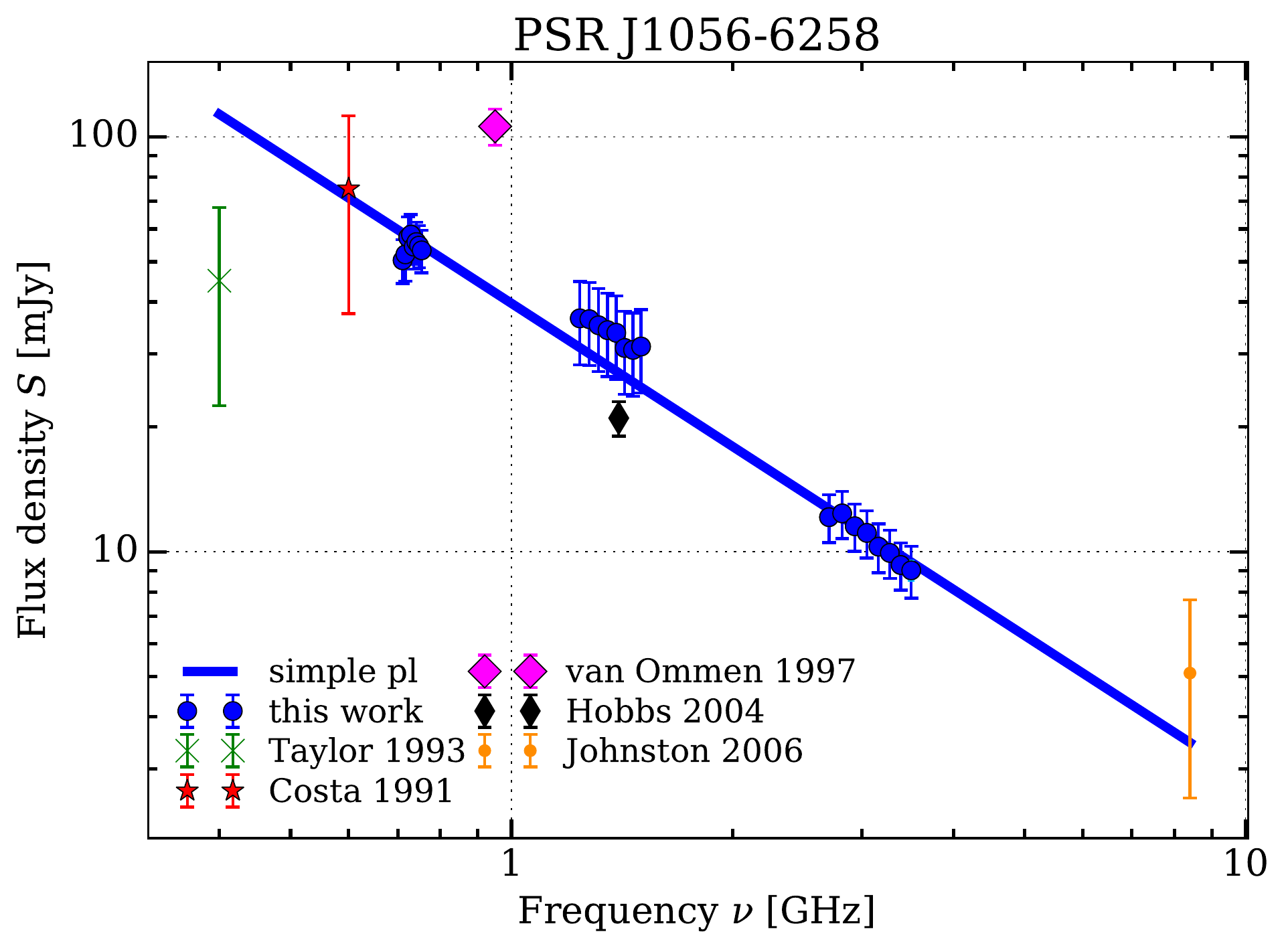}
	\includegraphics[width=\columnwidth]{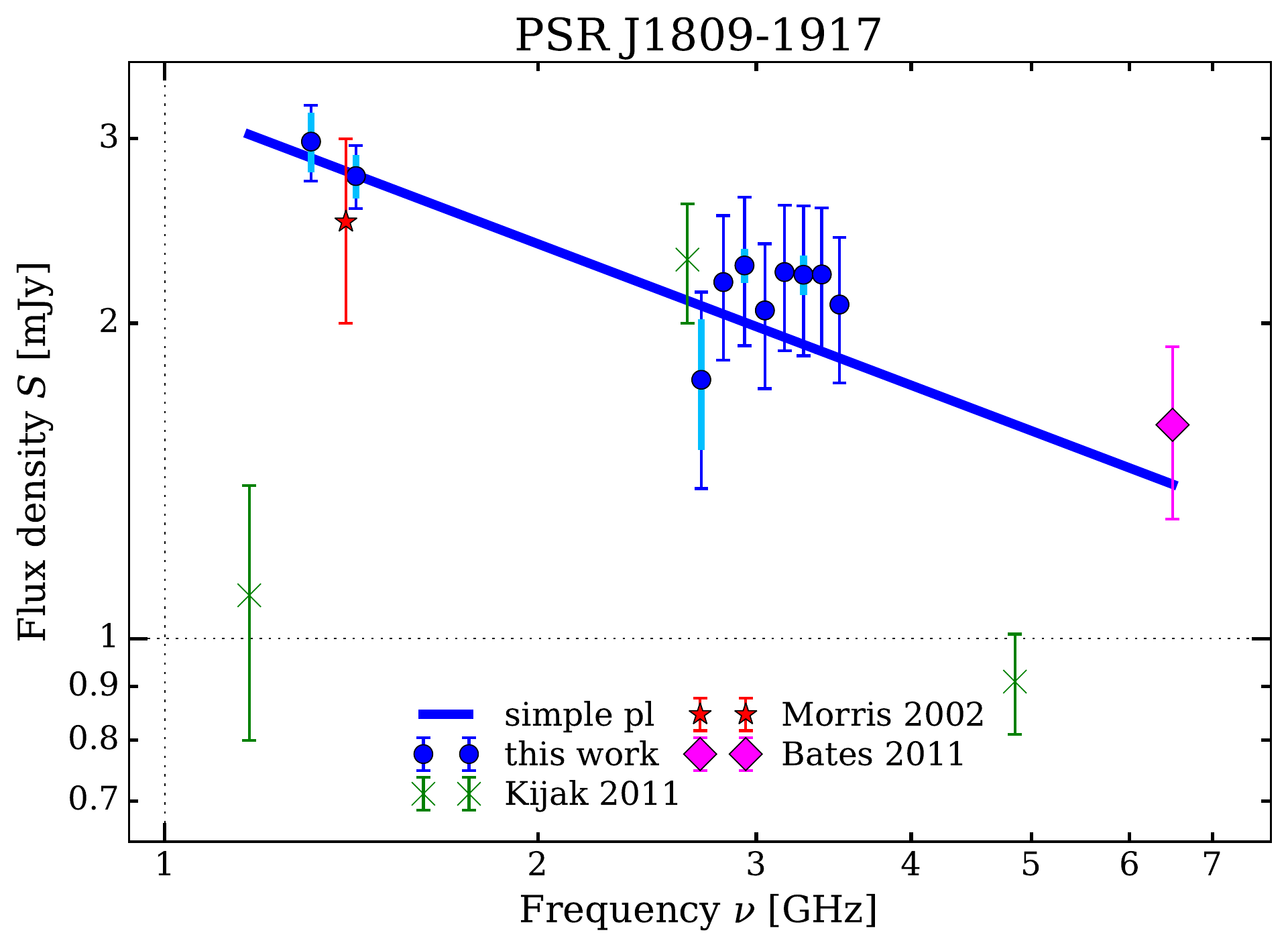}
	\includegraphics[width=\columnwidth]{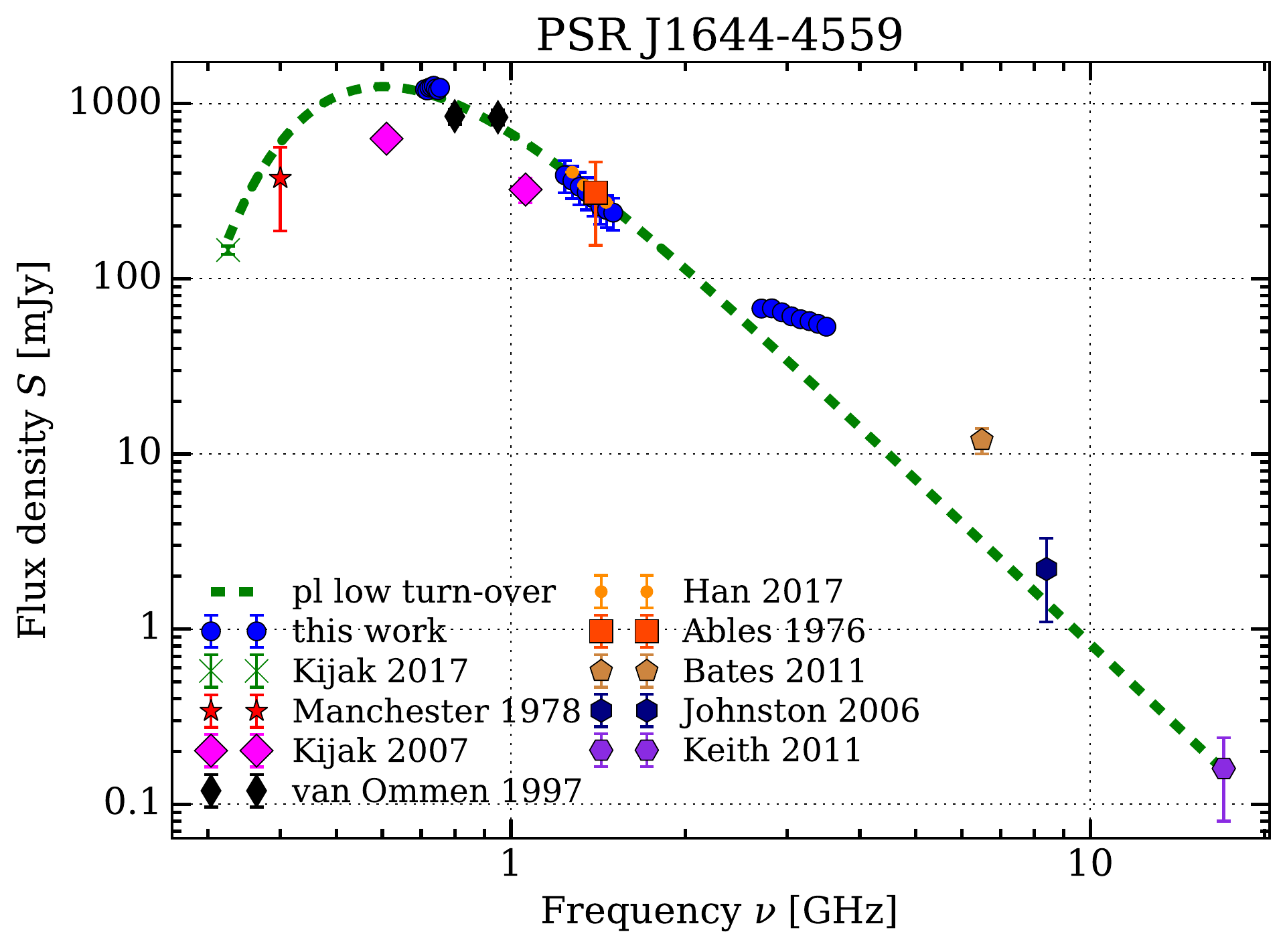}
	\caption{Top: Spectrum of PSR J1056--6258, which was previously declared to exhibit GPS. Middle: Spectrum of PSR J1809--1917, which similarly is consistent with a simple power law in the frequency range tested. Bottom: Spectrum of PSR J1644--4559, which seems to be complex, with a peak near 0.6~GHz indicating a GPS with a low peak frequency and a spectral flattening around 3.1~GHz.}
	\label{fig:GPSUnclear}
\end{figure}

We find that the spectra of two of the known GPS pulsars from the literature seem to be consistent with a simple power law and another seems to exhibit a complex morphology in addition to a GPS. We show their spectra in Fig.~\ref{fig:GPSUnclear}. PSR J1056--6258: We find that its spectrum is consistent with a simple power law with a candidate classification. Our measurements determine its spectrum between 0.7 and 3.1~GHz, where the GPS peak would be visible. At best there is a hint of a low-frequency turn-over. In total we have good frequency coverage from 0.4 to 8.4~GHz. PSR J1809--1917: Similarly, its spectrum is consistent with a simple power law with a candidate classification in the frequency range studied. The data from \citet{2011Kijak} near 1.2 and 4.8~GHz do not provide enough evidence for curvature. In fact, a simple power law also seems to be a better fit than the free-free absorption model in \citet{2017Kijak}. However, the pulsar might still show GPS at lower frequencies, once measurements below 1~GHz are available. PSR J1644--4559: Its spectrum is best fit by a power law with low-frequency turn-over around 600~MHz, suggesting a GPS at a relatively low frequency. Interestingly, our measurements between 2.6 and 3.6~GHz indicate a spectral flattening in that range, a spectral turn-up. The data from \citet{2011Bates} roughly agree with this trend. However, the data at 8.4 and 17~GHz seem to be consistent with a power law scaling. We also detect spectral flattening at high frequencies in other pulsars (Table~\ref{tab:BrokenPLPulsars}) and \citet{2015Dai} report steepening or flattening around 3.1~GHz for a small number of MSPs. We speculate that its spectrum might be complex, with a low frequency turn-over, a spectral break and turn-up at intermediate frequencies and a return to near power law scaling afterwards.

\subsubsection{Confirmations}
\label{sec:Confirmations}

\begin{figure*}
	\centering
	\includegraphics[width=0.93\columnwidth]{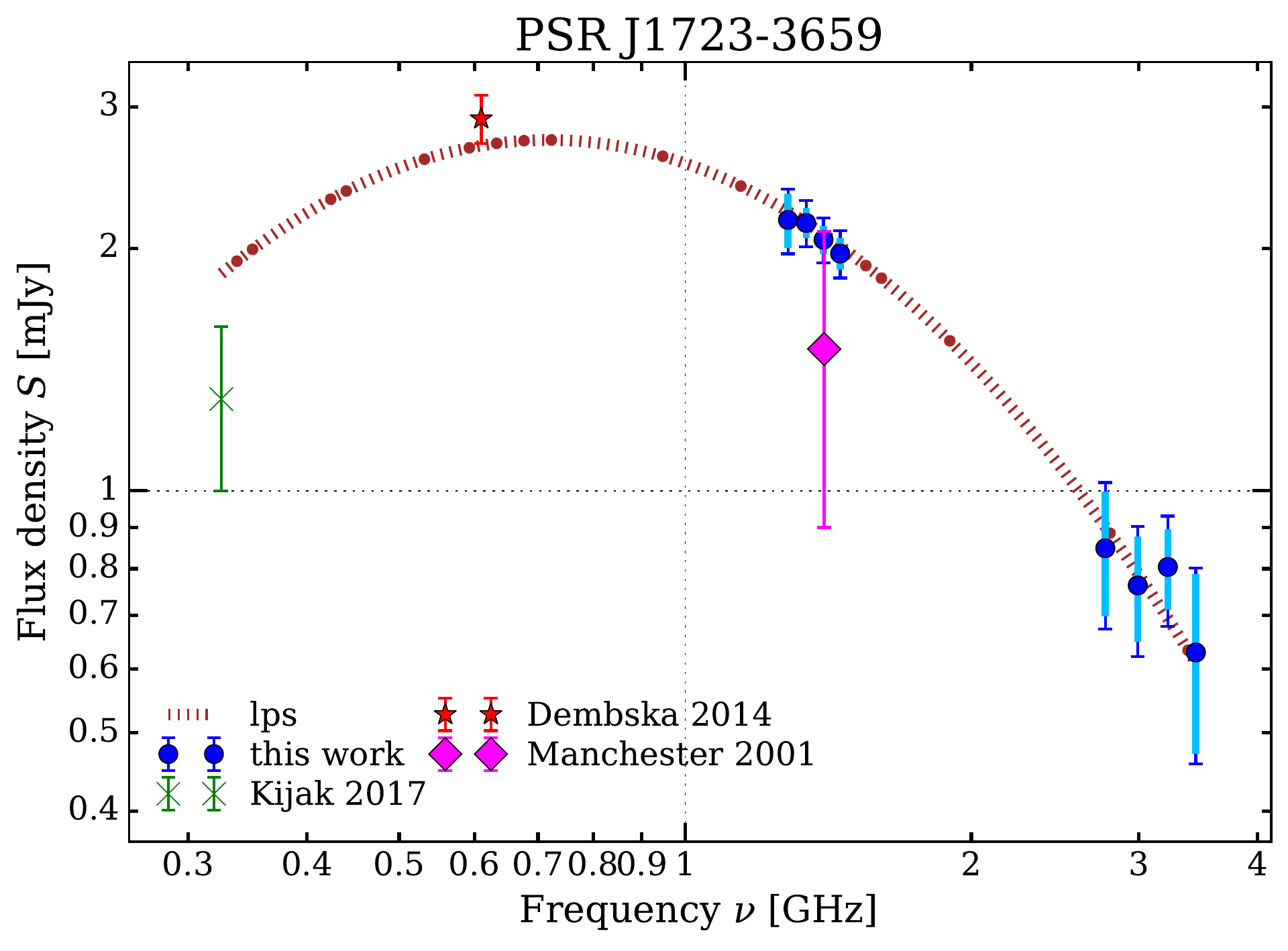}
	\includegraphics[width=0.93\columnwidth]{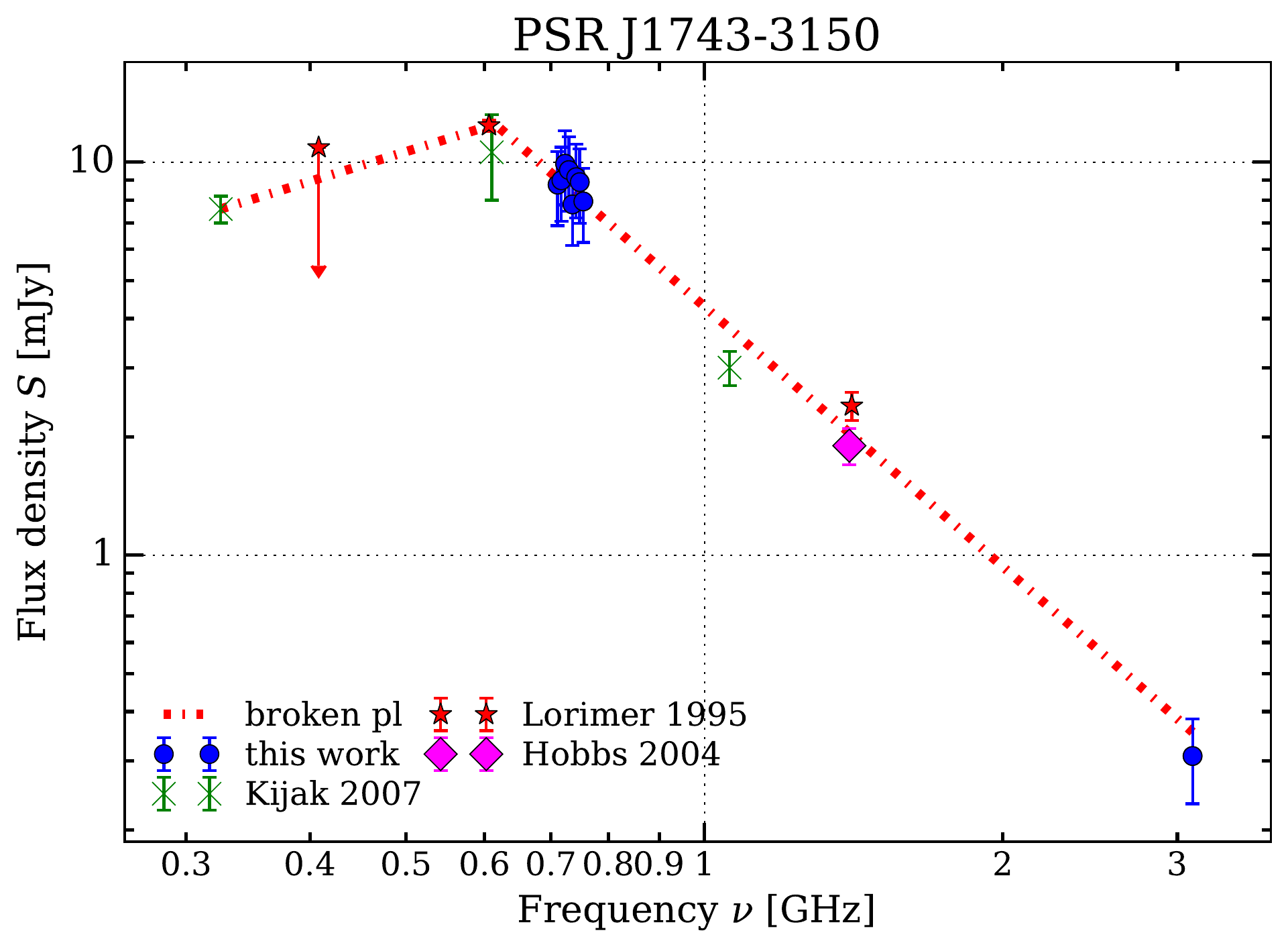}
	\includegraphics[width=0.93\columnwidth]{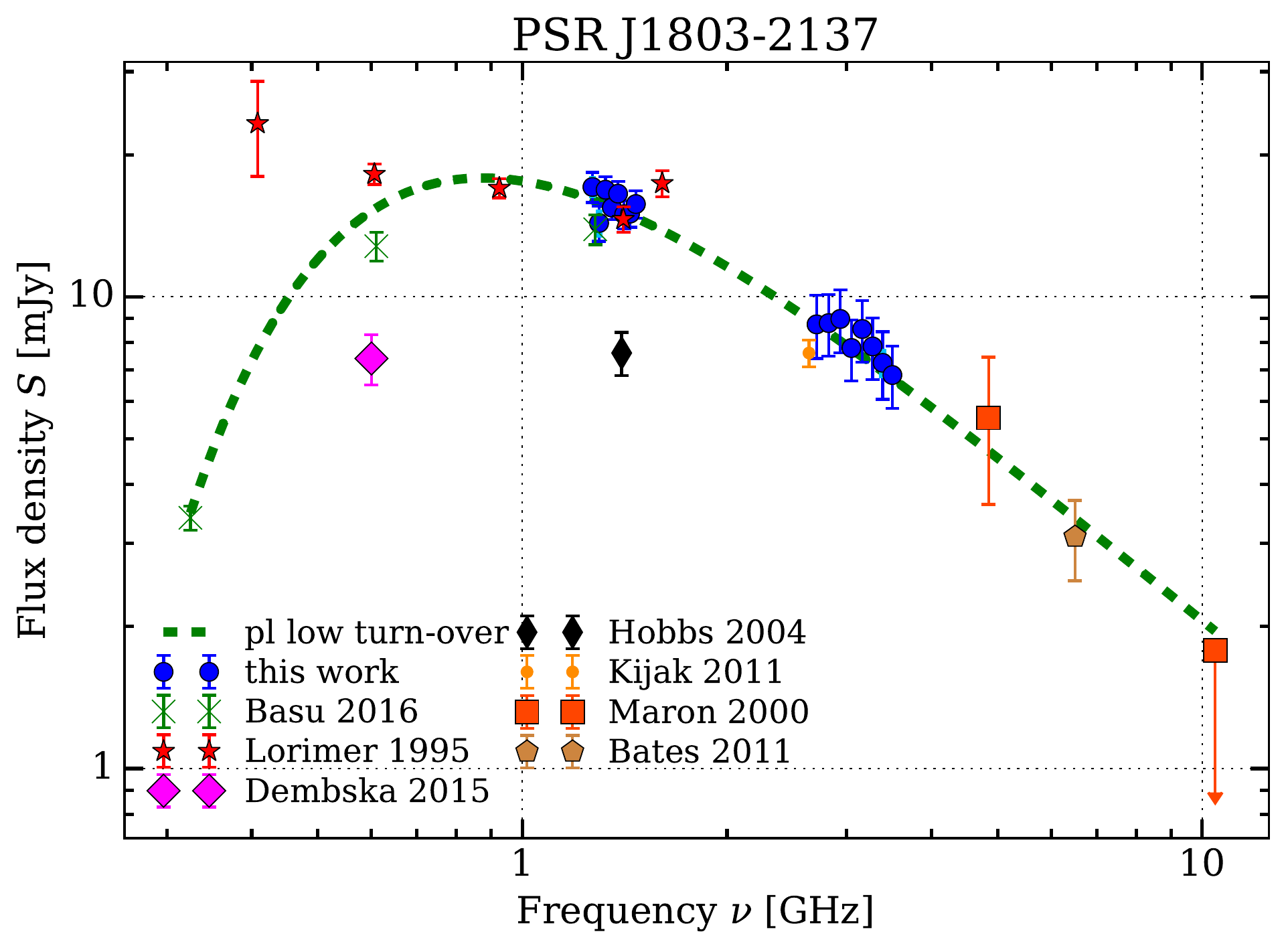}
	\includegraphics[width=0.93\columnwidth]{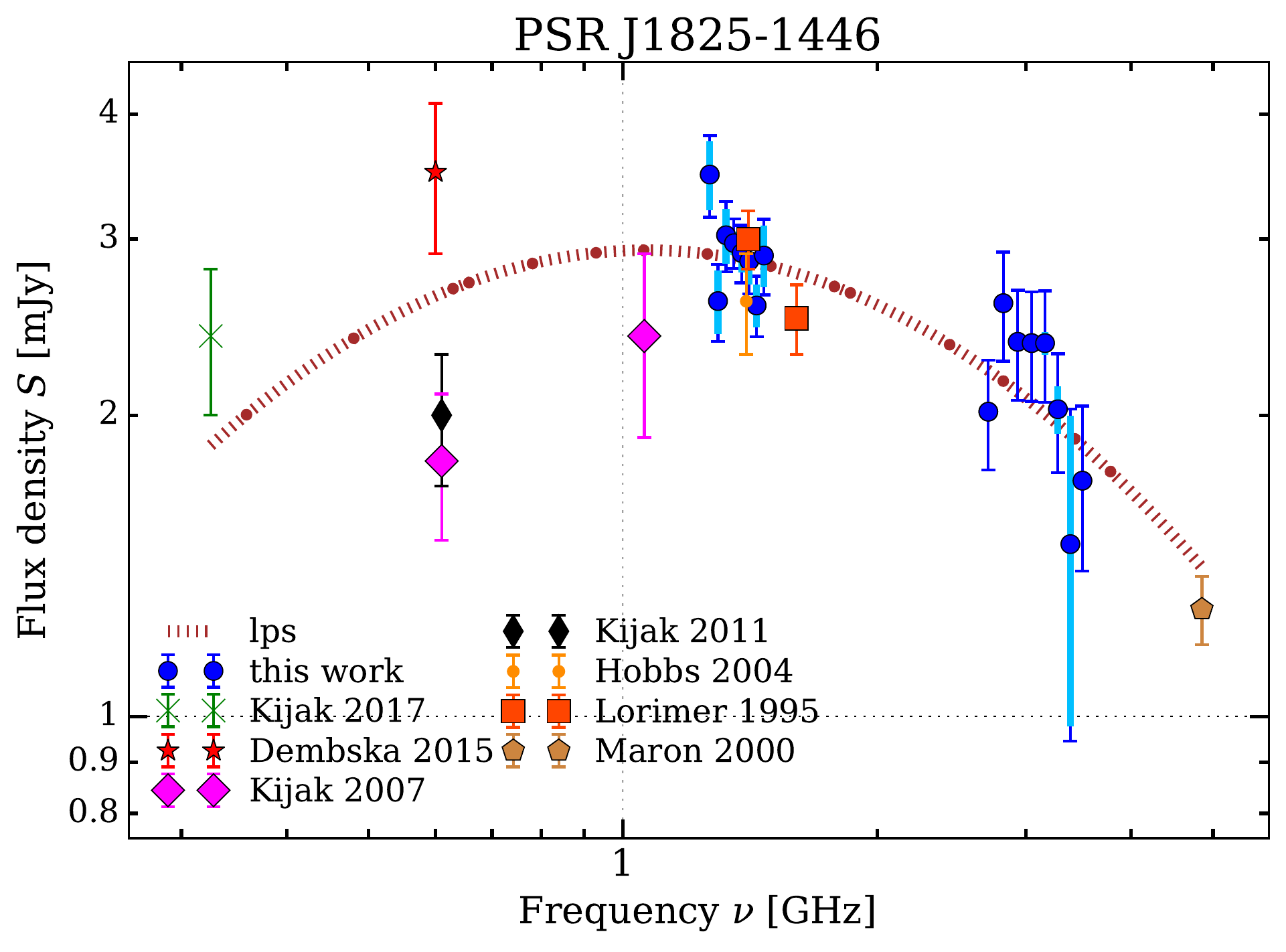}
	\includegraphics[width=0.93\columnwidth]{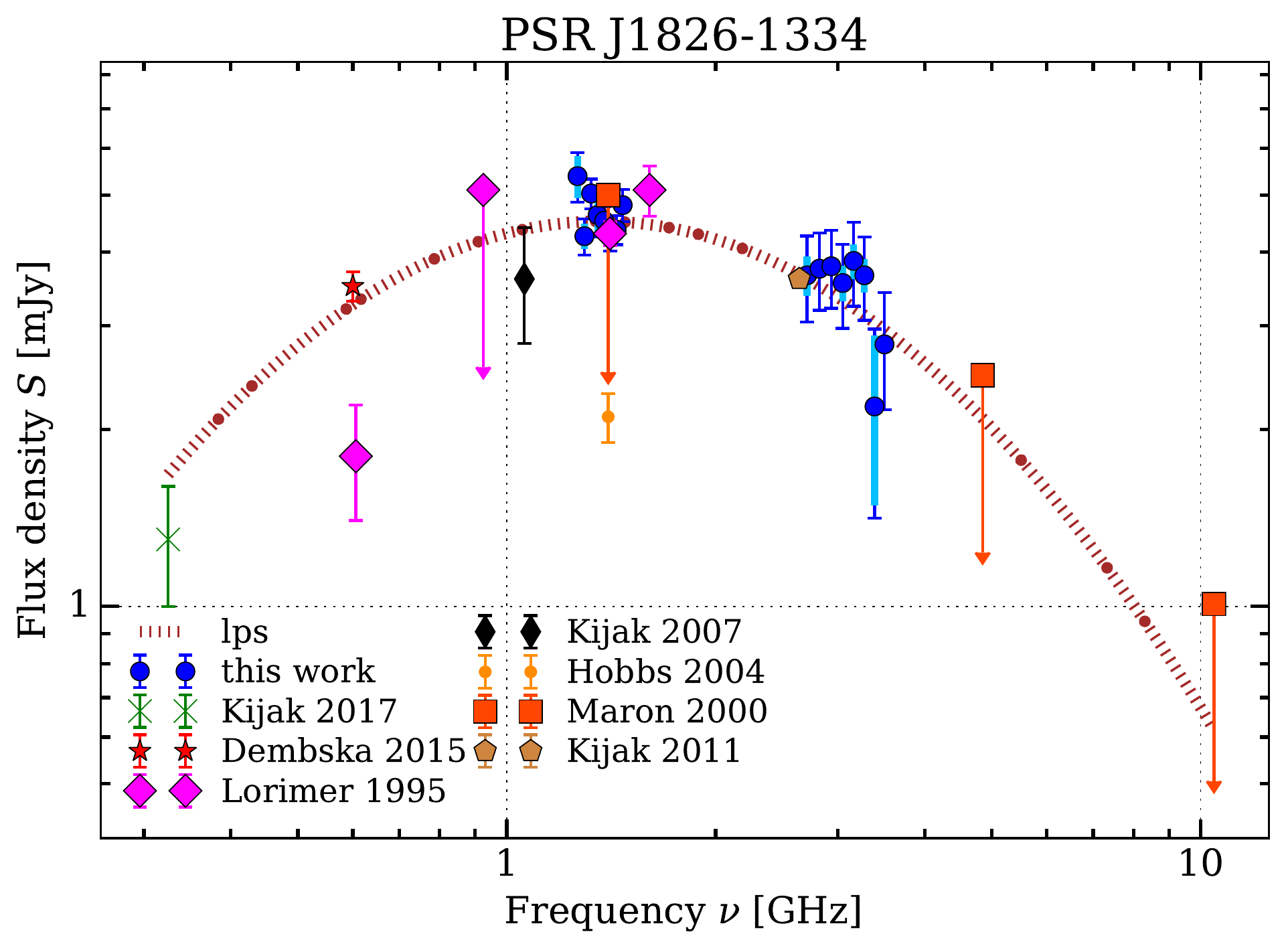}
	\includegraphics[width=0.93\columnwidth]{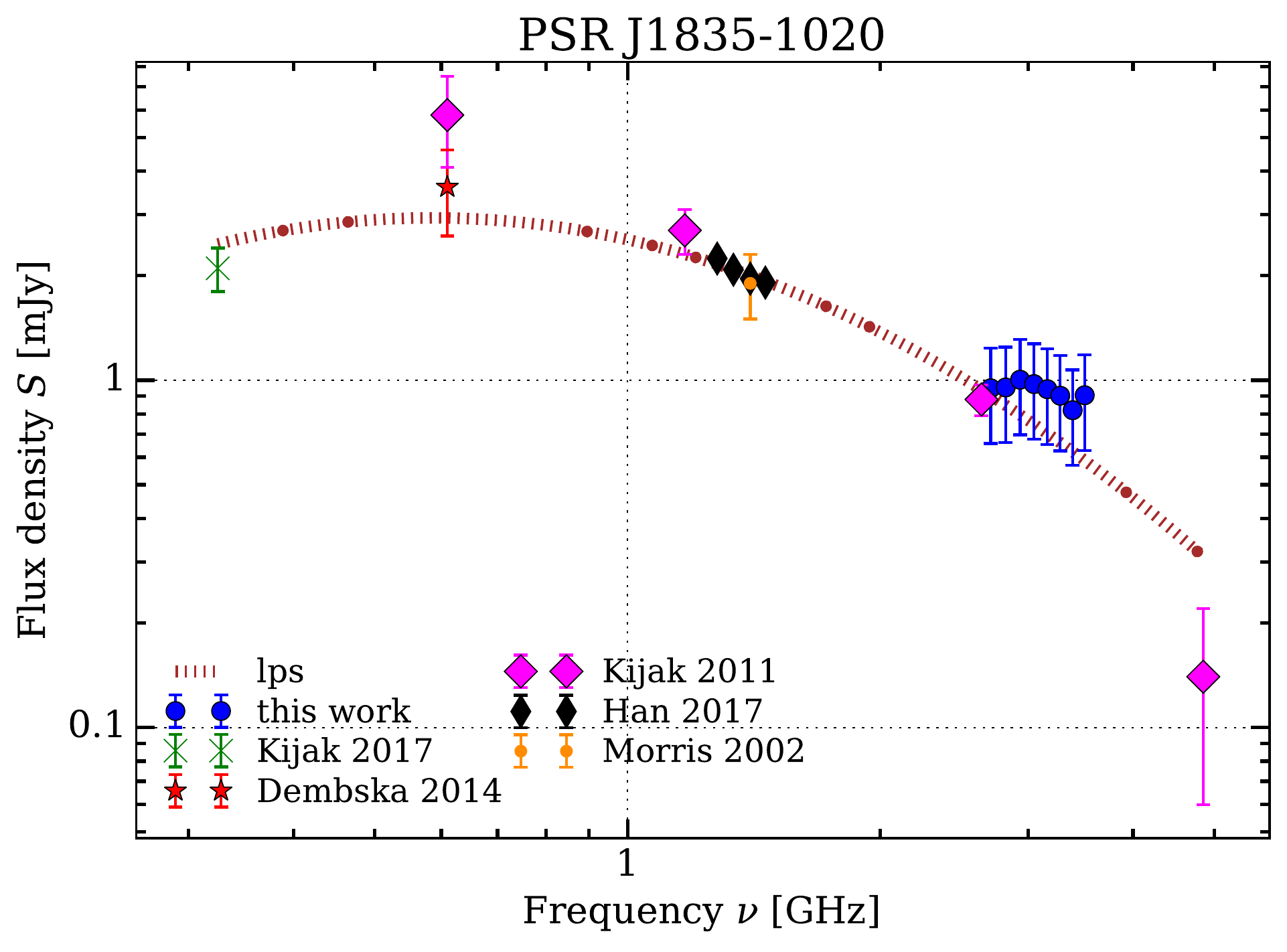}
	\includegraphics[width=0.93\columnwidth]{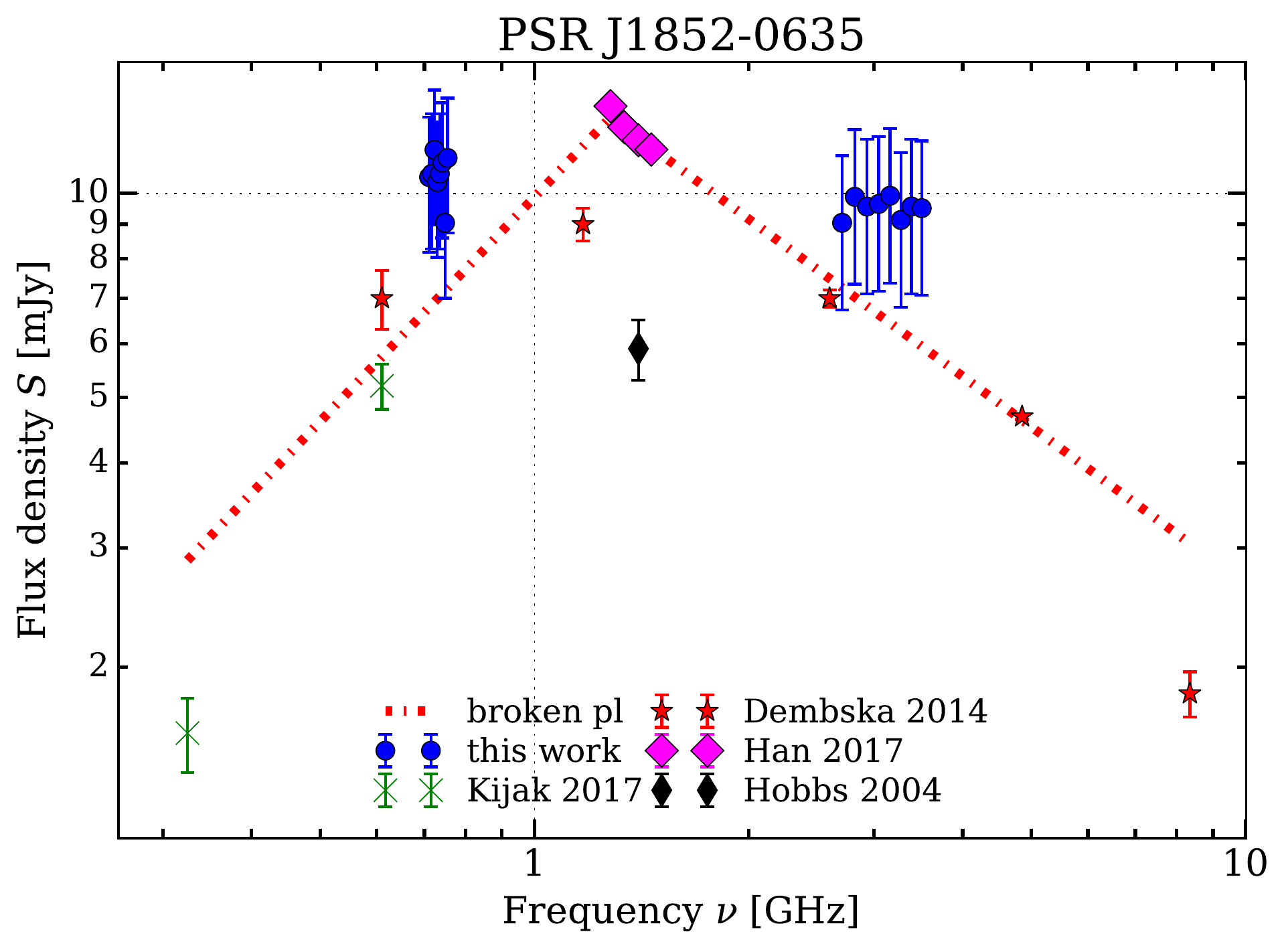}
	\caption{Spectra of the known GPS pulsars from the literature, for which we have new flux density measurements and whose GPS classifications we confirm.}
	\label{fig:GPSConfirmations}
\end{figure*}

We confirm the GPS classification for the pulsars in Fig.~\ref{fig:GPSConfirmations}. Some of the spectra peak at rather low frequencies near 600~MHz, in particular the ones of PSRs J1835--1020 and J1743--3150. PSR J1825--1446 and PSR J1826--1334: Their spectra peak around 1.3~GHz with a strong LPS classification. PSR J1852--0635: We determine a broken-power law as best-fitting model, with a clear classification category, rather than a power law with low-frequency turn-over, of which the free-free absorption model used by \citet{2017Kijak} is a special case. \citet{2016Rajwade} and \citet{2017Kijak} fit the spectra of a subset of these pulsars using the free-free absorption model, which we have considered separately in \S\ref{sec:FreeFreeAbsorption}. In this work we add a significant number of new flux density measurements for each pulsar and perform an objective and unbiased spectral classification among the models discussed in \S\ref{sec:SpectralModels}. An example is PSR J1723--3659, where we determine the spectral shape above 1.4~GHz.

\subsection{Correlations of spectral index with pulsar parameters}
\label{sec:CorrelationsOfSpectralIndexWithPulsarParameters}

\begin{table*}
\caption{Correlation of spectral index $\alpha$ with $\log_{10} \left| x \right|$ for different pulsar parameters $x$ and subsets of the total data set. $r_\text{s}$ is the Spearman rank correlation coefficient, $p$ is the corresponding p-value and $N$ the sample size. We show only a selection of all the correlations tested. The correlations with an absolute value of $r_\text{s}$ of at least $0.4$ and a p-value of less than $1 \: \%$ are marked in bold.}
\label{tab:CorrelationWithAlpha}
\begin{tabular}{llllllllllll}
\hline
set			& all	& in binary	& isolated	& MSP	& slow\\
\#pulsars 	& 276	& 10			& 266		& 9		& 267\\
\hline
$x$	& $r_\text{s}$ ($p, N$)	& $r_\text{s}$ ($p, N$)	& $r_\text{s}$ ($p, N$)	& $r_\text{s}$ ($p, N$)	& $r_\text{s}$ ($p, N$)\\
\hline
$\tilde{\nu}$		& $0.37$ (3.7e-10, $276$)	& $0.25$ (0.49, $10$)	& $0.36$ (1.3e-09, $266$)	& $0.28$ (0.46, $9$)	& $0.37$ (5.7e-10, $267$)\\
$\dot{\tilde{\nu}}$	& $\mathbf{0.4}$ (9.5e-12, 275)	& $0.18$ (0.63, $10$)	& $\mathbf{0.42}$ (9.3e-13, 265)	& $0.38$ (0.31, $9$)	& $\mathbf{0.43}$ (3.1e-13, 266)\\
$\dot{P}$	& $0.25$ (3.8e-05, $275$)	& $-0.018$ (0.96, $10$)	& $0.28$ (3.8e-06, $265$)	& $0.05$ (0.9, $9$)	& $0.28$ (4.9e-06, $266$)\\
$B_\text{LC}$	& $\mathbf{0.43}$ (1.1e-13, 275)	& $0.45$ (0.19, $10$)	& $\mathbf{0.42}$ (5.9e-13, 265)	& $0.23$ (0.55, $9$)	& $\mathbf{0.43}$ (2.5e-13, 266)\\
$\tau$	& $-0.35$ (1.9e-09, $275$)	& $-0.21$ (0.56, $10$)	& $-0.38$ (1.2e-10, $265$)	& $-0.5$ (0.17, $9$)	& $-0.39$ (5.9e-11, $266$)\\
$\dot{E}$	& $\mathbf{0.44}$ (4e-14, 275)	& $0.39$ (0.26, $10$)	& $\mathbf{0.43}$ (2.3e-13, 265)	& $0.33$ (0.38, $9$)	& $\mathbf{0.43}$ (1.4e-13, 266)\\
\hline
\end{tabular}
\end{table*}

\begin{figure}
	\centering
	\includegraphics[width=\columnwidth]{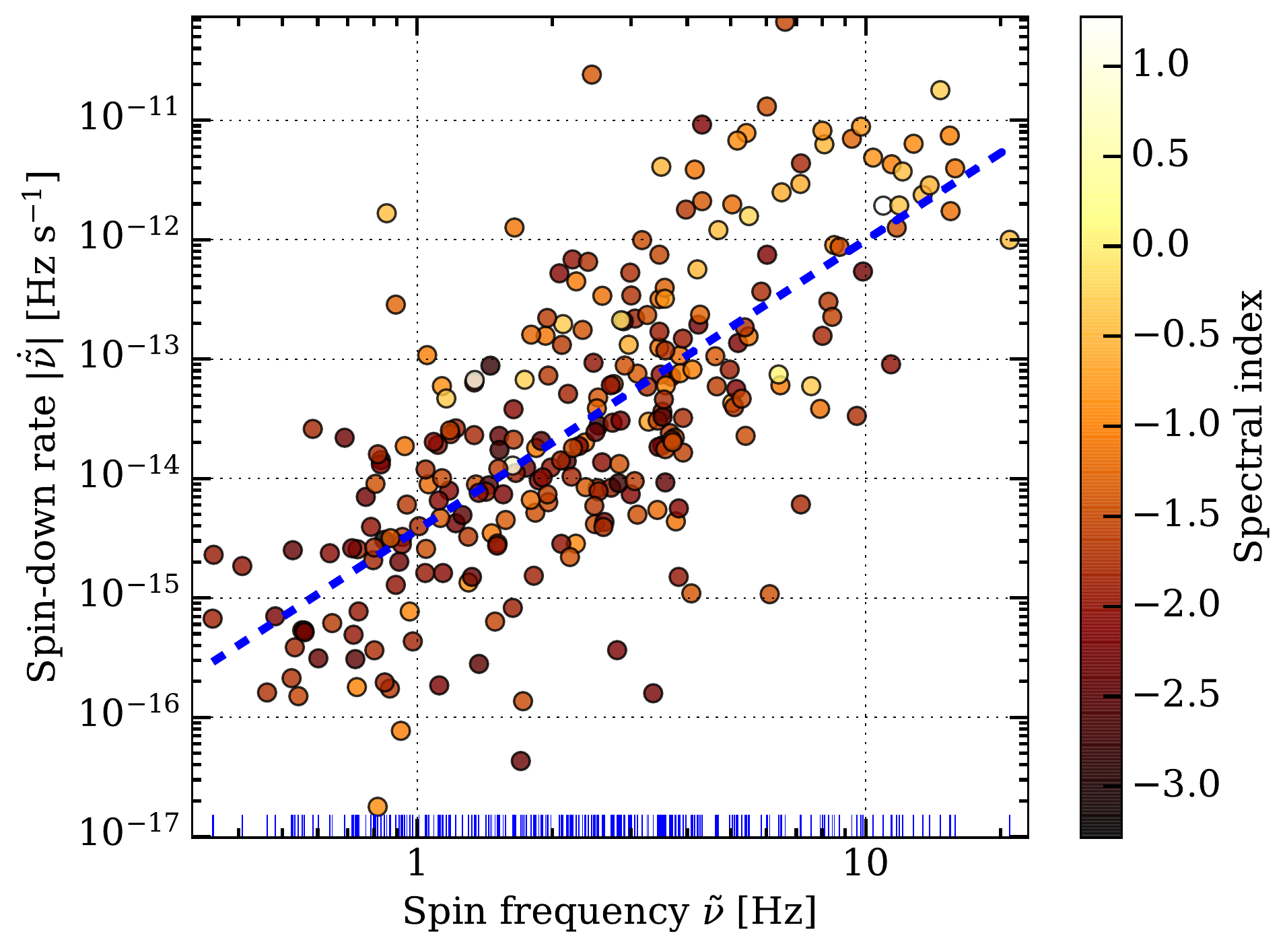}
	\includegraphics[width=\columnwidth]{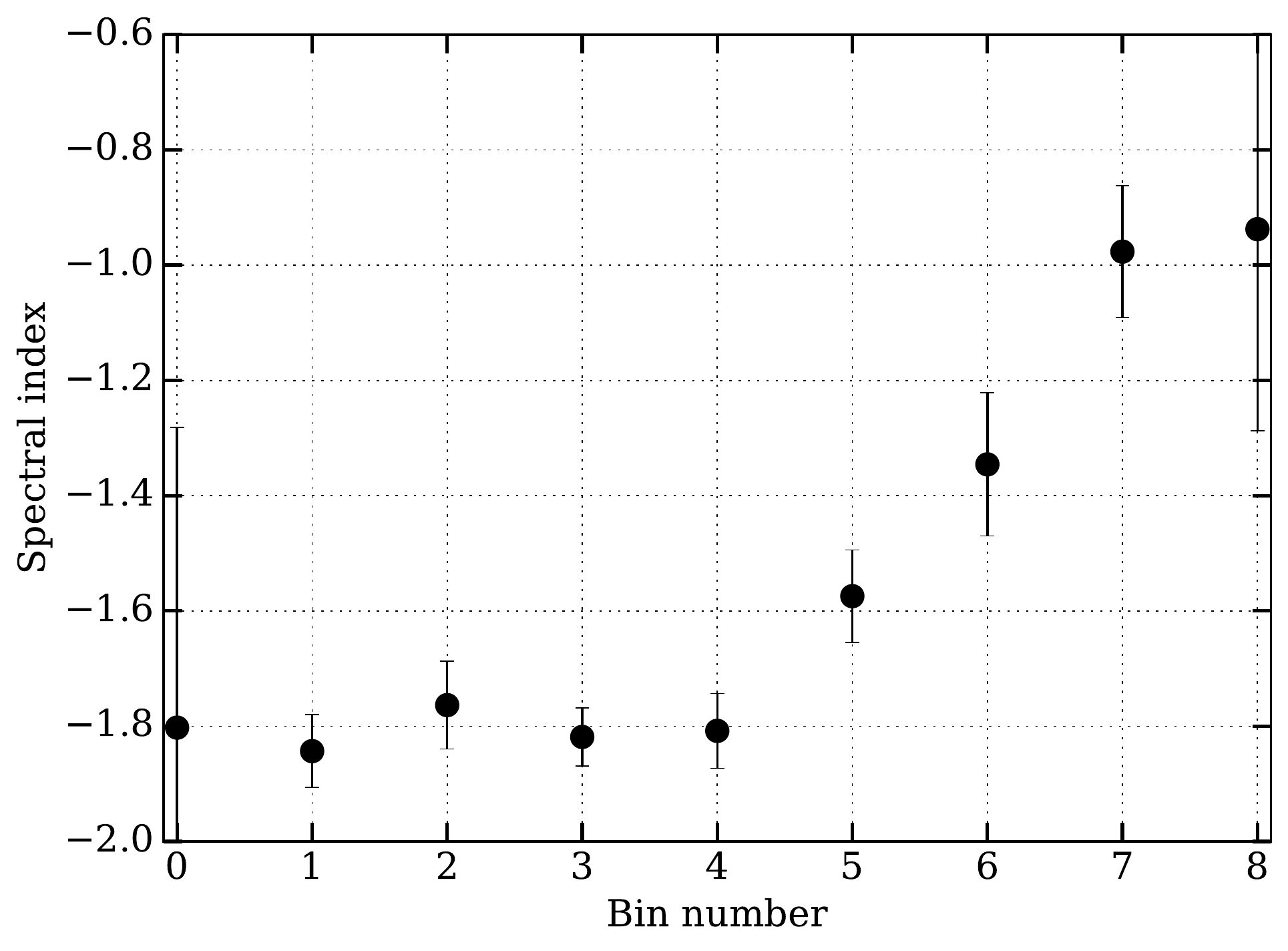}
	\caption{Top: Spectral index shown in the spin-down rate versus spin frequency plane for all pulsars that have simple power law spectra. The blue dashed line indicates the principal component, or the best power law fit. Bottom: Projection of the data from the top panel along the principle component line. Shown is the weighted mean spectral index computed in bins along the line arranged from bottom left to top right.}
  \label{fig:SpectralIndexInF0F1Plane}
\end{figure}

For all pulsars that have simple power law spectra we test for correlations between the measured spectral index $\alpha$ and various pulsar parameters. Specifically we test for a correlation between $\alpha$ and $\log_{10} \left|x \right| $, where $x$ is one of the pulsar parameters below. The parameters are spin frequency $\tilde{\nu}$, spin-down rate $\dot{\tilde{\nu}}$, second spin frequency derivative $\ddot{\tilde{\nu}}$, pulse period $P$, period derivative $\dot{P}$, the DM, the distance $d$ to the pulsar, the characteristic age $\tau_\text{c}$, the surface magnetic field $B_{\text{surf}}$, the magnetic field at the light cylinder radius $B_\text{LC}$, the W50 and W10 pulse widths, the duty cycles $\delta_{\text{W50}}$ and $\delta_{\text{W10}}$, the spin-down luminosity $\dot{E}$ and the total energy flux density $\dot{E}/d^2$ of the pulsar. In addition, we test the following intrinsic quantities, which are corrected for a kinematic term arising because of the large transverse velocity of a pulsar and are calculated using the intrinsic period derivative $\dot{P}_\text{i}$. Namely these are $\dot{P}_\text{i}$, $\tau_\text{c,i}$, $B_\text{surf,i}$, $\dot{E}_\text{i}$ and $\dot{E}_\text{i}/d^2$. All of these values were taken from the ATNF pulsar catalogue. Most of these quantities are covariant, because they depend on basic pulsar parameters such as spin frequency and spin-down rate.

We test the correlation by visual inspection and compute the Spearman rank correlation coefficient to characterise its strength. It is non-parametric and robust against outliers \citep{2014Ivezic}. We used the implementation in the \texttt{python} library \texttt{SciPy}\footnote{\url{http://www.scipy.org}} \citep{Scipy}. We first test all pulsars in our single power law data set, then the pulsars in binary systems and the isolated pulsars separately. Finally we test the MSPs ($P \leq 30 \: \text{ms}$) and the slow pulsars separately. A selection of the resulting correlation coefficients and corresponding p-values are shown in Table~\ref{tab:CorrelationWithAlpha} together with the number of pulsars $N$ for which the correlation was computed.

We find the highest correlation between spectral index and spin-down luminosity, magnetic field at the light-cylinder radius and spin-down rate. For the total data set (all) these have a moderate positive correlation with a maximum correlation coefficient of $0.44$. The probability that this could arise by chance is very low, see the quoted p-values. The energy flux density shows roughly the same positive correlation as $\dot{E}$ and there is a weaker negative one with the characteristic age $\tau$. Every other combination has a lower correlation. If we constrain the spectral indices to those that have uncertainties less than 0.5, the strength of the correlations increases further to $0.46$, $0.45$ and $0.46$ for $\dot{\tilde{\nu}}$, $B_\text{LC}$ and $\dot{E}$ for the 247 slow pulsars. Our conclusions, also the ones in the next section, stay the same. That means that the spectral index gets flatter with increasing absolute value of spin-down rate, magnetic field at the light cylinder and spin-down luminosity -- the faster the pulsar spins down and the faster it loses rotational energy, the flatter is its spectral index. For the intrinsic properties it is not possible to determine a clear correlation, as we do not have enough pulsars that have measured transverse velocities. The same is true for pulsars in binary systems and MSPs, as our sample size is not large enough to estimate it. The analysis of the isolated and slow pulsars show the same correlation behaviour, as the majority of the pulsars in our data set are of this type. We derive the following empirical relations between spectral index, spin frequency $\tilde{\nu}$ and spin-down rate $\dot{\tilde{\nu}}$ for the slow pulsars:
\begin{equation}
	\alpha (\tilde{\nu}) = (0.62 \pm 0.05) \: \log_{10} \left( \frac{\tilde{\nu}}{2.5 \: \text{Hz}} \right) - (1.65 \pm 0.02),
	\label{eq:ScalingLawAlphaNu}
\end{equation}

\begin{equation}
	\alpha (\dot{\tilde{\nu}}) = (0.23 \pm 0.02) \: \log_{10} \left( \frac{ | \dot{\tilde{\nu}} | }{2.2 \cdot 10^{-14} \: \text{Hz} \: \text{s}^{-1}} \right) - (1.67 \pm 0.02),
	\label{eq:ScalingLawAlphaNudot}
\end{equation}
where we show formal uncertainties from the fit at the $1 \sigma$ level. There is however a large scatter around the best fitting line. The measured spectral indices are shown in Fig.~\ref{fig:SpectralIndexInF0F1Plane} in the spin-down rate versus spin frequency plane. The spectral indices are shallower the higher the absolute value of the spin-down rate is and to lesser extent the higher the spin frequency is. The correlation between spin-down rate and spin frequency is unrelated to the spectral index measurements and is also present in the total ensemble of pulsars listed in the pulsar catalogue. We used principal component analysis to determine the principal component line. We then binned the spectral indices data along this line, which is shown in the bottom panel. We find that the mean spectral index increases roughly linearly with bin number above bin 4 and is nearly constant below that. It could also be that we see a transition to flatter spectral indices between normal pulsars and highly energetic ones around $|\dot{\tilde{\nu}}|$ of $10^{-13} \: \text{Hz} \: \text{s}^{-1}$.

\subsubsection{Which pulsar parameter is the dominant one?}

To find out which pulsar parameter is the dominant one in determining the spectral index, we computed the Spearman correlation coefficient between the spectral index $\alpha$ and the function:
\begin{equation}
	z = \log_{10} \left( \tilde{\nu}^{a} \: \left| \dot{\tilde{\nu}} \right|^b \right),
	\label{eq:F0F1Exponents}
\end{equation}
over a grid of $a$ and $b$ values, where $\tilde{\nu}$ is the spin frequency and $\dot{\tilde{\nu}}$ the spin-down rate. The resulting correlation coefficients have rotational symmetry in the $(a, b)$ plane and we re-parameterise the result in terms of the polar angle $\beta = \arctan{(b/a)}$ measured against the horizontal axis. Any combination of $(a, b)$ values that has the same ratio and therefore the same angle, has the same correlation coefficient. The maximum absolute correlation occurs for $b/a = 0.55$. That means that for any combination of $a$ and $b$ that fulfils this relationship the absolute correlation is maximised. The correlation coefficient is zero for $b/a = -0.29$, which is close to the angle corresponding to $B_\text{surf}$. While we cannot untangle the influence of $\tilde{\nu}$ and $\dot{\tilde{\nu}}$ on the spectral index, we can compare the $a, b$ exponents of all the pulsar parameters that we tested with this maximum value. We find that it is neither $\dot{P}$ ($-2, 1$), $B_\text{surf}$ ($-3/2, 1/2$), nor $\tau$ ($1, -1$) that determine the spectral index, as they all have negative $b/a$. On the other hand $B_\text{LC}$ ($3/2, 1/2$) and $\dot{E}$ ($1, 1$) are possible, with $B_\text{LC}$ being the closest and so the most likely parameter. However, the correlation coefficient for $\dot{E}$ is slightly higher.

\subsubsection{Robustness and significance of the correlations}
\label{sec:RobustnessAndSignificanceOfTheCorrelations}

We carried out extensive Monte Carlo simulations to assess the robustness of the correlations with respect to different sub-sets of the data and their significance in comparison with chance correlations. We confirm that the correlations are robust, significant and we estimate their uncertainties. We describe the details of the simulations in appendix \ref{sec:RobustnessAndSignificanceOfTheSpectralIndexCorrelations}.

\section{Discussion}
\label{sec:Discussion}

\subsection{Spectral indices and potential biases}
\label{sec:SpectralIndicesDiscussion}

\begin{table}
\caption{Top: Comparison between (weighted) mean spectral indices from this work and the literature. We consider only pulsars with simple power law spectra in all cases. Bottom: Intersection of pulsars studied in the various projects. $^\dagger$We could reconstruct 169 of the slow pulsars ($P > 20 \: \text{ms}$) examined by \citet{1998Toscano}.}
\label{tab:OverlapOfPulsarSets}
\centering
\begin{tabular}{lll}
\hline
set							& \#pulsars	& $\bar{\alpha}$\\
\hline
this work					& 276	& $-1.60 \pm 0.03$\\
\citet{1995Lorimer}			& 279	& $-1.6$\\
\citet{2000Maron}			& 263	& $-1.8 \pm 0.2$\\
\citet{1998Toscano}			& 216 ($169^\dagger$)	& $-1.72 \pm 0.04$\\
\citet{2000Malofeev}			& 175	& $-1.47 \pm 0.76$\\
\citet{2016Bilous}			& 48		& $-1.4$\\
\citet{2017Han}				& 228	& $-2.2$\\
\hline
\multicolumn{2}{l}{comparison}										& overlap\\
\multicolumn{2}{l}{ }												& [\#, \%]\\
\hline
\multicolumn{2}{l}{\citet{1995Lorimer} -- \citet{2000Maron}}			& 261, 94\\
\multicolumn{2}{l}{\citet{1995Lorimer} -- \citet{1998Toscano}}			& 20, 7\\
\multicolumn{2}{l}{this work -- \citet{1995Lorimer}}					& 54, 20\\
\multicolumn{2}{l}{this work -- \citet{2000Maron}}					& 53, 19\\
\multicolumn{2}{l}{this work -- \citet{1998Toscano}}					& 61, 22\\
\multicolumn{2}{l}{this work -- \citet{2017Han}}						& 55, 20\\
\hline
\end{tabular}
\end{table}

Other authors have determined the mean spectral index for the population of pulsars that they studied, for example \citet{1995Lorimer} found a mean spectral index of $-1.6$, \citet{2000Maron} found it to be $-1.8 \pm 0.2$ for single power law spectra. More recently \citet{2016Bilous} measured the spectral indices at low frequencies of 110 to 188~MHz and found a flatter mean spectral index of $-1.4$, which is comparable to what \citet{2000Malofeev} found ($-1.47 \pm 0.76$) for a similar frequency range. When considering flux density data in a narrow frequency range around 1.4~GHz, \citet{2017Han} found that the histogram peaks at a steeper spectral index of $-2.2$. In general it seems that all previous studies that included a large sample of pulsars (200 to 300 at least) over a large frequency range (about 300~MHz to 1.4~GHz or more) resulted in mean spectral indices of $-1.6$ to $-1.8$, see Table~\ref{tab:OverlapOfPulsarSets}. Our measurements are in good agreement with these. It is interesting that the mean spectral index that we find is very similar to the one calculated by \citet{1995Lorimer}. The question is, does the similarity in mean spectral index arise because these studies analysed the same or a similar set of pulsars? To answer this question we compute the intersection of observed pulsars in our data set with all the others. We find that it is only of the order $20 \: \%$. This leads us to conclude that the mean spectral index that we derived is independent of the set of pulsars studied.

We have also shown that the observed spectral index distribution is log-normal, which is different to what \citet{2013Bates} derived for the intrinsic spectral index distributions of three pulsar surveys, which they found to be Gaussian with mean spectral index of $-1.4$ and standard deviation of unity.

Another question is, have we introduced any biases in the result due to the selection of primarily bright, slow and isolated pulsars? To answer this we compared the spectral index distribution derived in this work with the one obtained using the pulsar catalogue alone. For the spectral index estimation using the catalogue we fit a power law to each pulsar that had at least three flux density measurements. We did not carry out any further spectral classification. The distribution contains 369 pulsars, is nearly symmetric and is consistent with being Gaussian with a mean spectral index of $-1.68$ and a median of $-1.72$. The addition of a significant number of flux density points at low and high frequencies from our survey and the literature (nearly all of our measurements are new ones) combined with a robust spectral classification has therefore skewed the distribution towards flatter indices. We assert that this is a more realistic representation of a sub-sample of the pulsar population and that further work at low and high frequencies will likely strengthen this trend. Another obvious bias is that our data set, as well as the pulsar catalogue, represents only the discovered sample of radio pulsars, not the population as a whole. That bias can be accounted for to some extent using population synthesis techniques, as implemented for example in the software \texttt{PSRPOP}, \texttt{PSRPOPPy} \citep{2006Lorimer, 2014Bates}. These can be used to study the underlying pulsar distribution based on the observed one. Such a study is beyond the scope of the current work.

\subsection{Spectral index correlations}
\label{sec:SpectralIndexCorrelationsDiscussion}

\citet{1995Lorimer} studied 343 pulsars in a combined data set and found a correlation of spectral index with period, and a broad inverse correlation between spectral index and characteristic age. Specifically they found a negative correlation between $\alpha$ and period for all pulsars studied with a Spearman rank correlation coefficient of $-0.15$ and for only normal pulsars of $-0.22$. MSPs showed a strong positive correlation between $\alpha$ and $P$ with a correlation coefficient of $+0.51$. For the correlation between $\alpha$ and characteristic age they found coefficients of $-0.22$ and $-0.19$ for all pulsars and normal pulsars respectively. They claimed that the characteristic age, rather than the period, is the key parameter that determines the value of the spectral index. On the other hand \citet{2000Maron} studied 281 pulsars and found no correlation between spectral index and period, period derivative, characteristic age and polarisation and profile type. In addition, \citet{1981Izvekova} found that the spectral index increases significantly with period at low radio frequencies between 61 and 408~MHz. Our analysis agrees with \citet{1995Lorimer} in the sense that we also find a negative correlation of spectral index with period and characteristic age, with a higher absolute correlation coefficient than in their work, see Table~\ref{tab:CorrelationWithAlpha}. We confirm that a negative correlation exists for normal pulsars. It also agrees with \citet{2017Han} in so far as they find the highest correlation with spin-down luminosity and potential drop in the polar gap ($0.32$); again we find a higher correlation. On the other hand, we find the highest absolute correlation between spectral index not with $P$ and $\tau$, but with spin-down luminosity, magnetic field at the light-cylinder radius and spin-down rate.

\subsection{Deviations from a simple power law spectrum}
\label{sec:DeviationFromASimplePowerLawSpectrum}

In this work the spectra of 73 pulsars ($21 \%$) deviate significantly from a simple power law. Three scenarios for the deviations can be conceived: 1) an environmental origin of the observed spectral features, i.e. pulsars emit radiation with featureless power law spectra and absorption processes in their environments are responsible for the spectral deviations. In that case we expect a correlation between the occurrence of spectral features and the presence of dense material near a pulsar, or along the line of sight. Observations of pulsars in high-density environments that have simple power law spectra provide constraints for that hypothesis. 2) The spectral features are intrinsic to the pulsar emission mechanism or due to (absorption) processes in the magnetosphere (e.g. synchrotron self-absorption), or the emission efficiency might change with radio frequency. We would expect to observe similar spectral features in all pulsars, given sufficient frequency coverage and measurement precision. In particular, pulsars with similarly well determined spectra should show similar features. 3) The emission physics is not generic, for example the emitted spectrum and its features could depend on spin frequency, beam geometry, or other pulsar parameters, and might change differently with radio frequency between pulsars.

We tested the hypotheses and find the following. All of the best-determined spectra ($\geq 40$ data points) deviate from a simple power law. Most have broken spectra, or power laws with low-frequency turn-over. This suggests that the spectral features are intrinsic. In addition, we searched for 2D spatial correlations between the pulsars with spectral deviations and Galactic supernova remnants (SNRs), molecular clouds (MCs), X-ray, GeV and TeV $\gamma$-ray sources from the literature \citep{2001Dame, 2008Wakely, 2014Green, 2015Acero, 2016Rosen, 2016Rice}. We use these as indicators for partially or fully ionized dense environments. 11 of the 73 pulsars have potential associations within $2 \arcmin$ or within the extent of literature sources, out of which 4 are with SNRs or MCs. These potential, purely 2D associations are: PSR J1707--4053 (hard cut-off) -- SNR G345.7--0.2, PSR J1745--3040 (LPS) -- SNR G358.5--0.9 and PSR J1857+0212 (LPS) -- SNR G35.6--0.4. PSR J1055-6028 (LPS, potential GPS) is spatially close to the edge of a MC. We repeated the analysis with the $15$ pulsars with the best-determined simple power law spectra ($p_\text{best} \geq 0.7$, $\geq 20$ data points). One pulsar has a potential X-ray association, but there are none with SNRs or MCs. This suggests that the spectral features are partially environmental in origin. However, the sample size of the best-determined power law spectra is small and our analysis does not include distance information, as the distances are either unknown or have large uncertainties. With regard to hypothesis 3, it is not unthinkable that spectral features depend on pulsar parameters, given that the spectral index shows a weak dependence with rotational parameters.

\section{Conclusions}
\label{sec:Conclusions}

In this work we present the largest sample of absolute calibrated pulsar flux density measurements to date. The data consist of new measurements at up to three centre frequencies (728, 1382 and 3100~MHz) of 441 radio pulsars, all taken with the Parkes radio telescope. We carefully calibrated the data and estimated their uncertainties. We studied the effects of interstellar scintillation using long-term pulsar observations that span 8.5~years and accounted for them in our measurements. We combined our measurements with spectral data from the literature and implemented an algorithm to objectively and robustly classify the pulsar spectra in an unbiased way into five different spectral classes. Our conclusions are the following:

\begin{enumerate}
	\item There is good overall agreement between the flux densities from this work and from the ATNF pulsar catalogue at 1.4~GHz. However, in some cases the catalogue data differ significantly. Most notably, the flux densities from the Parkes Multibeam Pulsar Survey \citep{2001Manchester, 2003Kramer, 2004Hobbs, 2006Lorimer} are in some cases about a factor of two lower than our measurements and other literature data.

	\item The vast majority of pulsars ($79 \%$) have featureless simple power law spectra in the frequency range studied.
	
	\item The observed spectral indices closely follow a shifted log-normal distribution. This has direct implications for population synthesis studies for future pulsar surveys. The weighted mean spectral index is $-1.60 \pm 0.03$ with a standard deviation of $0.54$, which is consistent with previous work. It is largely independent of the set of pulsars studied.
	
	\item The second largest class of pulsar spectra are the log-parabolic ones ($10 \%$), with roughly $60 \%$ of them peaking below 500~MHz, $30 \%$ peaking between 0.7 and 2~GHz and $10 \%$ having slightly concave spectra.
	
	\item The origin of the curvature in the LPS spectra is unclear. Only one of the pulsars is in a binary system, but five ($14 \%$) have known high-energy counterparts, making an origin due to (free-free) absorption in ionized high-density environments such as pulsar wind nebulae, supernova remnants, or HII regions, as suggested by \citet{2015Lewandowski}, \citet{2016Rajwade} and \citet{2017Kijak}, more likely. It could also be that we simply resolve an intrinsic turn-over at low and high frequencies in these sources.
	
	\item We identify 11 gigahertz-peaked spectrum pulsars, of which 3 are newly identified and 8 are confirmations of known GPS pulsars. 3 others show evidence of GPS, but require further low-frequency measurements to confirm the classification. Their spectra are best-fit by a LPS (8), broken power law (4) and power law with low-frequency turn-over (2).
	
	\item We confirm the classification for 8 of the 10 known GPS pulsars measured in this work. However, PSRs J1056--6258 and J1809--1917's spectra are consistent with simple power laws in the frequency range studied. In addition, PSR J1644--4559 might have a complex spectrum with a spectral flattening around 3.1~GHz and a turn-over at low frequencies.
	
	\item The spectra of 73 ($21 \%$) pulsars deviate from a simple power law. These include prominent and well studied pulsars such as the Vela pulsar (PSR J0835--4510), J1752--2806 and the MSPs J0437--4715, J0711--6830, J1024--0719 and J1045--4509.
	
	\item The physical reason for the occurrence of spectral features is uncertain. Our analysis suggests that the observed features are partially intrinsic to the emission mechanism or (absorption) processes in the magnetosphere and partially environmental.
	
	\item Following the prescription given by \citet{2013Kontorovich}, we derived unexpectedly low emission heights for three pulsars of 15 to 27~km ($< 0.1 \%$ of the light-cylinder radius), which could point to a problem in the spectral fitting of these sources or in the adopted emission model.
	
	\item The highest correlation between spectral index is with spin-down luminosity, magnetic field at the light cylinder radius and spin-down rate. This suggests that the rate of loss of rotational energy and/or the magnetic field strength at the light-cylinder are the key parameters that predict the spectral index.
\end{enumerate}

\section*{Acknowledgements}

The Parkes radio telescope is part of the Australia Telescope National Facility which is funded by the Australian Government for operation as a National Facility managed by CSIRO. Parts of this research were conducted by the Australian Research Council Centre of Excellence for All-sky Astrophysics (CAASTRO), through project number CE110001020. Work at NRL is supported by NASA. We thank the anonymous referee for constructive comments that have improved the paper.

%%%%%%%%%%%%%%%%%%%%%%%%%%%%%%%%%%%%%%%%%%%%%%%%%%



\bibliographystyle{mnras}
\bibliography{Spectral_properties_paper}

\begin{thebibliography}{}
\makeatletter
\relax
\def\mn@urlcharsother{\let\do\@makeother \do\$\do\&\do\#\do\^\do\_\do\%\do\~}
\def\mn@doi{\begingroup\mn@urlcharsother \@ifnextchar [ {\mn@doi@}
  {\mn@doi@[]}}
\def\mn@doi@[#1]#2{\def\@tempa{#1}\ifx\@tempa\@empty \href
  {http://dx.doi.org/#2} {doi:#2}\else \href {http://dx.doi.org/#2} {#1}\fi
  \endgroup}
\def\mn@eprint#1#2{\mn@eprint@#1:#2::\@nil}
\def\mn@eprint@arXiv#1{\href {http://arxiv.org/abs/#1} {{\tt arXiv:#1}}}
\def\mn@eprint@dblp#1{\href {http://dblp.uni-trier.de/rec/bibtex/#1.xml}
  {dblp:#1}}
\def\mn@eprint@#1:#2:#3:#4\@nil{\def\@tempa {#1}\def\@tempb {#2}\def\@tempc
  {#3}\ifx \@tempc \@empty \let \@tempc \@tempb \let \@tempb \@tempa \fi \ifx
  \@tempb \@empty \def\@tempb {arXiv}\fi \@ifundefined
  {mn@eprint@\@tempb}{\@tempb:\@tempc}{\expandafter \expandafter \csname
  mn@eprint@\@tempb\endcsname \expandafter{\@tempc}}}

\bibitem[\protect\citeauthoryear{{Acero} et~al.,}{{Acero}
  et~al.}{2015}]{2015Acero}
{Acero} F.,  et~al., 2015, \mn@doi [\apjs] {10.1088/0067-0049/218/2/23}, \href
  {http://adsabs.harvard.edu/abs/2015ApJS..218...23A} {218, 23}

\bibitem[\protect\citeauthoryear{{Armstrong}, {Rickett}  \&
  {Spangler}}{{Armstrong} et~al.}{1995}]{1995Armstrong}
{Armstrong} J.~W.,  {Rickett} B.~J.,   {Spangler} S.~R.,  1995, \mn@doi [\apj]
  {10.1086/175515}, \href {http://adsabs.harvard.edu/abs/1995ApJ...443..209A}
  {443, 209}

\bibitem[\protect\citeauthoryear{{Arons} \& {Scharlemann}}{{Arons} \&
  {Scharlemann}}{1979}]{1979Arons}
{Arons} J.,  {Scharlemann} E.~T.,  1979, \mn@doi [\apj] {10.1086/157250}, \href
  {http://adsabs.harvard.edu/abs/1979ApJ...231..854A} {231, 854}

\bibitem[\protect\citeauthoryear{{Baars}, {Genzel}, {Pauliny-Toth}  \&
  {Witzel}}{{Baars} et~al.}{1977}]{1977Baars}
{Baars} J.~W.~M.,  {Genzel} R.,  {Pauliny-Toth} I.~I.~K.,   {Witzel} A.,  1977,
  \aap, \href {http://adsabs.harvard.edu/abs/1977A%26A....61...99B} {61, 99}

\bibitem[\protect\citeauthoryear{{Bailes} et~al.,}{{Bailes}
  et~al.}{2017}]{2017Bailes}
{Bailes} M.,  et~al., 2017, preprint, \href
  {http://adsabs.harvard.edu/abs/2017arXiv170809619B} {} (\mn@eprint {arXiv}
  {1708.09619})

\bibitem[\protect\citeauthoryear{{Baring}}{{Baring}}{2004}]{2004Baring}
{Baring} M.~G.,  2004, \mn@doi [Advances in Space Research]
  {10.1016/j.asr.2003.08.020}, \href
  {http://adsabs.harvard.edu/abs/2004AdSpR..33..552B} {33, 552}

\bibitem[\protect\citeauthoryear{{Bartel}, {Sieber}  \& {Wielebinski}}{{Bartel}
  et~al.}{1978}]{1978Bartel}
{Bartel} N.,  {Sieber} W.,   {Wielebinski} R.,  1978, \aap, \href
  {http://adsabs.harvard.edu/abs/1978A%26A....68..361B} {68, 361}

\bibitem[\protect\citeauthoryear{{Basu}, {Ro{\.z}ko}, {Lewandowski}, {Kijak}
  \& {Dembska}}{{Basu} et~al.}{2016}]{2016Basu}
{Basu} R.,  {Ro{\.z}ko} K.,  {Lewandowski} W.,  {Kijak} J.,   {Dembska} M.,
  2016, \mn@doi [\mnras] {10.1093/mnras/stw394}, \href
  {http://adsabs.harvard.edu/abs/2016MNRAS.458.2509B} {458, 2509}

\bibitem[\protect\citeauthoryear{{Bates} et~al.,}{{Bates}
  et~al.}{2011}]{2011Bates}
{Bates} S.~D.,  et~al., 2011, \mn@doi [\mnras]
  {10.1111/j.1365-2966.2010.17790.x}, \href
  {http://adsabs.harvard.edu/abs/2011MNRAS.411.1575B} {411, 1575}

\bibitem[\protect\citeauthoryear{{Bates}, {Lorimer}  \& {Verbiest}}{{Bates}
  et~al.}{2013}]{2013Bates}
{Bates} S.~D.,  {Lorimer} D.~R.,   {Verbiest} J.~P.~W.,  2013, \mn@doi [\mnras]
  {10.1093/mnras/stt257}, \href
  {http://adsabs.harvard.edu/abs/2013MNRAS.431.1352B} {431, 1352}

\bibitem[\protect\citeauthoryear{{Bates}, {Lorimer}, {Rane}  \&
  {Swiggum}}{{Bates} et~al.}{2014}]{2014Bates}
{Bates} S.~D.,  {Lorimer} D.~R.,  {Rane} A.,   {Swiggum} J.,  2014, \mn@doi
  [\mnras] {10.1093/mnras/stu157}, \href
  {http://adsabs.harvard.edu/abs/2014MNRAS.439.2893B} {439, 2893}

\bibitem[\protect\citeauthoryear{{Bell} et~al.,}{{Bell}
  et~al.}{2016}]{2016Bell}
{Bell} M.~E.,  et~al., 2016, \mn@doi [\mnras] {10.1093/mnras/stw1293}, \href
  {http://adsabs.harvard.edu/abs/2016MNRAS.tmp..958B} {}

\bibitem[\protect\citeauthoryear{{Bhat}, {Rao}  \& {Gupta}}{{Bhat}
  et~al.}{1999}]{1999Bhat_1}
{Bhat} N.~D.~R.,  {Rao} A.~P.,   {Gupta} Y.,  1999, \mn@doi [ApJS]
  {10.1086/313198}, \href {http://adsabs.harvard.edu/abs/1999ApJS..121..483B}
  {121, 483}

\bibitem[\protect\citeauthoryear{{Bhat}, {Cordes}, {Camilo}, {Nice}  \&
  {Lorimer}}{{Bhat} et~al.}{2004}]{2004Bhat}
{Bhat} N.~D.~R.,  {Cordes} J.~M.,  {Camilo} F.,  {Nice} D.~J.,   {Lorimer}
  D.~R.,  2004, \mn@doi [\apj] {10.1086/382680}, \href
  {http://adsabs.harvard.edu/abs/2004ApJ...605..759B} {605, 759}

\bibitem[\protect\citeauthoryear{{Bilous} et~al.,}{{Bilous}
  et~al.}{2016}]{2016Bilous}
{Bilous} A.~V.,  et~al., 2016, \mn@doi [\aap] {10.1051/0004-6361/201527702},
  \href {http://adsabs.harvard.edu/abs/2016A%26A...591A.134B} {591, A134}

\bibitem[\protect\citeauthoryear{Burnham, Anderson  \& Huyvaert}{Burnham
  et~al.}{2010}]{2010Burnham}
Burnham K.~P.,  Anderson D.~R.,   Huyvaert K.~P.,  2010, \mn@doi [Behavioral
  Ecology and Sociobiology] {10.1007/s00265-010-1029-6}, 65, 23

\bibitem[\protect\citeauthoryear{{Cheng}, {Ho}  \& {Ruderman}}{{Cheng}
  et~al.}{1986}]{1986Cheng}
{Cheng} K.~S.,  {Ho} C.,   {Ruderman} M.,  1986, \mn@doi [\apj]
  {10.1086/163829}, \href {http://adsabs.harvard.edu/abs/1986ApJ...300..500C}
  {300, 500}

\bibitem[\protect\citeauthoryear{{Cordes}}{{Cordes}}{1978}]{1978Cordes}
{Cordes} J.~M.,  1978, \mn@doi [\apj] {10.1086/156218}, \href
  {http://adsabs.harvard.edu/abs/1978ApJ...222.1006C} {222, 1006}

\bibitem[\protect\citeauthoryear{{Cordes} \& {Lazio}}{{Cordes} \&
  {Lazio}}{1991}]{1991Cordes}
{Cordes} J.~M.,  {Lazio} T.~J.,  1991, \mn@doi [\apj] {10.1086/170261}, \href
  {http://adsabs.harvard.edu/abs/1991ApJ...376..123C} {376, 123}

\bibitem[\protect\citeauthoryear{{Cordes} \& {Lazio}}{{Cordes} \&
  {Lazio}}{2002}]{2002Cordes}
{Cordes} J.~M.,  {Lazio} T.~J.~W.,  2002, ArXiv Astrophysics e-prints, \href
  {http://adsabs.harvard.edu/abs/2002astro.ph..7156C} {}

\bibitem[\protect\citeauthoryear{{Cordes} \& {Rickett}}{{Cordes} \&
  {Rickett}}{1998}]{1998Cordes}
{Cordes} J.~M.,  {Rickett} B.~J.,  1998, \mn@doi [\apj] {10.1086/306358}, \href
  {http://adsabs.harvard.edu/abs/1998ApJ...507..846C} {507, 846}

\bibitem[\protect\citeauthoryear{{Dai} et~al.,}{{Dai} et~al.}{2015}]{2015Dai}
{Dai} S.,  et~al., 2015, \mn@doi [\mnras] {10.1093/mnras/stv508}, \href
  {http://adsabs.harvard.edu/abs/2015MNRAS.449.3223D} {449, 3223}

\bibitem[\protect\citeauthoryear{{Dame}, {Hartmann}  \& {Thaddeus}}{{Dame}
  et~al.}{2001}]{2001Dame}
{Dame} T.~M.,  {Hartmann} D.,   {Thaddeus} P.,  2001, \mn@doi [\apj]
  {10.1086/318388}, \href {http://adsabs.harvard.edu/abs/2001ApJ...547..792D}
  {547, 792}

\bibitem[\protect\citeauthoryear{{Dembska}, {Kijak}, {Jessner}, {Lewandowski},
  {Bhattacharyya}  \& {Gupta}}{{Dembska} et~al.}{2014}]{2014Dembska}
{Dembska} M.,  {Kijak} J.,  {Jessner} A.,  {Lewandowski} W.,  {Bhattacharyya}
  B.,   {Gupta} Y.,  2014, \mn@doi [\mnras] {10.1093/mnras/stu1905}, \href
  {http://adsabs.harvard.edu/abs/2014MNRAS.445.3105D} {445, 3105}

\bibitem[\protect\citeauthoryear{{Dembska}, {Kijak}, {Koralewska},
  {Lewandowski}, {Melikidze}  \& {Ro{\.z}ko}}{{Dembska}
  et~al.}{2015a}]{2015DembskaBinary}
{Dembska} M.,  {Kijak} J.,  {Koralewska} O.,  {Lewandowski} W.,  {Melikidze}
  G.,   {Ro{\.z}ko} K.,  2015a, \mn@doi [\apss] {10.1007/s10509-015-2447-8},
  \href {http://adsabs.harvard.edu/abs/2015Ap%26SS.359...31D} {359, 31}

\bibitem[\protect\citeauthoryear{{Dembska}, {Basu}, {Kijak}  \&
  {Lewandowski}}{{Dembska} et~al.}{2015b}]{2015Dembska}
{Dembska} M.,  {Basu} R.,  {Kijak} J.,   {Lewandowski} W.,  2015b, \mn@doi
  [\mnras] {10.1093/mnras/stv333}, \href
  {http://adsabs.harvard.edu/abs/2015MNRAS.449.1869D} {449, 1869}

\bibitem[\protect\citeauthoryear{{Gil}, {Lyubarsky}  \& {Melikidze}}{{Gil}
  et~al.}{2004}]{2004Gil}
{Gil} J.,  {Lyubarsky} Y.,   {Melikidze} G.~I.,  2004, \mn@doi [\apj]
  {10.1086/379972}, \href {http://adsabs.harvard.edu/abs/2004ApJ...600..872G}
  {600, 872}

\bibitem[\protect\citeauthoryear{{Goldreich} \& {Julian}}{{Goldreich} \&
  {Julian}}{1969}]{1969Goldreich}
{Goldreich} P.,  {Julian} W.~H.,  1969, \mn@doi [\apj] {10.1086/150119}, \href
  {http://adsabs.harvard.edu/abs/1969ApJ...157..869G} {157, 869}

\bibitem[\protect\citeauthoryear{{Green}}{{Green}}{2014}]{2014Green}
{Green} D.~A.,  2014, Bulletin of the Astronomical Society of India, \href
  {http://adsabs.harvard.edu/abs/2014BASI...42...47G} {42, 47}

\bibitem[\protect\citeauthoryear{{Han}, {Wang}, {Xu}  \& {Han}}{{Han}
  et~al.}{2017}]{2017Han}
{Han} J.,  {Wang} C.,  {Xu} J.,   {Han} J.,  2017, preprint, \href
  {http://adsabs.harvard.edu/abs/2017arXiv170305988H} {} (\mn@eprint {arXiv}
  {1703.05988})

\bibitem[\protect\citeauthoryear{{Haslam}, {Salter}, {Stoffel}  \&
  {Wilson}}{{Haslam} et~al.}{1982}]{1982Haslam}
{Haslam} C.~G.~T.,  {Salter} C.~J.,  {Stoffel} H.,   {Wilson} W.~E.,  1982,
  \aaps, \href {http://adsabs.harvard.edu/abs/1982A%26AS...47....1H} {47, 1}

\bibitem[\protect\citeauthoryear{{Hassall} et~al.,}{{Hassall}
  et~al.}{2012}]{2012Hassall}
{Hassall} T.~E.,  et~al., 2012, \mn@doi [\aap] {10.1051/0004-6361/201218970},
  \href {http://adsabs.harvard.edu/abs/2012A%26A...543A..66H} {543, A66}

\bibitem[\protect\citeauthoryear{{Hewish}, {Bell}, {Pilkington}, {Scott}  \&
  {Collins}}{{Hewish} et~al.}{1968}]{1968Hewish}
{Hewish} A.,  {Bell} S.~J.,  {Pilkington} J.~D.~H.,  {Scott} P.~F.,   {Collins}
  R.~A.,  1968, \mn@doi [\nat] {10.1038/217709a0}, \href
  {http://adsabs.harvard.edu/abs/1968Natur.217..709H} {217, 709}

\bibitem[\protect\citeauthoryear{{Hobbs} et~al.,}{{Hobbs}
  et~al.}{2004}]{2004Hobbs}
{Hobbs} G.,  et~al., 2004, \mn@doi [\mnras] {10.1111/j.1365-2966.2004.08042.x},
  \href {http://adsabs.harvard.edu/abs/2004MNRAS.352.1439H} {352, 1439}

\bibitem[\protect\citeauthoryear{{Hobbs} et~al.,}{{Hobbs}
  et~al.}{2011}]{2011Hobbs}
{Hobbs} G.,  et~al., 2011, \mn@doi [\pasa] {10.1071/AS11016}, \href
  {http://adsabs.harvard.edu/abs/2011PASA...28..202H} {28, 202}

\bibitem[\protect\citeauthoryear{{Hotan}, {van Straten}  \&
  {Manchester}}{{Hotan} et~al.}{2004}]{2004Hotan}
{Hotan} A.~W.,  {van Straten} W.,   {Manchester} R.~N.,  2004, \mn@doi [\pasa]
  {10.1071/AS04022}, \href {http://adsabs.harvard.edu/abs/2004PASA...21..302H}
  {21, 302}

\bibitem[\protect\citeauthoryear{Huber}{Huber}{1964}]{1964Huber}
Huber P.~J.,  1964, \mn@doi [Ann. Math. Statist.] {10.1214/aoms/1177703732},
  35, 73

\bibitem[\protect\citeauthoryear{{Ivezi{\'c}}, {Connelly}, {VanderPlas}  \&
  {Gray}}{{Ivezi{\'c}} et~al.}{2014}]{2014Ivezic}
{Ivezi{\'c}} {\v Z}.,  {Connelly} A.~J.,  {VanderPlas} J.~T.,   {Gray} A.,
  2014, {Statistics, Data Mining, and Machine Learning in Astronomy}

\bibitem[\protect\citeauthoryear{{Izvekova}, {Kuzmin}, {Malofeev}  \&
  {Shitov}}{{Izvekova} et~al.}{1981}]{1981Izvekova}
{Izvekova} V.~A.,  {Kuzmin} A.~D.,  {Malofeev} V.~M.,   {Shitov} I.~P.,  1981,
  \mn@doi [\apss] {10.1007/BF00654022}, \href
  {http://adsabs.harvard.edu/abs/1981Ap%26SS..78...45I} {78, 45}

\bibitem[\protect\citeauthoryear{{James} \& {Roos}}{{James} \&
  {Roos}}{1975}]{1975James}
{James} F.,  {Roos} M.,  1975, \mn@doi [Computer Physics Communications]
  {10.1016/0010-4655(75)90039-9}, \href
  {http://adsabs.harvard.edu/abs/1975CoPhC..10..343J} {10, 343}

\bibitem[\protect\citeauthoryear{{Johnston}, {Manchester}, {Lyne}, {D'Amico},
  {Bailes}, {Gaensler}  \& {Nicastro}}{{Johnston} et~al.}{1996}]{1996Johnston}
{Johnston} S.,  {Manchester} R.~N.,  {Lyne} A.~G.,  {D'Amico} N.,  {Bailes} M.,
   {Gaensler} B.~M.,   {Nicastro} L.,  1996, \mn@doi [\mnras]
  {10.1093/mnras/279.3.1026}, \href
  {http://adsabs.harvard.edu/abs/1996MNRAS.279.1026J} {279, 1026}

\bibitem[\protect\citeauthoryear{{Johnston}, {Karastergiou}  \&
  {Willett}}{{Johnston} et~al.}{2006}]{2006Johnston}
{Johnston} S.,  {Karastergiou} A.,   {Willett} K.,  2006, \mn@doi [\mnras]
  {10.1111/j.1365-2966.2006.10440.x}, \href
  {http://adsabs.harvard.edu/abs/2006MNRAS.369.1916J} {369, 1916}

\bibitem[\protect\citeauthoryear{Jones, Oliphant, Peterson  et~al.}{Jones
  et~al.}{2001}]{Scipy}
Jones E.,  Oliphant T.,  Peterson P.,   et~al., 2001, {SciPy}: Open source
  scientific tools for {Python}, \url {http://www.scipy.org/}

\bibitem[\protect\citeauthoryear{{Karastergiou}, {Johnston}  \&
  {Manchester}}{{Karastergiou} et~al.}{2005}]{2005Karastergiou}
{Karastergiou} A.,  {Johnston} S.,   {Manchester} R.~N.,  2005, \mn@doi
  [\mnras] {10.1111/j.1365-2966.2005.08909.x}, \href
  {http://adsabs.harvard.edu/abs/2005MNRAS.359..481K} {359, 481}

\bibitem[\protect\citeauthoryear{{Keane}, {Bhattacharyya}, {Kramer}, {Stappers}
   et~al.}{{Keane} et~al.}{2015}]{2015Keane}
{Keane} E.,  {Bhattacharyya} B.,  {Kramer} M.,  {Stappers} B.,   et~al., 2015,
  Advancing Astrophysics with the Square Kilometre Array (AASKA14), \href
  {http://adsabs.harvard.edu/abs/2015aska.confE..40K} {p.~40}

\bibitem[\protect\citeauthoryear{{Keith}, {Johnston}, {Kramer}, {Weltevrede},
  {Watters}  \& {Stappers}}{{Keith} et~al.}{2008}]{2008Keith}
{Keith} M.~J.,  {Johnston} S.,  {Kramer} M.,  {Weltevrede} P.,  {Watters}
  K.~P.,   {Stappers} B.~W.,  2008, \mn@doi [\mnras]
  {10.1111/j.1365-2966.2008.13711.x}, \href
  {http://adsabs.harvard.edu/abs/2008MNRAS.389.1881K} {389, 1881}

\bibitem[\protect\citeauthoryear{{Keith}, {Johnston}, {Levin}  \&
  {Bailes}}{{Keith} et~al.}{2011}]{2011Keith}
{Keith} M.~J.,  {Johnston} S.,  {Levin} L.,   {Bailes} M.,  2011, \mn@doi
  [\mnras] {10.1111/j.1365-2966.2011.19041.x}, \href
  {http://adsabs.harvard.edu/abs/2011MNRAS.416..346K} {416, 346}

\bibitem[\protect\citeauthoryear{{Kijak}, {Gupta}  \& {Krzeszowski}}{{Kijak}
  et~al.}{2007}]{2007Kijak}
{Kijak} J.,  {Gupta} Y.,   {Krzeszowski} K.,  2007, \mn@doi [\aap]
  {10.1051/0004-6361:20066125}, \href
  {http://adsabs.harvard.edu/abs/2007A%26A...462..699K} {462, 699}

\bibitem[\protect\citeauthoryear{{Kijak}, {Dembska}, {Lewandowski}, {Melikidze}
   \& {Sendyk}}{{Kijak} et~al.}{2011a}]{2011KijakBinary}
{Kijak} J.,  {Dembska} M.,  {Lewandowski} W.,  {Melikidze} G.,   {Sendyk} M.,
  2011a, \mn@doi [\mnras] {10.1111/j.1745-3933.2011.01155.x}, \href
  {http://adsabs.harvard.edu/abs/2011MNRAS.418L.114K} {418, L114}

\bibitem[\protect\citeauthoryear{{Kijak}, {Lewandowski}, {Maron}, {Gupta}  \&
  {Jessner}}{{Kijak} et~al.}{2011b}]{2011Kijak}
{Kijak} J.,  {Lewandowski} W.,  {Maron} O.,  {Gupta} Y.,   {Jessner} A.,
  2011b, \mn@doi [\aap] {10.1051/0004-6361/201014274}, \href
  {http://adsabs.harvard.edu/abs/2011A%26A...531A..16K} {531, A16}

\bibitem[\protect\citeauthoryear{{Kijak}, {Basu}, {Lewandowski}, {Ro{\.z}ko}
  \& {Dembska}}{{Kijak} et~al.}{2017}]{2017Kijak}
{Kijak} J.,  {Basu} R.,  {Lewandowski} W.,  {Ro{\.z}ko} K.,   {Dembska} M.,
  2017, \mn@doi [\apj] {10.3847/1538-4357/aa6ff2}, \href
  {http://adsabs.harvard.edu/abs/2017ApJ...840..108K} {840, 108}

\bibitem[\protect\citeauthoryear{{Koester} \& {Reimers}}{{Koester} \&
  {Reimers}}{2000}]{2000Koester}
{Koester} D.,  {Reimers} D.,  2000, \aap, \href
  {http://adsabs.harvard.edu/abs/2000A%26A...364L..66K} {364, L66}

\bibitem[\protect\citeauthoryear{{Kontorovich} \& {Flanchik}}{{Kontorovich} \&
  {Flanchik}}{2013}]{2013Kontorovich}
{Kontorovich} V.~M.,  {Flanchik} A.~B.,  2013, \mn@doi [\apss]
  {10.1007/s10509-013-1369-6}, \href
  {http://adsabs.harvard.edu/abs/2013Ap%26SS.345..169K} {345, 169}

\bibitem[\protect\citeauthoryear{{Kramer}, {Wielebinski}, {Jessner}, {Gil}  \&
  {Seiradakis}}{{Kramer} et~al.}{1994}]{1994Kramer}
{Kramer} M.,  {Wielebinski} R.,  {Jessner} A.,  {Gil} J.~A.,   {Seiradakis}
  J.~H.,  1994, \aaps, \href
  {http://adsabs.harvard.edu/abs/1994A%26AS..107..515K} {107}

\bibitem[\protect\citeauthoryear{{Kramer} et~al.,}{{Kramer}
  et~al.}{2003}]{2003Kramer}
{Kramer} M.,  et~al., 2003, \mn@doi [\mnras]
  {10.1046/j.1365-8711.2003.06637.x}, \href
  {http://adsabs.harvard.edu/abs/2003MNRAS.342.1299K} {342, 1299}

\bibitem[\protect\citeauthoryear{{Kuehr}, {Witzel}, {Pauliny-Toth}  \&
  {Nauber}}{{Kuehr} et~al.}{1981}]{1981Kuehr}
{Kuehr} H.,  {Witzel} A.,  {Pauliny-Toth} I.~I.~K.,   {Nauber} U.,  1981,
  \aaps, \href {http://adsabs.harvard.edu/abs/1981A%26AS...45..367K} {45, 367}

\bibitem[\protect\citeauthoryear{{Lane}, {Clarke}, {Taylor}, {Perley}  \&
  {Kassim}}{{Lane} et~al.}{2004}]{2004Lane}
{Lane} W.~M.,  {Clarke} T.~E.,  {Taylor} G.~B.,  {Perley} R.~A.,   {Kassim}
  N.~E.,  2004, \mn@doi [\aj] {10.1086/379858}, \href
  {http://adsabs.harvard.edu/abs/2004AJ....127...48L} {127, 48}

\bibitem[\protect\citeauthoryear{{Lawson}, {Mayer}, {Osborne}  \&
  {Parkinson}}{{Lawson} et~al.}{1987}]{1987Lawson}
{Lawson} K.~D.,  {Mayer} C.~J.,  {Osborne} J.~L.,   {Parkinson} M.~L.,  1987,
  \mn@doi [\mnras] {10.1093/mnras/225.2.307}, \href
  {http://adsabs.harvard.edu/abs/1987MNRAS.225..307L} {225, 307}

\bibitem[\protect\citeauthoryear{{Levin} et~al.,}{{Levin}
  et~al.}{2013}]{2013Levin}
{Levin} L.,  et~al., 2013, \mn@doi [\mnras] {10.1093/mnras/stt1103}, \href
  {http://adsabs.harvard.edu/abs/2013MNRAS.434.1387L} {434, 1387}

\bibitem[\protect\citeauthoryear{{Lewandowski}, {Ro{\.z}ko}, {Kijak}  \&
  {Melikidze}}{{Lewandowski} et~al.}{2015}]{2015Lewandowski}
{Lewandowski} W.,  {Ro{\.z}ko} K.,  {Kijak} J.,   {Melikidze} G.~I.,  2015,
  \mn@doi [\apj] {10.1088/0004-637X/808/1/18}, \href
  {http://adsabs.harvard.edu/abs/2015ApJ...808...18L} {808, 18}

\bibitem[\protect\citeauthoryear{{Liddle}}{{Liddle}}{2007}]{2007Liddle}
{Liddle} A.~R.,  2007, \mn@doi [\mnras] {10.1111/j.1745-3933.2007.00306.x},
  \href {http://adsabs.harvard.edu/abs/2007MNRAS.377L..74L} {377, L74}

\bibitem[\protect\citeauthoryear{{Lorimer} \& {Kramer}}{{Lorimer} \&
  {Kramer}}{2012}]{2012Lorimer}
{Lorimer} D.~R.,  {Kramer} M.,  2012, {Handbook of Pulsar Astronomy}

\bibitem[\protect\citeauthoryear{{Lorimer}, {Yates}, {Lyne}  \&
  {Gould}}{{Lorimer} et~al.}{1995}]{1995Lorimer}
{Lorimer} D.~R.,  {Yates} J.~A.,  {Lyne} A.~G.,   {Gould} D.~M.,  1995, \mnras,
  \href {http://ads.nao.ac.jp/abs/1995MNRAS.273..411L} {273, 411}

\bibitem[\protect\citeauthoryear{{Lorimer} et~al.,}{{Lorimer}
  et~al.}{2006}]{2006Lorimer}
{Lorimer} D.~R.,  et~al., 2006, \mn@doi [\mnras]
  {10.1111/j.1365-2966.2006.10887.x}, \href
  {http://adsabs.harvard.edu/abs/2006MNRAS.372..777L} {372, 777}

\bibitem[\protect\citeauthoryear{{Lyne} \& {Manchester}}{{Lyne} \&
  {Manchester}}{1988}]{1988Lyne}
{Lyne} A.~G.,  {Manchester} R.~N.,  1988, \mn@doi [\mnras]
  {10.1093/mnras/234.3.477}, \href
  {http://adsabs.harvard.edu/abs/1988MNRAS.234..477L} {234, 477}

\bibitem[\protect\citeauthoryear{{Malofeev} \& {Malov}}{{Malofeev} \&
  {Malov}}{1980}]{1980Malofeev}
{Malofeev} V.~M.,  {Malov} I.~F.,  1980, \sovast, \href
  {http://adsabs.harvard.edu/abs/1980SvA....24...54M} {24, 90}

\bibitem[\protect\citeauthoryear{{Malofeev}, {Malov}  \&
  {Shchegoleva}}{{Malofeev} et~al.}{2000}]{2000Malofeev}
{Malofeev} V.~M.,  {Malov} O.~I.,   {Shchegoleva} N.~V.,  2000, \mn@doi
  [Astronomy Reports] {10.1134/1.163868}, \href
  {http://adsabs.harvard.edu/abs/2000ARep...44..436M} {44, 436}

\bibitem[\protect\citeauthoryear{{Manchester} et~al.,}{{Manchester}
  et~al.}{2001}]{2001Manchester}
{Manchester} R.~N.,  et~al., 2001, \mn@doi [\mnras]
  {10.1046/j.1365-8711.2001.04751.x}, \href
  {http://adsabs.harvard.edu/abs/2001MNRAS.328...17M} {328, 17}

\bibitem[\protect\citeauthoryear{{Manchester}, {Hobbs}, {Teoh}  \& et.
  al.}{{Manchester} et~al.}{2005}]{2005Manchester}
{Manchester} R.~N.,  {Hobbs} G.~B.,  {Teoh} A.,   et. al. 2005, \mn@doi [\aj]
  {10.1086/428488}, \href {http://adsabs.harvard.edu/abs/2005AJ....129.1993M}
  {129, 1993}

\bibitem[\protect\citeauthoryear{{Manchester} et~al.,}{{Manchester}
  et~al.}{2013}]{2013Manchester}
{Manchester} R.~N.,  et~al., 2013, \mn@doi [\pasa] {10.1017/pasa.2012.017},
  \href {http://adsabs.harvard.edu/abs/2013PASA...30...17M} {30, e017}

\bibitem[\protect\citeauthoryear{{Maron}, {Kijak}, {Kramer}  \&
  {Wielebinski}}{{Maron} et~al.}{2000}]{2000Maron}
{Maron} O.,  {Kijak} J.,  {Kramer} M.,   {Wielebinski} R.,  2000, \mn@doi
  [\aaps] {10.1051/aas:2000298}, \href
  {http://adsabs.harvard.edu/abs/2000A%26AS..147..195M} {147, 195}

\bibitem[\protect\citeauthoryear{{Melrose} \& {Yuen}}{{Melrose} \&
  {Yuen}}{2016}]{2016Melrose}
{Melrose} D.~B.,  {Yuen} R.,  2016, \mn@doi [Journal of Plasma Physics]
  {10.1017/S0022377816000398}, \href
  {http://adsabs.harvard.edu/abs/2016JPlPh..82b6302M} {82, 635820202}

\bibitem[\protect\citeauthoryear{{Morris} et~al.,}{{Morris}
  et~al.}{2002}]{2002Morris}
{Morris} D.~J.,  et~al., 2002, \mn@doi [\mnras]
  {10.1046/j.1365-8711.2002.05551.x}, \href
  {http://adsabs.harvard.edu/abs/2002MNRAS.335..275M} {335, 275}

\bibitem[\protect\citeauthoryear{{Murphy} et~al.,}{{Murphy}
  et~al.}{2017}]{2017Murphy}
{Murphy} T.,  et~al., 2017, \mn@doi [\pasa] {10.1017/pasa.2017.13}, \href
  {http://adsabs.harvard.edu/abs/2017PASA...34...20M} {34, e020}

\bibitem[\protect\citeauthoryear{{Nita} \& {Gary}}{{Nita} \&
  {Gary}}{2010}]{2010Nita}
{Nita} G.~M.,  {Gary} D.~E.,  2010, \mn@doi [\pasp] {10.1086/652409}, \href
  {http://adsabs.harvard.edu/abs/2010PASP..122..595N} {122, 595}

\bibitem[\protect\citeauthoryear{{Perera}, {Stappers}, {Weltevrede}, {Lyne}  \&
  {Rankin}}{{Perera} et~al.}{2016}]{2016Perera}
{Perera} B.~B.~P.,  {Stappers} B.~W.,  {Weltevrede} P.,  {Lyne} A.~G.,
  {Rankin} J.~M.,  2016, \mn@doi [\mnras] {10.1093/mnras/stv2403}, \href
  {http://adsabs.harvard.edu/abs/2016MNRAS.455.1071P} {455, 1071}

\bibitem[\protect\citeauthoryear{{Rajwade}, {Lorimer}  \& {Anderson}}{{Rajwade}
  et~al.}{2016}]{2016Rajwade}
{Rajwade} K.,  {Lorimer} D.~R.,   {Anderson} L.~D.,  2016, \mn@doi [\mnras]
  {10.1093/mnras/stv2334}, \href
  {http://adsabs.harvard.edu/abs/2016MNRAS.455..493R} {455, 493}

\bibitem[\protect\citeauthoryear{{Rice}, {Goodman}, {Bergin}, {Beaumont}  \&
  {Dame}}{{Rice} et~al.}{2016}]{2016Rice}
{Rice} T.~S.,  {Goodman} A.~A.,  {Bergin} E.~A.,  {Beaumont} C.,   {Dame}
  T.~M.,  2016, \mn@doi [\apj] {10.3847/0004-637X/822/1/52}, \href
  {http://adsabs.harvard.edu/abs/2016ApJ...822...52R} {822, 52}

\bibitem[\protect\citeauthoryear{{Rickett}}{{Rickett}}{1990}]{1990Rickett}
{Rickett} B.~J.,  1990, \mn@doi [\araa] {10.1146/annurev.aa.28.090190.003021},
  \href {http://adsabs.harvard.edu/abs/1990ARA%26A..28..561R} {28, 561}

\bibitem[\protect\citeauthoryear{{Romani}, {Narayan}  \& {Blandford}}{{Romani}
  et~al.}{1986}]{1986Romani}
{Romani} R.~W.,  {Narayan} R.,   {Blandford} R.,  1986, \mn@doi [\mnras]
  {10.1093/mnras/220.1.19}, \href
  {http://adsabs.harvard.edu/abs/1986MNRAS.220...19R} {220, 19}

\bibitem[\protect\citeauthoryear{{Rosen} et~al.,}{{Rosen}
  et~al.}{2016}]{2016Rosen}
{Rosen} S.~R.,  et~al., 2016, \mn@doi [\aap] {10.1051/0004-6361/201526416},
  \href {http://adsabs.harvard.edu/abs/2016A%26A...590A...1R} {590, A1}

\bibitem[\protect\citeauthoryear{{Ruderman} \& {Sutherland}}{{Ruderman} \&
  {Sutherland}}{1975}]{1975Ruderman}
{Ruderman} M.~A.,  {Sutherland} P.~G.,  1975, \mn@doi [\apj] {10.1086/153393},
  \href {http://adsabs.harvard.edu/abs/1975ApJ...196...51R} {196, 51}

\bibitem[\protect\citeauthoryear{{Sieber}}{{Sieber}}{1973}]{1973Sieber}
{Sieber} W.,  1973, \aap, \href
  {http://adsabs.harvard.edu/abs/1973A%26A....28..237S} {28, 237}

\bibitem[\protect\citeauthoryear{{Sieber}}{{Sieber}}{1982}]{1982Sieber}
{Sieber} W.,  1982, \aap, \href
  {http://adsabs.harvard.edu/abs/1982A%26A...113..311S} {113, 311}

\bibitem[\protect\citeauthoryear{{Stinebring}, {Smirnova}, {Hankins}, {Hovis},
  {Kaspi}, {Kempner}, {Myers}  \& {Nice}}{{Stinebring}
  et~al.}{2000}]{2000Stinebring}
{Stinebring} D.~R.,  {Smirnova} T.~V.,  {Hankins} T.~H.,  {Hovis} J.~S.,
  {Kaspi} V.~M.,  {Kempner} J.~C.,  {Myers} E.,   {Nice} D.~J.,  2000, \mn@doi
  [\apj] {10.1086/309201}, \href
  {http://adsabs.harvard.edu/abs/2000ApJ...539..300S} {539, 300}

\bibitem[\protect\citeauthoryear{{Stovall} et~al.,}{{Stovall}
  et~al.}{2015}]{2015Stovall}
{Stovall} K.,  et~al., 2015, \mn@doi [\apj] {10.1088/0004-637X/808/2/156},
  \href {http://adsabs.harvard.edu/abs/2015ApJ...808..156S} {808, 156}

\bibitem[\protect\citeauthoryear{{Sturrock}}{{Sturrock}}{1971}]{1971Sturrock}
{Sturrock} P.~A.,  1971, \mn@doi [\apj] {10.1086/150865}, \href
  {http://adsabs.harvard.edu/abs/1971ApJ...164..529S} {164, 529}

\bibitem[\protect\citeauthoryear{{Toscano}, {Bailes}, {Manchester}  \&
  {Sandhu}}{{Toscano} et~al.}{1998}]{1998Toscano}
{Toscano} M.,  {Bailes} M.,  {Manchester} R.~N.,   {Sandhu} J.~S.,  1998,
  \mn@doi [\apj] {10.1086/306282}, \href
  {http://adsabs.harvard.edu/abs/1998ApJ...506..863T} {506, 863}

\bibitem[\protect\citeauthoryear{{Tsai}, {Simonetti}, {Akukwe}, {Bear},
  {Gough}, {Shawhan}  \& {Kavic}}{{Tsai} et~al.}{2016}]{2016Tsai}
{Tsai} Jr. .,  {Simonetti} J.~H.,  {Akukwe} B.,  {Bear} B.,  {Gough} J.~D.,
  {Shawhan} P.,   {Kavic} M.,  2016, \mn@doi [\aj]
  {10.3847/0004-6256/151/2/28}, \href
  {http://adsabs.harvard.edu/abs/2016AJ....151...28T} {151, 28}

\bibitem[\protect\citeauthoryear{{Wakely} \& {Horan}}{{Wakely} \&
  {Horan}}{2008}]{2008Wakely}
{Wakely} S.~P.,  {Horan} D.,  2008, International Cosmic Ray Conference, \href
  {http://adsabs.harvard.edu/abs/2008ICRC....3.1341W} {3, 1341}

\bibitem[\protect\citeauthoryear{{Weltevrede} et~al.,}{{Weltevrede}
  et~al.}{2010}]{2010Weltevrede}
{Weltevrede} P.,  et~al., 2010, \mn@doi [\pasa] {10.1071/AS09054}, \href
  {http://adsabs.harvard.edu/abs/2010PASA...27...64W} {27, 64}

\bibitem[\protect\citeauthoryear{{Xilouris}, {Kramer}, {Jessner}, {Wielebinski}
   \& {Timofeev}}{{Xilouris} et~al.}{1996}]{1996Xilouris}
{Xilouris} K.~M.,  {Kramer} M.,  {Jessner} A.,  {Wielebinski} R.,   {Timofeev}
  M.,  1996, \aap, \href {http://adsabs.harvard.edu/abs/1996A%26A...309..481X}
  {309, 481}

\bibitem[\protect\citeauthoryear{{Yakovleva} \& {Kulberg}}{{Yakovleva} \&
  {Kulberg}}{2013}]{2013Yakovleva}
{Yakovleva} T.~V.,  {Kulberg} N.~S.,  2013, \mn@doi [American Journal of
  Theoretical and Applied Statistics] {10.11648/j.ajtas.20130203.15}, 2, 67

\bibitem[\protect\citeauthoryear{{Zakharenko} et~al.,}{{Zakharenko}
  et~al.}{2013}]{2013Zakharenko}
{Zakharenko} V.~V.,  et~al., 2013, \mn@doi [\mnras] {10.1093/mnras/stt470},
  \href {http://adsabs.harvard.edu/abs/2013MNRAS.431.3624Z} {431, 3624}

\bibitem[\protect\citeauthoryear{{van Kerkwijk}, {Bassa}, {Jacoby}  \&
  {Jonker}}{{van Kerkwijk} et~al.}{2005}]{2005vanKerkwijk}
{van Kerkwijk} M.~H.,  {Bassa} C.~G.,  {Jacoby} B.~A.,   {Jonker} P.~G.,  2005,
  in {Rasio} F.~A.,  {Stairs} I.~H.,  eds,  Astronomical Society of the Pacific
  Conference Series Vol. 328, Binary Radio Pulsars. p.~357 (\mn@eprint {}
  {astro-ph/0405283})

\bibitem[\protect\citeauthoryear{{van Ommen}, {D'Alessandro}, {Hamilton}  \&
  {McCulloch}}{{van Ommen} et~al.}{1997}]{1997vanOmmen}
{van Ommen} T.~D.,  {D'Alessandro} F.,  {Hamilton} P.~A.,   {McCulloch} P.~M.,
  1997, \mn@doi [\mnras] {10.1093/mnras/287.2.307}, \href
  {http://adsabs.harvard.edu/abs/1997MNRAS.287..307V} {287, 307}

\makeatother
\end{thebibliography}

%%%%%%%%%%%%%%%%%%%%%%%%%%%%%%%%%%%%%%%%%%%%%%%%%%

%%%%%%%%%%%%%%%%% APPENDICES %%%%%%%%%%%%%%%%%%%%%

\clearpage
\appendix

\section{Influence of scintillation on observed pulsar flux densities}
\label{sec:InfluenceOfScinitillationOnObservedPulsarFluxDensities}

We summarise below the formulas from the literature that we use in our theoretical simulation of the influence of scintillation on measured pulsar flux densities in \S\ref{sec:EstimatingTheInfluenceTheoretically}. The variability in flux density can be characterised by the modulation index $m_\nu = \frac{\sigma_{\text{S}, \nu}}{\bar{S_\nu}}$, where $\sigma_{\text{S}, \nu}$ is the standard deviation and $\bar{S_\nu}$ is the mean flux density computed over all measurements for a given pulsar at a centre frequency $\nu$ \citep{2012Lorimer}. There are two scintillation regimes, strong and and weak scintillation, depending on the value of the scintillation strength:
\begin{equation}
	u = \sqrt{\frac{\nu}{\Delta \nu_\text{DISS}}},
	\label{eq:ScintillationStrength}
\end{equation}
where $\nu$ is the observing frequency and $\Delta \nu_\text{DISS}$ is the diffractive scintillation bandwidth, which is the frequency range over which diffractive scintillation is correlated \citep{2002Cordes}. The total modulation index is given by:
\begin{equation}
	m_\text{tot}^2 = \begin{cases}
		m_\text{DISS}^2 + m_\text{RISS}^2 + m_\text{DISS} \: m_\text{RISS}		& \text{if} \: \: u > 1\\
		m_\text{weak}^2														& \text{if} \: \: u < 1\\
  \end{cases}.
  \label{eq:TotalModulationIndex}
\end{equation}
In the case of strong scintillation ($u > 1$) the total modulation index is a combination of the effects of strong diffractive (DISS) and strong refractive interstellar scintillation (RISS) and can be larger than unity \citep{1990Rickett}. For the diffractive part, the modulation index decreases from unity as more scintles are averaged, both in the frequency and time domain; that is,
\begin{equation}
	m_{\text{DISS}} = 1/\sqrt{N_f \: N_t},
	\label{eq:DISSModulationIndex}
\end{equation}
where the number of scintles
\begin{equation}
	N_f \approx 1 + \eta \frac{B}{\Delta \nu_\text{DISS}}; \ \ \ \ N_t \approx 1 + \eta \frac{t}{\Delta t_\text{DISS}}.
	\label{eq:NumberOfScintlesDISS}
\end{equation}
Here $B$ is the observing bandwidth, $\Delta \nu_\text{DISS}$ the diffractive scintillation bandwidth, $t$ the observing time, $\Delta t_\text{DISS}$ the diffractive scintillation time and $\eta = 0.15$ \citep{1991Cordes, 2014Bates}. The modulation index due to RISS can be computed as:
\begin{equation}
	m_\text{RISS} = u^{-1/3}.
	\label{eq:RISSModulationIndex}
\end{equation}
Typically, DISS affects pulsar flux densities on timescales of minutes to hours and RISS affects them on timescales of months to years and becomes measurable only if DISS has been accounted for. Nearly all of our pulsar observations occurred in the strong regime. For weak scintillation ($u < 1$) the modulation index is:
\begin{equation}
	m_\text{weak} = u^{5/6}.
	\label{eq:ModulationIndexWeakScintillation}
\end{equation}
$\Delta \nu_\text{DISS}$ and $\Delta t_\text{DISS}$ depend on the exact line-of-sight to the pulsar through the Galaxy, i.e. its Galactic coordinates, its DM, velocity and the observing frequency.

We compute the scintillation bandwidth and time using the \texttt{NE2001} galaxy model and calculate the expected total modulation index using Eq.~\ref{eq:TotalModulationIndex} for each pulsar, observing frequency and bandwidth. The calculation considers the correct number of frequency sub-bands and observation times to replicate the data. We use the transverse velocity listed in the pulsar catalogue for pulsars for which it is known, otherwise we assume a default value of $100 \: \text{km} \: \text{s}^{-1}$.

\section{Robustness and significance of the spectral index correlations}
\label{sec:RobustnessAndSignificanceOfTheSpectralIndexCorrelations}

Given that $35 \: \%$ of the pulsars are from radio timing observations for the Fermi $\gamma$-ray mission (P574) and that these are nearly all high-$\dot{E}$ pulsars, we tested how robust our findings are with respect to this subset. We calculated the correlation independently for the P574 pulsars and the rest (P875). In the P574 data set most of the correlations are present, in particular the ones with $\dot{\tilde{\nu}}$, $B_\text{LC}$ and $\dot{E}$, but with a reduced significance. In the P875 data set, that contains more pulsars, the correlations are less significant, but visible for some parameters, such as $B_\text{LC}$. We studied the correlations in randomly drawn subsets of the whole data set to understand how the mixture of data affected our conclusions. We focussed on the three most significant correlations, the ones with spin-down rate, $B_\text{LC}$ and $\dot{E}$ for the slow pulsars. We resampled the data without replacement $10^5$ times in each run of varying sample size, calculated the correlation with spectral index and analysed the resulting distribution. The correlations are largely independent of the choice of subset. The uncertainties reduce towards higher sample sizes as expected. In a second test we performed a bootstrap estimation of the uncertainties of the correlation coefficients, i.e. we resampled the whole data set $10^5$ times with replacement and determined the $1 \sigma$ uncertainties from the distribution. The results are $0.43_{-0.05}^{+0.05}$, $0.43_{-0.05}^{+0.05}$ and $0.43_{-0.05}^{+0.05}$ for $\dot{\tilde{\nu}}$, $B_\text{LC}$ and $\dot{E}$ respectively. This agrees well with our previous estimates.

In addition to using the p-values given above as a measure of the significance of the correlations compared with the uncorrelated case, we carried out a Monte Carlo simulation to assess their significance. In particular we drew random samples of spectral index from a shifted log-normal distribution with values of $0.2$, $-4.6$ and $3.0$ for the shape, location and scale parameter, which should resemble the spectral index distribution, see \S\ref{sec:SimplePowerlawSpectra}. For the second simulated data set, which should resemble $\log_{10} \left| x \right|$, we drew random values from a Gaussian distribution with a mean of $-13.6$ and standard deviation of $1.3$ for $\dot{\tilde{\nu}}$, $2.3$ and $1.1$ for $B_\text{LC}$ and $33.4$ and $1.5$ for $\dot{E}$. The sample size was 276 in each case. These values were extracted from fits of a Gaussian distribution to the data for the pulsars with simple power law spectra in our sample. We found that the $\log_{10} \left| x \right|$ distributions are nearly Gaussian with S--W test values of $0.99 \: (0.04)$, $0.99 \: (0.002)$ and $0.99 \: (0.01)$ respectively, although formally we have to reject the normality hypothesis in all cases. The values in brackets are the corresponding p-values. These simulated data sets are uncorrelated. We then tested for chance correlations between the simulated spectral index sample and each of the other ones using the Spearman correlation coefficient. We did that for $10^{10}$ simulated random samples each and defined the random chance probability for correlation as the number of samples for which the Spearman correlation coefficient $r_\text{s}$ reached at least a given value $a$ in a one-tailed test, i.e. by considering the absolute value only. We fit a quadratic function to the chance probabilities:
\begin{equation}
	p_\text{sim} = p ( \left| r_\text{s} \right| \geq a)
	\label{eq:CorrelationChanceProbability}
\end{equation}
from the simulation in logarithmic space and extrapolated $p_\text{sim}$ to a selection of $a$ values that matched the measured correlation values above. The resulting probabilities are $1.65 \cdot 10^{-11}$ (for $\alpha$ vs. $\dot{\tilde{\nu}}$, $a = 0.40$), $2.56 \cdot 10^{-13}$ (for $\alpha$ vs. $B_\text{LC}$, $a = 0.43$) and $2.30 \cdot 10^{-13}$ (for $\alpha$ vs. $\dot{E}$, $a = 0.44$). The values agree with the p-values of the corresponding pulsar parameter correlations in Table~\ref{tab:CorrelationWithAlpha} to within 40 to $80 \: \%$, with the difference arising most likely because of the extrapolation. That means that the p-values are trustworthy and that a correlation between $\alpha$ and the other pulsar parameters by chance is extremely unlikely. We conclude that the correlations that we see in our data are real.

\section{Spectral data}

\begin{table*}
\caption{Band-integrated flux densities at 728, 1382 and 3100 MHz, the spectral classification, the frequency range $\Delta \nu$ the classification was performed over and an eventual spectral index was determined for, the spectral index $\alpha$ for the pulsars that have simple power law spectra and the robust modulation index $m_\text{r}$ at all three centre frequencies for pulsars of which we have at least six measurement epochs. The flux density uncertainties include scintillation and a systematic contribution, in addition to the statistical uncertainty. Upper limits are reported at the $3 \sigma$ level and all other uncertainties at the $1 \sigma$ level.}
\label{tab:FluxDensities}
\begin{tabular}{llllllllll}
\hline
PSRJ	& $S_{728}$ 	& $S_{1382}$ 	& $S_{3100}$	& class    & $\Delta \nu$	    & $\alpha$	& $m_\text{r, 728}$   & $m_\text{r, 1382}$	& $m_\text{r, 3100}$\\
	& [mJy]		& [mJy]		& [mJy]		&		    & [GHz]	    &	    &	&   &\\
\hline
J0401--7608 & $4 \pm 2$ & -- & $0.8 \pm 0.5$ & pl & 0.4 -- 3.4 & $-1.5 \pm 0.5$ & -- & -- & --\\
J0437--4715 & $300 \pm 200$ & $160 \pm 20$ & $30 \pm 30$ & broken pl & 0.08 -- 17.0 & -- & -- & -- & --\\
J0511--6508 & $1.3 \pm 0.7$ & $0.6 \pm 0.2$ & $<0.1$ & -- & 0.7 -- 1.4 & -- & -- & -- & --\\
J0519--6932 & -- & $0.14 \pm 0.04$ & -- & -- & 1.4 -- 1.4 & -- & -- & -- & --\\
J0536--7543 & $11 \pm 7$ & -- & $3 \pm 2$ & pl & 0.4 -- 3.5 & $-1.8 \pm 0.4$ & -- & -- & --\\
J0543+2329 & $25 \pm 2$ & $10.7 \pm 0.7$ & $3.0 \pm 0.2$ & broken pl & 0.1 -- 32.0 & -- & $0.18_{-0.08}^{+0.2}$ & $0.41_{-0.04}^{+0.09}$ & $0.23_{-0.07}^{+0.3}$\\
\end{tabular}
\end{table*}

\begin{table*}
\contcaption{}
\begin{tabular}{llllllllll}
\hline
PSRJ	& $S_{728}$ 	& $S_{1382}$ 	& $S_{3100}$	& class    & $\Delta \nu$	    & $\alpha$	& $m_\text{r, 728}$   & $m_\text{r, 1382}$	& $m_\text{r, 3100}$\\
	& [mJy]		& [mJy]		& [mJy]		&		    & [GHz]	    &	    &	&   &\\
\hline
J0953+0755 & $600 \pm 600$ & $100 \pm 40$ & $20 \pm 20$ & low turn-over & 0.02 -- 10.6 & -- & -- & -- & --\\
J0954--5430 & $0.7 \pm 0.1$ & $0.51 \pm 0.03$ & $0.38 \pm 0.04$ & pl & 0.7 -- 3.4 & $-0.3 \pm 0.2$ & -- & $0.43_{-0.07}^{+0.04}$ & $0.29_{-0.08}^{+0.2}$\\
J0959--4809 & $13 \pm 3$ & $4.0 \pm 0.8$ & $0.5 \pm 0.2$ & lps & 0.2 -- 3.3 & -- & -- & -- & --\\
J1001--5507 & $41 \pm 5$ & $10 \pm 3$ & $2.7 \pm 0.4$ & broken pl & 0.1 -- 3.5 & -- & -- & -- & --\\
J1001--5559 & $3.4 \pm 0.8$ & $1.4 \pm 0.4$ & $0.11 \pm 0.03$ & pl & 0.7 -- 3.1 & $-2.4 \pm 0.4$ & -- & -- & --\\
J1003--4747 & $4.8 \pm 0.9$ & $1.43 \pm 0.08$ & $0.3 \pm 0.04$ & pl & 0.3 -- 3.1 & $-1.4 \pm 0.1$ & $0.5_{-0.3}^{+0.2}$ & $0.32_{-0.03}^{+0.07}$ & $0.29_{-0.07}^{+0.2}$\\
\end{tabular}
\end{table*}

\begin{table*}
\contcaption{}
\begin{tabular}{llllllllll}
\hline
PSRJ	& $S_{728}$ 	& $S_{1382}$ 	& $S_{3100}$	& class    & $\Delta \nu$	    & $\alpha$	& $m_\text{r, 728}$   & $m_\text{r, 1382}$	& $m_\text{r, 3100}$\\
	& [mJy]		& [mJy]		& [mJy]		&		    & [GHz]	    &	    &	&   &\\
\hline
J1156--5909 & $<40.0$ & -- & $<0.01$ & -- & 0.4 -- 0.4 & -- & -- & -- & --\\
J1157--6224 & $54 \pm 9$ & $14 \pm 3$ & $1.8 \pm 0.4$ & pl & 0.4 -- 3.5 & $-2.4 \pm 0.1$ & -- & -- & --\\
J1159--7910 & -- & $0.7 \pm 0.2$ & -- & pl & 0.4 -- 1.5 & $-2.0 \pm 1.0$ & -- & -- & --\\
J1202--5820 & $9 \pm 2$ & $3.0 \pm 0.8$ & $0.7 \pm 0.2$ & pl & 0.4 -- 3.3 & $-1.8 \pm 0.3$ & -- & -- & --\\
J1210--5559 & $10 \pm 2$ & $2.3 \pm 0.6$ & $0.27 \pm 0.07$ & pl & 0.4 -- 3.3 & $-2.4 \pm 0.3$ & -- & -- & --\\
J1216--6223 & $<0.4$ & $0.23 \pm 0.01$ & $0.061 \pm 0.006$ & -- & 1.4 -- 3.1 & -- & -- & $0.19_{-0.04}^{+0.05}$ & --\\
\end{tabular}
\end{table*}

\begin{table*}
\contcaption{}
\begin{tabular}{llllllllll}
\hline
PSRJ	& $S_{728}$ 	& $S_{1382}$ 	& $S_{3100}$	& class    & $\Delta \nu$	    & $\alpha$	& $m_\text{r, 728}$   & $m_\text{r, 1382}$	& $m_\text{r, 3100}$\\
	& [mJy]		& [mJy]		& [mJy]		&		    & [GHz]	    &	    &	&   &\\
\hline
J1515--5720 & $1.04 \pm 0.07$ & $0.25 \pm 0.01$ & $0.08 \pm 0.01$ & pl & 0.7 -- 3.1 & $-2.0 \pm 0.2$ & $0.13_{-0.03}^{+0.2}$ & $0.25_{-0.04}^{+0.04}$ & --\\
J1522--5829 & $30 \pm 6$ & -- & $5.0 \pm 1.0$ & broken pl & 0.6 -- 8.4 & -- & -- & -- & --\\
J1524--5625 & $1.8 \pm 0.5$ & $1.28 \pm 0.07$ & $0.55 \pm 0.03$ & pl & 0.7 -- 3.4 & $-0.9 \pm 0.2$ & -- & $0.18_{-0.01}^{+0.02}$ & $0.13_{-0.05}^{+0.04}$\\
J1524--5706 & $0.17 \pm 0.04$ & $0.45 \pm 0.02$ & $0.14 \pm 0.02$ & pl & 0.7 -- 3.1 & $-1.3 \pm 0.4$ & -- & $0.21_{-0.02}^{+0.04}$ & $0.4_{-0.2}^{+0.2}$\\
J1527--5552 & $<200.0$ & -- & $<8.0$ & -- & 0.4 -- 1.4 & -- & -- & -- & --\\
J1530--5327 & $1.6 \pm 0.4$ & $0.92 \pm 0.06$ & $0.41 \pm 0.06$ & pl & 0.7 -- 3.1 & $-0.9 \pm 0.4$ & -- & $0.47_{-0.04}^{+0.05}$ & $0.5_{-0.2}^{+0.2}$\\
\end{tabular}
\end{table*}

\begin{table*}
\contcaption{}
\begin{tabular}{llllllllll}
\hline
PSRJ	& $S_{728}$ 	& $S_{1382}$ 	& $S_{3100}$	& class    & $\Delta \nu$	    & $\alpha$	& $m_\text{r, 728}$   & $m_\text{r, 1382}$	& $m_\text{r, 3100}$\\
	& [mJy]		& [mJy]		& [mJy]		&		    & [GHz]	    &	    &	&   &\\
\hline
J1658--4958 & $3.7 \pm 0.8$ & -- & $0.23 \pm 0.06$ & lps & 0.7 -- 3.3 & -- & -- & -- & --\\
J1700--3312 & $3.9 \pm 0.9$ & -- & $0.31 \pm 0.08$ & pl & 0.4 -- 3.1 & $-1.8 \pm 0.3$ & -- & -- & --\\
J1701--3726 & $14 \pm 2$ & -- & $0.9 \pm 0.2$ & pl & 0.6 -- 3.4 & $-2.0 \pm 0.2$ & -- & -- & --\\
J1701--4533 & $5.2 \pm 0.7$ & -- & $0.6 \pm 0.1$ & pl & 0.6 -- 3.3 & $-1.4 \pm 0.2$ & -- & -- & --\\
J1702--4128 & $<1.0$ & $1.17 \pm 0.07$ & $0.94 \pm 0.06$ & pl & 1.4 -- 6.5 & $-0.2 \pm 0.2$ & -- & $0.27_{-0.02}^{+0.04}$ & $0.14_{-0.04}^{+0.03}$\\
J1702--4306 & -- & $0.46 \pm 0.04$ & -- & -- & 1.3 -- 1.4 & -- & -- & -- & --\\
\end{tabular}
\end{table*}

\begin{table*}
\contcaption{}
\begin{tabular}{llllllllll}
\hline
PSRJ	& $S_{728}$ 	& $S_{1382}$ 	& $S_{3100}$	& class    & $\Delta \nu$	    & $\alpha$	& $m_\text{r, 728}$   & $m_\text{r, 1382}$	& $m_\text{r, 3100}$\\
	& [mJy]		& [mJy]		& [mJy]		&		    & [GHz]	    &	    &	&   &\\
\hline
J1759--2205 & -- & -- & $0.5 \pm 0.1$ & pl & 0.4 -- 4.9 & $-1.8 \pm 0.2$ & -- & -- & --\\
J1759--3107 & $4.0 \pm 0.7$ & -- & $0.38 \pm 0.09$ & pl & 0.7 -- 3.3 & $-1.6 \pm 0.2$ & -- & -- & --\\
J1801--2154 & -- & $0.21 \pm 0.02$ & $0.069 \pm 0.009$ & -- & 1.4 -- 3.1 & -- & -- & -- & --\\
J1801--2304 & -- & $7.0 \pm 0.5$ & $1.6 \pm 0.1$ & pl & 1.3 -- 6.5 & $-1.7 \pm 0.1$ & -- & $0.19_{-0.03}^{+0.05}$ & --\\
J1801--2451 & -- & $1.46 \pm 0.09$ & $0.22 \pm 0.03$ & pl & 0.6 -- 3.1 & $-0.8 \pm 0.2$ & -- & $0.14_{-0.05}^{+0.05}$ & --\\
J1801--2920 & $7.0 \pm 1.0$ & -- & $0.42 \pm 0.08$ & pl & 0.6 -- 3.3 & $-2.0 \pm 0.2$ & -- & -- & --\\
\end{tabular}
\end{table*}

\begin{table*}
\contcaption{}
\begin{tabular}{llllllllll}
\hline
PSRJ	& $S_{728}$ 	& $S_{1382}$ 	& $S_{3100}$	& class    & $\Delta \nu$	    & $\alpha$	& $m_\text{r, 728}$   & $m_\text{r, 1382}$	& $m_\text{r, 3100}$\\
	& [mJy]		& [mJy]		& [mJy]		&		    & [GHz]	    &	    &	&   &\\
\hline
J1839--0643 & $<2.0$ & -- & $0.56 \pm 0.09$ & pl & 1.3 -- 3.3 & $-1.5 \pm 0.3$ & -- & -- & --\\
J1839--0905 & -- & $0.22 \pm 0.02$ & $0.11 \pm 0.02$ & -- & 1.4 -- 3.1 & -- & -- & $0.15_{-0.07}^{+0.08}$ & --\\
J1840--0809 & $5.0 \pm 0.8$ & -- & $0.8 \pm 0.1$ & pl & 0.7 -- 3.5 & $-1.5 \pm 0.1$ & -- & -- & --\\
J1840--0815 & $9 \pm 2$ & -- & $0.5 \pm 0.1$ & pl & 0.7 -- 3.3 & $-2.3 \pm 0.2$ & -- & -- & --\\
J1841--0425 & -- & $3.3 \pm 0.2$ & $0.72 \pm 0.09$ & pl & 0.6 -- 3.4 & $-1.7 \pm 0.1$ & -- & $0.04_{-0.01}^{+0.01}$ & --\\
J1841--0524 & -- & $0.2 \pm 0.04$ & $0.048 \pm 0.007$ & -- & 1.4 -- 3.1 & -- & -- & $0.43_{-0.3}^{+0.04}$ & --\\
\end{tabular}
\end{table*}

\begin{table*}
\contcaption{}
\begin{tabular}{llllllllll}
\hline
PSRJ	& $S_{728}$ 	& $S_{1382}$ 	& $S_{3100}$	& class    & $\Delta \nu$	    & $\alpha$	& $m_\text{r, 728}$   & $m_\text{r, 1382}$	& $m_\text{r, 3100}$\\
	& [mJy]		& [mJy]		& [mJy]		&		    & [GHz]	    &	    &	&   &\\
\hline
J2053--7200 & $11 \pm 7$ & -- & $0.3 \pm 0.2$ & broken pl & 0.1 -- 3.3 & -- & -- & -- & --\\
J2108--3429 & $2.0 \pm 1.0$ & -- & -- & -- & 0.4 -- 0.7 & -- & -- & -- & --\\
J2145--0750 & $10 \pm 10$ & $0.8 \pm 0.4$ & $2 \pm 2$ & pl & 0.1 -- 5.0 & $-1.8 \pm 0.1$ & -- & -- & --\\
J2222--0137 & $1.0 \pm 1.0$ & -- & $0.5 \pm 0.7$ & -- & 0.7 -- 3.1 & -- & -- & -- & --\\
J2317+1439 & -- & -- & $<0.02$ & pl & 0.1 -- 1.4 & $-1.3 \pm 0.4$ & -- & -- & --\\
\hline
\end{tabular}
\end{table*}

%%%%%%%%%%%%%%%%%%%%%%%%%%%%%%%%%%%%%%%%%%%%%%%%%%

% Don't change these lines
\bsp	% typesetting comment
\label{lastpage}
\end{document}